\documentclass[11pt]{article}
\usepackage{graphicx}
\usepackage{tikz}
\usepackage[english]{babel}
\usepackage[T1]{fontenc}
\usepackage{hyperref}
\usepackage{amstext,amsmath}
\usepackage{dcolumn,booktabs}
\usepackage{amsfonts}
\usepackage{float}
\usepackage{comment}
\usepackage{listings}
\usepackage{url}
\usepackage{longtable}
\usepackage[tableposition=b]{caption}
\usepackage{color}
\usepackage{graphicx}
\usepackage[margin=1in]{geometry}
\usepackage{subcaption}
\usepackage{amssymb}
\usepackage{etex}
\usepackage{pstricks}
\usepackage{pst-plot}
\usepackage{pstricks-add}
\usepackage[round]{natbib}   
\usepackage[T1]{fontenc}
\usepackage{setspace}
\usepackage{bbm}
\usepackage{graphicx,floatpag,fancyhdr}
\fancypagestyle{plainlower}{	
	\fancyhf{}
	\fancyfoot[C]{\raisebox{-6\baselineskip}{\thepage}}
	
}
\usepackage{lscape} 
\usepackage{placeins}
\usepackage{tabularx}
\defcitealias{hkp}{hkp, 2021}

\newlength{\bibitemsep}
\newlength{\bibparskip}
\let\oldthebibliography\thebibliography
\renewcommand\thebibliography[1]{%
	\oldthebibliography{#1}%
	\setlength{\parskip}{\bibitemsep}%
	\setlength{\itemsep}{\bibparskip}%
}

\graphicspath{{/Users/ingridhagele/Dropbox/My_Documents_Dropbox/Research/2_GenderGap/2021_Paper/2022_07_arxiv/Graphs/}}

\usepackage{listings}
\usepackage[toc,page]{appendix}

\usepackage{amsfonts, amsmath, amsthm, amssymb}

\theoremstyle{definition}

\usepackage{hyperref}
\begin{document}
\title{Talent Hoarding in Organizations}
\author{Ingrid Haegele\footnote{Ludwig-Maximilians University; ingrid.haegele@econ.lmu.de. I thank David Card, Patrick Kline, and Christopher Walters for their advice and encouragement.  This paper has benefited tremendously from comments and suggestions from  Sydnee Caldwell, Florian Englmaier, Robert Gibbons, Lisa Kahn, Supreet Kaur, Danielle Li, Conrad Miller, Evan Rose, Jesse Rothstein, Steve Tadelis, Dmitry Taubinsky, Ricardo Perez-Truglia, Fabian Waldinger,  Francis Wong, and numerous seminar and conference participants. I thank Leon Jacob and the employees of my institutional partner for their feedback and support. This work was supported by the Alfred P. Sloan Foundation Pre-doctoral Fellowship on the Economics of an Aging Workforce awarded through the NBER,  IRLE Student Research Grant, O-LAB Labor Science Initiative Grant, Strandberg Grant for Gender in Economic Research, UC Berkeley Matrix Dissertation Fellowship, and Washington Center for Equitable Growth Doctoral Grant.}}
	\date{July 2025}
	
	\maketitle
	\thispagestyle{empty}
  Most organizations rely on managers to identify talented workers. However, managers who are evaluated on team performance have an incentive to hoard workers. This study provides the first empirical evidence of talent hoarding using personnel records and survey evidence from a large firm. Talent hoarding is self-reported by three-fourths of managers, is detectable in manager ratings of worker talent, and occurs more frequently under stronger hoarding incentives, proxied by performance-related pay, team size, and talent visibility. Using quasi-random exposure to talent hoarding, I show that hoarding deters internal job applications, inhibiting career progression and altering talent allocation in the firm.

	\clearpage
	\pagenumbering{arabic}
	\newpage
	
	\section{Introduction}
	\vspace{-0.3cm}
	Firms must continually decide how to allocate workers to jobs, a process which has critical implications for productivity (\citealp{rosen1982}, \citealp{holmstromtirole}). Because it is difficult to perfectly observe worker ability, most firms rely on managers to identify talented workers who can be promoted to higher-level positions. However, when a talented worker leaves their team for a new position,  team performance suffers. Since managers are rewarded based on team performance and firms cannot perfectly monitor manager actions, the conflicting interests of manager and firm create the potential for moral hazard (\citealp{holmstrom1979}). A growing body of evidence documents that workers in high-level positions have large impacts on firm performance (\citealp{BloomVanReenen}, \citealp{lazear2015value}), implying that managers may create significant efficiency costs if they hoard talented workers rather than promote them.
	
	Ample anecdotal and survey evidence points to widespread talent hoarding in organizations. In a global survey, half of organizations report that managers hoard talent by discouraging worker mobility (\citealp{i4cp}).  A US-based survey finds that workers in one-third of firms feel the need to keep internal applications secret from their managers out of fear of retaliation (\citealp{kornferry}). In Germany, 83\% of the top publicly listed companies cite managerial talent hoarding as a key friction in their organization (\citetalias{hkp}).\footnote{Talent hoarding also occurs in science and academia. \cite{zuckerman} documents how Katalin Karik\'o, a seminal developer of mRNA vaccines and 2023 Nobel prize winner, experienced talent hoarding when she decided to leave her lab for a new position. Her advisor Robert Suhadolnik vowed ``to do whatever he could to stop his prot\'eg\'ee from leaving...he made it clear she had two career choices. `You can work in my lab or go home,' he told her. Suhadolnik followed through on this threat, telling a local immigration office that she was living in the country illegally and should be deported.''} Despite the apparent prevalence of talent hoarding and its likely detrimental consequences, very little empirical evidence on talent hoarding exists in economics. Studying talent hoarding empirically is challenging. Managers often hoard talent through hidden actions that are difficult to observe, even in rich datasets. Furthermore, identifying the causal impacts of talent hoarding requires plausibly exogenous variation in hoarding. 
	
	This study provides the first empirical evidence on talent hoarding in organizations. I combine rich personnel records from a large manufacturing firm with survey evidence from both managers and workers in the firm. I document widespread managerial incentives to hoard talent: the majority of managers perceive a conflict of interest with respect to developing talent, because more developed workers are more likely to leave the team. Talent hoarding is reported by 75\% of managers. Using a direct measure of talent hoarding derived from the firm's personnel records, I test  predictions from my conceptual framework and confirm that stronger hoarding incentives lead to more talent hoarding. I leverage quasi-random variation in workers' exposure to talent hoarding to demonstrate the impacts of talent hoarding on workers' career outcomes. Hoarding significantly deters workers' applications for internal positions, many of which would have been successful. Because talent hoarding has a larger impact on high-quality workers, hoarding reduces the quality of the applicant pool, thus restricting the firm's ability to effectively allocate talent within the organization.

	I develop a simple conceptual framework that formalizes the misaligned incentives that give rise to talent hoarding. Since managers are both tasked with identifying productive workers and are rewarded based on team productivity, managers have an incentive to hoard talent by preventing workers from seeking new positions outside the team. This framework yields empirical predictions for both manager and worker behavior. For example, when gains to hoarding are larger or worker departure from the team is costlier, managers engage in more hoarding. High-productivity workers who experience more hoarding are more deterred from applying, which negatively impacts internal applicant pools and has the potential to distort the allocation of talent within the firm.  
	
	To empirically test for talent hoarding, I collect a unique combination of personnel records, internal job application data, and new survey evidence from a large manufacturing firm that employs over 200,000 workers. I focus my analysis on the firm's largest internal labor market, consisting of over 30,000 white-collar and management employees in Germany. I demonstrate that the firm is similar to other large firms in terms of its workforce and organizational design. As in many other large firms, employees who want to switch to a new position or to be promoted are required to apply for the position. Since most workers in the same team work in similar positions, workers who seek different positions, including promotions, typically need to apply to jobs outside the team.
	
	The data I assemble offer three key advantages. First, collecting new survey evidence from managers in the firm allows me to directly document that managers face misaligned incentives and act accordingly. Second, the firm's personnel records allow me to construct a measure of talent hoarding using the difference between managers' private and public ratings of worker talent. Without such data, directly measuring managerial talent hoarding is empirically challenging because hoarding by definition involves hidden actions. Third, by combining personnel records with the universe of application and hiring decisions at the firm, I am able to document the extent to which talent hoarding deters workers' high-stakes applications that would otherwise have been successful. 
	
	Survey responses from 62\% of managers in the firm reveal that incentives to hoard talent are widespread. The majority of managers (55\%) explicitly acknowledge that supporting workers' career development entails a conflict of interest because more developed workers are more likely to leave the team. This conflict of interest relates to the misaligned incentives managers face: while almost all managers (96\%) believe they would have a large impact on workers' careers, only 36\% perceive this impact to be valued by the firm. A further disincentive to develop talent is the risk of workers leaving the team: 68\% of managers state that they would be more likely to support workers' career development if it would be easier to replace lost talent. The survey demonstrates that managerial talent hoarding actions, which are reported by 75\% of managers, are a common response to these prevailing managerial incentives.
	
	Managers' revealed actions provide further evidence that managers purposefully hoard talent. To capture talent hoarding in the firm data, I construct a direct hoarding measure based on managers' systematic suppression of worker visibility using the difference between the private performance ratings and public potential ratings that managers assign to workers. The ratings differ in the extent of their visibility outside of a worker's team, meaning that managers can suppress worker visibility by lowering the public rating relative to the private rating. Such suppression of visibility is cited as a key form of talent hoarding in both managers' and workers' survey responses. Testing predictions from my conceptual framework based on three proxy measures for managerial hoarding incentives (i.e., performance-related pay, team size, and talent visibility), I demonstrate that managers are indeed more likely to engage in talent hoarding if they face greater incentives to hoard.  Additional evidence from alternative measures of talent hoarding based on managers restricting access to high-visibility trainings and managers' explicit hoarding statements from the survey further corroborates this finding.

These manager-level findings establish a clear link between managerial incentives and hoarding actions. I then examine whether these behaviors affect workers. To test whether managerial hoarding has meaningful consequences for worker outcomes, 
I use quasi-random variation in workers’ exposure to talent hoarding based on manager rotations.   For workers whose manager learns that they will move to a new position on a different team, this creates a temporary window of time during which their manager does not have an incentive to hoard talent. Consistent with the cessation of talent hoarding, I find that manager rotations increase worker applications by 78\%. To support the interpretation of rotation effects as being driven by talent hoarding, I test predictions from the conceptual framework, which predicts that workers with high productivity, those whose replacement is costlier, and those working under managers with lower utility costs of hoarding should exhibit larger responses to manager rotations. Coupled with a series of tests for potentially confounding mechanisms, such as worker-manager specific match effects and finite project timelines, these empirical exercises support my interpretation based on talent hoarding and motivate the use of manager rotations to measure the impacts of talent hoarding on worker outcomes. 
	
Talent hoarding has important consequences for workers' careers and the firm's ability to allocate talent. To quantify how successful deterred applications would have been, I instrument for workers' applications with manager rotations. Marginal applicants (identified via standard IV assumptions), who would not have applied in absence of a manager rotation and only apply because their manager rotates, face a 49.1\% likelihood of landing a new position. This large and positive hiring likelihood suggests that many deterred applicants would have been successful and is particularly striking given that the average hiring likelihood is 27.6\%. The positive selection of marginal applicants is further corroborated by a complier analysis \citep{abadie2003}, which implies that talent hoarding has negative impacts on the composition of the applicant pool. Using workers' performance ratings, I present evidence suggesting that a substantial share of deterred applicants would have been highly productive in their new positions. Together, my findings suggest that in the presence of talent hoarding, firms may forgo productivity gains from hoarded workers who are not assigned to positions in which they would have been productive.
	
Talent hoarding may also exacerbate between-group inequality at the firm. A survey of the firm's employees suggests that male and female workers may react differently to  talent hoarding. In line with the literature on gender differences in preferences (\citealp{bertrand2011}), the survey finds that women in the firm  are 22\% more likely to place a high value on preserving a good relationship with their manager and are 26\% more likely to rely on managers' career guidance when making career decisions. The effects of manager rotations corroborate that talent hoarding has differential effects by gender, affecting women at a higher part of the quality distribution compared to men. In particular, marginal female applicants are more qualified in terms of their educational qualifications and past performance. They are also more positively selected with respect to their hiring probability for higher-level positions. When comparing potential outcomes for marginal applicants, I find that both men and women experience higher earnings if they choose to apply in the absence of talent hoarding; however, the larger gains realized by women lead to a reduction in the pay gap of 86\%. 
	
A number of factors suggest that talent hoarding is very likely to manifest similarly in other organizations. The firm I study is similar to other large firms in Germany both in terms of the characteristics of its workforce and its internal organization, in that it is standard that managers are tasked with identifying talented workers, but are neither monitored nor rewarded in doing so (\citetalias{hkp}). Companies across the world report that talent hoarding is commonplace, creates barriers to talent allocation, and occurs through many of the same managerial behaviors that are documented in this study (\citealp{i4cp}, \citealp{kornferry}, \citealp{forbes2015}, \citealp{sulllivan2017}).
	
	This study contributes to two broad strands of research on organizations. Prior theoretical research has hypothesized that managers engage in self-interested behavior (\citealp{holmstromtirole}), largely focusing on managers' misaligned incentives in the context of biased performance evaluations (\citealp{milgromroberts}, \citealp{prendergastfavoritism},  \citealp{fairburnmalcomson}). While research studying internal labor markets has documented the importance of incentive provision in organizations (\citealp{gibbons1999careers}, \citealp{prendergast1999}), little attention has been paid to how managers' incentive problems may affect the efficiency of job assignments. One notable exception is theoretical work by \cite{friebelraith} who show that different organizational designs may change managers' incentives to train subordinates and accurately represent their abilities. My study provides the first empirical demonstration of a costly moral hazard problem among managers that likely affects the efficient allocation of talent within organizations. 
	
	Second, a large empirical literature in economics studies the impacts of managers on firm outcomes. The majority of this literature has focused on upper management, and in particular on documenting the impacts of CEOs on firm performance (\citealp{bertrandschoar}, \citealp{bennedsen2007inside}, \citealp{malmendier2009superstar}, \citealp{bennedsen}). An emerging body of work uses detailed data on managers and workers to show that even managers at lower levels of the firm hierarchy have large impacts on worker outcomes, including worker productivity (\citealp{lazear2015value}, \citealp{frederiksen2020supervisors}, \citealp{fenizia2019}), turnover (\citealp{hoffmantadelis}), and career progression (\citealp{KunzeMiller}, \citealp{CullenPerezTruglia}, \citealp{Benson2021potential}). This study adds to these findings by uncovering talent hoarding as an important mechanism that influences managers' impacts on firms and workers. By demonstrating that talent hoarding has meaningful impacts on career progression, this study also contributes to both theoretical and empirical work seeking to understand the dynamics of internal labor markets (\citealp{waldman1984job}, \citealp{milgromoster}, \citealp{bgh}, \citealp{bensonpeter}, \citealp{Huitfeldt2021}).

	The rest of the paper proceeds as follows. Section \ref{sec:background} provides anecdotal evidence and introduces a simple framework that offers a formal definition of talent hoarding. Section \ref{sec:settingdata} describes the institutional setting and data.  Section \ref{sec:managers} focuses on managers and provides direct evidence on the hoarding incentives they face and the actions that follow. Section \ref{sec:instrument} uses quasi-random variation centered around manager rotations to document the impact of managerial talent hoarding on workers. Section \ref{sec:results_misallocation} presents evidence on the consequences of talent hoarding for workers' careers and the firm's ability to allocate talent. Section \ref{sec:conclusion} concludes. 
	\vspace{-0.3cm}
	\section{Background and Conceptual Framework}
	\label{sec:background}
	\vspace{-0.3cm}
	This section presents anecdotal evidence of how talent hoarding occurs in practice. I conduct a large-scale survey at the firm I study, which illustrates the reasons why and the actions through which managers hoard talent. Building on this intuition, a simple conceptual framework offers a formal definition of talent hoarding as well as a set of predictions that guides my empirical analysis.
	
	What is talent hoarding? A 2017 Wall Street Journal article describes talent hoarding as ``manager’s natural tendency to hold on to top performers instead of working to promote them or transfer them to other areas of the company.'' (\citealp{wallstreet2017}). Media outlets also describe how managers hoard talent. A 2015 Forbes article observes that managers who hoard talent ``never recommend...people for a promotion in another department.'' (\citealp{forbes2015}). The industry publication Talent Management \& HR writes in 2017 that ``hoarding managers, in order to reduce the internal visibility of their top team members, may purposely restrict them from serving on task forces and outside-of-function committees'' (\citealp{sulllivan2017}). Other  hoarding strategies are described as underrating potential for higher-level positions or threatening workers who try to leave the team. 
	
	Talent hoarding is widespread and occurs in a variety of settings. In a survey of 665 global organizations, covering both the private and public sector, half of organizations report that their managers hoard talent by discouraging worker mobility (\citealp{i4cp}). In Germany, 83\% of the top publicly listed companies cite talent hoarding as a crucial friction in their organization (\citetalias{hkp}). Managers' talent hoarding actions seem consequential.  In one-third of US firms, workers feel the need to keep applications secret from their managers out of fear of retaliation (\citealp{kornferry}).
	\vspace{-0.3cm}
	\subsection{Evidence from an Employee Survey within the Firm}
	\label{sec:survey}
	\vspace{-0.2cm}
	To provide the first detailed evidence on the dynamics of talent hoarding, I conduct a large-scale survey at the firm I study. Employees described challenges regarding their internal career progression both in the form of  multiple-choice answers  and in free-text responses.\footnote{For questions requiring free-text responses, I exclude respondents from the analysis who did not provide any answer.} The survey received over 15,000 responses, yielding a 50\% response rate. Section \ref{sec:employee_svy_descr} provides details about the survey implementation and response patterns. Appendix Section \ref{sec:app_survey} presents the survey instrument.  
	
	Respondents report a variety of different actions through which managers hoard talent, which include suppressing public signals of worker ability, restricting access to high-visibility projects or trainings, and explicitly discouraging workers from applying to internal positions. Appendix Table \ref{table:survey_talenthoarding} provides selected quotations. In addition, when asked to state the biggest challenge to their internal career progression, the modal answer (provided by 22\% of workers who respond to this question) is managers' limited support for career progression, such as refusal to assist in career planning and restricted access to trainings that would increase workers' visibility outside the team. 
	
	Managers' actions appear to strongly influence workers' application decisions. 41\% of respondents indicate that they are afraid to apply to internal positions, fearing negative repercussions if managers find out about the application. 25\% of workers state that they feel the need to ask for managers' permission before applying for an internal job opening, even though the firm's internal policies are meant to enable workers to initiate an application on their own. 16\% of respondents indicate that applying away from the team is seen as disloyal. These findings suggest that fear of retaliation represents a key dimension through which talent hoarding deters worker applications. 
	\vspace{-0.3cm}
	\subsection{A Simple Framework of Talent Hoarding}
	\label{sec:framework}
	\vspace{-0.2cm}
	To formally define talent hoarding, consider a firm that employs two types of agents, managers $m$ and workers $i$. For simplicity, teams are composed of one manager and one worker. Workers are characterized by a marginal productivity $\alpha_i$ drawn from some known distribution $G$. The firm seeks to efficiently allocate talent to maximize total firm productivity by choosing which workers to promote to managerial positions. Consistent with \cite{rosen1982}, productivity is maximized when high-ability workers are assigned to high-level positions. Thus, in the absence of any constraints, the firm would fill a new managerial vacancy with the most productive worker and would fill the worker's vacated position by hiring a worker from outside of the firm (i.e., a random draw from $G$). 
	
	In practice, firms neither perfectly observe worker productivity, nor do they know which workers would accept a promotion. Accordingly, managers are tasked with identifying high-productivity workers and encouraging them to seek promotions. However, if a high-productivity worker is promoted and leaves their team, that team incurs a productivity loss. Managers, who observe the productivity of their workers, are compensated according to team productivity, creating a conflict of interest between the firm and managers. Talent hoarding is defined as the actions taken by managers that lower the likelihood that a worker applies for and receives a promotion.\footnote{Respondents to the survey discussed in Section \ref{sec:survey} report that managers can deter workers from applying by explicitly discouraging or threatening them, underrating worker ability, and restricting access to high-visibility projects or training. In theory, the firm could offer managers a promotion-contingent contract to resolve the misaligned incentives. In practice, firms generally do not compensate managers for promoting their workers, plausibly because of the practical challenges associated with these contracts (discussed in detail in \citealp{friebelraith}).}
	
	The one-period framework proceeds as follows. A managerial vacancy opens exogenously. $M$ managers observe the productivity of the worker on their teams and decide to what extent to engage in talent hoarding. Based on managers' choices, workers decide whether to apply for a promotion. The firm observes noisy signals of worker productivity (e.g., by conducting interviews) and chooses the worker with the highest signal to fill the vacancy. The promoted worker's previous position is replaced with an outside hire. Team productivity is realized and managers are compensated.
	
	Let $\beta\in [0,\infty)$ index the degree of talent hoarding chosen by a manager, with 0 representing no talent hoarding.\footnote{Survey responses presented in Appendix Table \ref{table:survey_talenthoarding} indicate that suppression of potential ratings and pressure to refrain from applying are common examples of talent hoarding that can be represented by $\beta$. In Section \ref{sec:th_measure}, I construct a direct measure of talent hoarding by comparing the private performance ratings to the public potential ratings that managers assign to workers, where $\beta$ can be interpreted as reflecting the disparity between these signals.} Let $q(\alpha_i,\beta)$ denote the equilibrium probability that a worker with productivity $\alpha_i$ gets promoted, conditional on applying for a promotion. This promotion probability increases with worker productivity ($\frac{\partial q}{\partial \alpha_i}>0 $), but decreases in the level of talent hoarding ($\frac{\partial q}{\partial \beta}<0$), and reflects the noisy signal received by the firm. Thus, one can interpret talent hoarding as impacting workers through the likelihood that they get promoted.\footnote{In practice, workers report that managers diminish their visibility, thus lowering their promotion prospects. In theory, talent hoarding can also operate through the cost of applying, which would yield similar predictions.} Furthermore, I assume that the effect of talent hoarding on the conditional promotion probability is larger for more productive workers ($\frac{\partial^2 q}{\partial \beta \partial \alpha_i}<0$). This assumption relates to the noisy signals of applicant productivity observed by the firm. Intuitively, since low-productivity workers are less likely than high-productivity workers to generate a favorable signal, there is less scope for talent hoarding to lower the likelihood that a low-productivity worker gets chosen among multiple applicants. One example in which this would be the case is if a firm employs a two-part screening strategy in which it chooses a subset of applicants to interview based on the CVs of all applicants. A very low-productivity worker may never clear the bar to be interviewed. Thus, the manager can do little to further lower their hiring likelihood.
	
	Workers decide whether to apply by weighing expected costs and benefits. Let $b$ denote the benefits of a successful application and $c$ denote the costs of applying. Workers apply if
	\vspace{-2mm}
	\begin{align}\label{eq:wrk_rule}
		q(\alpha_i,\beta) b &\geq c + \varepsilon_i
	\end{align}
	
	\noindent where $\varepsilon_i \sim \Psi$ captures worker-specific heterogeneity, with $\Psi$ known to the manager. Therefore, from the manager's perspective, the probability that the worker leaves the team is given by 
	\vspace{-2mm}
	\begin{align*}
		p(\alpha_i,\beta) = q(\alpha_i, \beta)\Psi\big(q(\alpha_i,\beta)b -c\big)
	\end{align*}
	
	\noindent It follows that talent hoarding reduces the probability that workers leave the team (i.e., $\frac{\partial p}{\partial \beta}<0$).\footnote{For simplicity, this framework does not distinguish between different worker types. Talent hoarding may exacerbate between-group differences in promotions if it has a differential effect on workers' application decisions.}
	
	Managers optimize by choosing their level of talent hoarding $\beta$. If a worker leaves the team for a promotion, the firm hires a worker of unknown ability ($\alpha_j \sim G(.)$ with $E[\alpha_j]=\bar{\alpha}$) from outside the firm. Consequently, a high-productivity worker getting promoted out of the team is likely to decrease team productivity. Without compensation for promoting workers, managers have an incentive to engage in talent hoarding by reducing workers' likelihood of promotion. However, managers incur increasing and convex costs from talent hoarding, which vary in their magnitude by manager according to the parameter $\phi_m>0$.\footnote{The utility costs that managers experience when hoarding talent can be interpreted as consequence of manager altruism, in line with \cite{hoffman2018discretion} who motivate why managers might value making hiring decisions in the interest of the firm and not in their best self-interest. Alternatively, such costs could arise from the probability of detection, for instance in the form of reputation costs.} Thus, managers solve the following problem:
	\vspace{-2mm}
	\begin{align*} \max_{\beta} \hspace{1mm} (1-p(\alpha_i,\beta))\alpha_i  +  p(\alpha_i,\beta)\bar{\alpha} - \frac{\phi_m}{2}\big(p(\alpha_i,0)-p(\alpha_i,\beta)\big)^2 
	\end{align*}	
	
	\noindent This optimization problem yields the following first-order condition, which provides a formal definition of talent hoarding as well as predictions with respect to the realized level of talent hoarding: 
	\begin{align*} 
		p(\alpha_i,0)- p(\alpha_i,\beta^*) = \frac{1}{\phi_m} (\alpha_i-\bar{\alpha})
	\end{align*} 
	
	\noindent	\textbf{Definition.}		(Talent hoarding) When worker $i$'s productivity exceeds the expected productivity of an outside hire, the manager optimally hoards talent by choosing $\beta^*>0$. As a result, the likelihood that a worker leaves the team is lower relative to $\beta=0$. \\
	\vspace{-0.2cm}
	
	\noindent This definition yields following predictions  for managers' talent hoarding actions: 
	
		\vspace{0.3cm}
	
	\noindent	\textbf{Prediction 1.}	
	(Worker heterogeneity) \\ High-productivity workers (i.e., high $\alpha_i$) experience more talent hoarding.\footnote{In many related models (e.g., \citealp{garicano}), ability is multi-dimensional. To the extent that productivity on the worker’s current team $\alpha_i$ is positively related with performance external to the current team (e.g., leading a team), managers will have an incentive to hoard talent that conflicts with the firm’s objectives.}
	
	\vspace{0.3cm}
	
	\noindent	\textbf{Prediction 2.}	
	(Team heterogeneity) \\
	Workers that are more difficult to replace (i.e., if $\bar{\alpha}$ is lower) experience more talent hoarding.
	
	\vspace{0.3cm}
	
	\noindent\textbf{Prediction 3.}		
	(Manager heterogeneity) \\
	Talent hoarding is greater among managers with low utility costs of hoarding (i.e., low $\phi_m$). \\
	\vspace{-0.3cm}
	
	\noindent In addition, workers' decision rule implies that talent hoarding reduces the number of applicants and the quality of the applicant pool, limiting the firm's ability to promote a high-productivity worker and generating misallocation of talent.
	\vspace{0.3cm}
	
	\noindent	\textbf{Prediction 4.}	
	(Number of applicants) \\
	If workers face less talent hoarding, they are more likely to apply for a promotion. 
	
	Pr[$i$ applies$\vert \beta= \beta_1$]>Pr[$i$ applies$\vert \beta= \beta_2$] for $\beta_1 < \beta_2$ \\
	
	\noindent	\textbf{Prediction 5.} (Composition of applicants)	\\
	Talent hoarding has larger impacts for high-productivity workers. Therefore, higher levels of talent hoarding lead to a lower-quality applicant pool:
	
	If $\alpha_1 < \alpha_2$ and $\beta_1 < \beta_2$ $\implies$ $\frac{\text{Pr}[i \text{ applies} \vert \alpha_2, \beta_1]}{\text{Pr}[i \text{ applies} \vert \alpha_1, \beta_1]} > \frac{\text{Pr}[i \text{ applies} \vert \alpha_2, \beta_2]}{\text{Pr}[i \text{ applies} \vert \alpha_1, \beta_2]}$ \\
	\vspace{-0.3cm}
	
	\noindent Appendix Section \ref{sec:appendix_framework} contains formal derivations of these predictions. 
		\vspace{-0.3cm}
	\subsection{Implications for Empirical Tests}
	\vspace{-0.3cm}
	The predictions that follow from the conceptual framework suggest that the existence of talent hoarding can be inferred empirically from both manager and worker actions. 
	
	First, the framework predicts that managers should engage more in talent hoarding if they face stronger incentives to hoard, for instance because of larger financial gains from hoarding. This prediction motivates empirical tests that relate managers' revealed hoarding actions to variation in their hoarding incentives. This approach has the advantage that, relative to analyzing worker behavior, it is less likely to be confounded by potential worker inaction that may be driven by factors unrelated to talent hoarding. However, this test requires a direct measure of managerial talent hoarding actions, which is empirically challenging to construct given that talent hoarding involves hidden actions by definition. Moreover, since managerial hoarding can take many forms, any one measure can only capture a subset of talent hoarding actions.
	
    Second, the framework predicts that workers should be deterred from applying to positions outside the team if they are subject to talent hoarding. Accordingly, an empirical test based on workers requires both observing revealed worker actions and variation in workers' exposure to managerial hoarding. Using high-stakes worker actions, such as internal applications or job transitions,\footnote{Although the language used in the framework is centered around applications for promotions, the intuition and predictions also pertain to any type of internal application to a position outside of the team. From the manager's perspective, both promotions and lateral moves have the same impacts on team productivity. From the firm's perspective, deterring both types of transitions can create efficiency costs.} has the advantage of being easier to observe in administrative data than managerial talent hoarding actions. In addition, such a test is not necessarily restricted to one specific measure of managerial hoarding behavior. 
	\vspace{-0.3cm}
	\section{Setting and Data}
	\label{sec:settingdata}
	\vspace{-0.3cm}
	My analysis relies on a unique combination of personnel records, internal job application data, and survey evidence from a large manufacturing firm. 
	\vspace{-0.3cm}
	\subsection{Firm Overview}
	\vspace{-0.2cm}
	I collect rich data on over 30,000 white-collar and management employees from a large manufacturing firm.  This anonymous firm is one of the largest manufacturers in Europe and employs over 200,000 workers around the world, the plurality of which work in Germany. 
	
	I restrict my sample to Germany because it represents the largest internal labor market at the firm. The firm operates in many other countries, including the United States. The firm's establishments outside of Germany share many features in common with those in Germany, including organizational design and internal labor market policies (e.g., application systems, widespread use of performance and potential ratings). Because I am interested in career progression to higher-level positions, my analysis focuses on employees in white-collar and management positions (i.e., employees that are either already in or could ultimately be promoted to managerial positions).  There are over 200 occupations represented in this sample, ranging from technical positions in engineering to support functions in HR and finance.
	
    Table \ref{table:sample_descriptive} describes my analysis sample, which consists of over 300,000 employee-by-quarter observations from 2015 to 2018.\footnote{To maintain confidentiality, I do not disclose the exact number of employees in my sample.} Women represent 21\% of employees in the sample, stemming from the underrepresentation of women in technical occupations. Employees' educational qualifications are high, a result of restricting the sample to white-collar and management employees. The average employee holds a Bachelor's degree and 92\% of employees work full-time. Tenures at the firm are long, with an average of 13 years, highlighting the importance of internal career progression for employees' long-term income. Managers (i.e., those that lead a team) comprise 19\% of the sample. 
	
    The demographics of the employees at the firm are comparable to other large manufacturing firms in Germany. In Appendix Table  \ref{table:bibb_2018}, I compare employees in my sample to those employees in large manufacturing firms in the BiBB, a representative survey of the German workforce conducted in 2018. I find very similar patterns with respect to most employee characteristics (e.g., gender, age, German citizenship, and marital status). 
	
    The firm also resembles other large firms with respect to the design of its internal labor market (\citetalias{hkp}). As in most large German firms, employees who want to switch to a new position or to be promoted are required to apply for the position using a centralized online job portal at the firm. All job openings are posted to the job portal, where openings are visible to all employees. Applications through the portal typically take less than five minutes to complete. While employees can choose to apply to multiple positions at the same time, the median applicant applies to only one position in a given quarter. Callback and hiring decisions are also recorded in the job portal. 
	
	The internal labor market is comprised of competitive openings for new positions, much like the external labor market.  Only a quarter of applications are successful and the internal labor market is both spatially and interpersonally diffuse. The firm operates in over 50 cities in 250 establishments throughout Germany and one-third of internal applications are for positions in a different city. For the vast majority of applications, applicants have not previously worked with the hiring manager of the position they are applying for, indicating that application and hiring decisions are distinct in the internal labor market. 
	
    Because most teams are small, consisting of one manager and six workers on average, and because most workers in a team work in very similar positions, workers who want to move up the job ladder or seek different types of jobs must leave their team.  At the firm, 97\% of applications are to positions outside of a worker's current team. Thus, managers who encourage their workers to develop their career lose team members. While workers can independently apply to any internal job opening, managers' are tasked with assisting employees in their career planning (e.g., through regular conversations, mentoring activities, assignments to trainings). However, these activities are neither directly incentivized nor monitored at the firm. 
	
	The performance of all employees, including managers themselves, is regularly evaluated by their direct supervisor. Individual contributors are only rated based on their individual achievements, but manager performance ratings take into account the performance of the team. While performance ratings have modest impacts on bonus payments for workers at lower levels, the performance of the manager and her team has substantial impacts on manager pay. For instance, for managers who hold executive positions (e.g., department head or above), bonus payments represent a quarter of their total compensation. Therefore, the composition and performance of the team can have important effects on manager compensation.
	\vspace{-0.3cm}
	\subsection{Personnel Records and Application Data}
	\label{sec:firmdata}
	\vspace{-0.2cm}
	My data combine the firm's internal personnel records from 1998 to 2019 with the universe of application and hiring decisions from 2015 to 2019. The personnel records capture over 30,000 employees, and the application data cover over 16,000 job openings and over 200,000 external and internal applicants. In my main analysis sample, I restrict to employees active at the firm from 2015 to 2018, for whom I can observe outcomes through 2019, yielding a sample of over 300,000 employee-by-quarter observations. Appendix Section \ref{sec:appendix_data_construction} contains details on data construction. 
	
	The personnel records contain a large set of employee characteristics, such as age, educational qualifications,  and tenure. The  records also contain detailed position characteristics, such as position titles, leadership responsibility, and the reporting distance to the CEO. I supplement this data with payroll data, capturing employees' working hours, earnings, and bonus payments. The richness of these data allow me to account for factors unrelated to talent hoarding that may influence workers' career progression. Unless otherwise noted, all analyses include the following set of controls: female, age, German citizenship, educational qualifications, marital status, family status, parental leave, firm tenure, function, division,  location, full-time status, hours, and number of direct reports. 
	
	In addition, the personnel records capture workers' assignments to teams and managers.  I use this information to characterize manager behavior and construct measures of manager quality based on past outcomes of managers' team members (e.g., absenteeism, turnover).  Because these data capture team assignments over time, I am also able to construct measures of managers' formal ties to other units at the firm, by measuring whether they have previously worked in that unit.  Finally, I collect information from the firm's talent management system that allow me to measure manager actions. These include worker evaluations, such as performance and potential ratings, which are conducted by a worker's direct supervisor. The data also capture manager actions with respect to assigning workers to internal trainings that likely increase visibility outside the team. 
	
	When testing how managerial talent hoarding affects worker actions, I focus on worker applications as a key outcome.  Because all job openings are posted to the centralized portal and all applications are required to be submitted through the portal, I am able to construct a panel dataset of employees' application histories at the firm from 2015 to 2018. Through the job portal, I also observe the outcome of each application in terms of rejections, interview callbacks, and subsequent hiring outcomes. Separately measuring applications and hiring outcomes is important given that workers' forgone job switches could be explained by low hiring success unrelated to talent hoarding. 
    \vspace{-0.4cm}
	\subsection{Survey of Employees}
	\label{sec:employee_svy_descr}
    \vspace{-0.2cm}
To capture worker perceptions of career progression at the firm, I conducted a large-scale survey with the employees in my sample.
Employees described challenges regarding their internal career progression both  in multiple-choice answers and in the form of free-text responses. The survey implementation followed the standard protocol that the firm uses for internal surveys. The survey was approved by the firm's data protection officer and the workers' council and was conducted through the firm's online survey tool. Employees were invited via e-mail by the firm's HR department. The invitation introduced the survey as eliciting employee feedback about the internal labor market to improve HR practices and provided employees with a participation link.  See Appendix Section \ref{sec:app_survey} for information on more details on the implementation and the survey instrument. 

 The survey received over 15,000 responses, yielding a 50\% response rate. Respondents are similar to the overall population in terms of their demographics (Appendix Table \ref{table:survey_X}).\footnote{Because of strict data protection regulations in Germany, I am not able to link the survey responses to the administrative employee records at the individual level. This limitation precludes a direct comparison of respondents to non-respondents in the administrative data.}   As is common for surveys within organizations, the survey did not include financial incentives. However, the survey invitation highlighted that employees’ responses would be used to improve
the internal labor market. To increase trust in the survey, the survey invitation mentioned that the survey was approved by the workers’ council. Additionally, before the survey launch, all HR employees in Germany were briefed and instructed to address any employee inquiries, ensuring clarity and transparency about the survey’s purpose and approval.

	\vspace{-0.4cm}
	\subsection{Survey of Managers}
	\label{sec:manager_svy_descr}
	\vspace{-0.2cm}
	In addition to the employee survey that elicited worker perceptions, I conducted a survey of managers at the firm. The primary objective of the survey was to directly elicit the extent to which managers perceive misaligned incentives with respect to the development of talent in the firm. This survey offers the advantage that it does not rely on worker reports, allowing me to more directly elicit the extent of managerial talent hoarding in the firm.
    
    Following the firm's standard survey protocol, managers in the firm were invited to participate in the survey via e-mail by the firm's HR department. Respondents were prompted to provide their perspectives on managers' role in developing talent at the firm based on their own experiences as managers. The survey received responses from 62\% of managers, yielding over 3,000 responses.    Managers were asked to respond to a series of questions including explicit statements with respect to potential talent hoarding.  See Appendix Section 	\ref{sec:app_survey_manager} for the survey instrument and the  distribution of the responses.  

Several features of the survey were designed to foster trust in the survey instrument, and to frame the survey as an opportunity to provide feedback to the firm's HR department. First, the survey invitation emphasized anonymity, explicitly assuring that responses could not be linked to individual managers (Appendix Figure \ref{fig:invitation_manager}). Rather, it framed the survey as an opportunity for managers to provide confidential input on improving HR practices. Additionally, the invitation noted that the survey had been approved by the firm's workers' council, signaling that it had been vetted to protect participants' confidentiality. Second, before the survey launch, HR employees were briefed and instructed to address manager inquiries, reinforcing the survey’s legitimacy and approval by the firm. Third, the firm’s leadership council, a prominent and influential internal body advocating for managers’ interests, publicly endorsed the survey and encouraged managers to participate. The high response rate of 62\% of managers at the firm indicates strong engagement and suggests that managers responded favorably to these design features. Moreover, the fact that the majority of respondents (62\%) voluntarily provided written suggestions on how the firm could better support leaders suggests that managers saw the survey as a meaningful channel for providing anonymous feedback.

The fact that the survey invitation explicitly stated that participation was anonymous and could neither be linked to individual managers and was not permitted to be used for monitoring, suggests that respondents' concerns about reputational risks of admitting to talent hoarding in the survey were small. A related concern is that social desirability bias may lead managers to underreport talent hoarding because it is perceived as detrimental to the firm. To minimize such bias, the survey invitation and instrument (Appendix   \ref{sec:app_survey_manager_implementation}) both avoided emphasizing talent hoarding, and the bulk of questions elicit managers' perceptions of their firm's general policies and leadership behaviors, rather than managers' own actions.  In light of the negative view of talent hoarding, any remaining social desirability bias would likely lead to an underestimation of its prevalence (see Appendix Section \ref{sec:appendix_benchmark} for a more detailed discussion). 

A final concern regarding the interpretation of the survey results is that respondents may be a selected sample of all managers. Since the survey invitation did not explicitly mention talent hoarding, it is unlikely that managers deliberately sorted into the survey by their propensity to hoard talent. An analysis of respondent characteristics shows that survey participants are largely representative of the overall sample in terms of their demographics, experience, and functional area (Appendix Table \ref{table:surveymanager_X}). In addition, when comparing managers who responded before a reminder was sent to those who responded after (likely more reluctant participants), I find that the demographics of early and late respondents are similar, further suggesting that non-random selection in the survey is unlikely to meaningfully affect my results (Appendix Table \ref{table:surveymanager_reminderX}). 
	\vspace{-0.3cm}
	\section{Evidence from Managers: Hoarding Incentives and Actions}
	\label{sec:managers}
	\vspace{-0.2cm}
	The conceptual framework in Section \ref{sec:framework} predicts that managers should hoard workers on their team  when they face incentives to retain talent that conflict with broader organizational goals. In this setting, such misaligned incentives are likely to be particularly severe when a manager's compensation depends heavily on team performance, when a worker’s departure would be particularly costly (e.g., in small teams), or when hoarding talent is less costly because individual contributions are hard to observe (e.g., in low-visibility functions). This section uses a survey of managers in the firm to provide direct evidence that managers perceive misaligned incentives that lead to talent hoarding. I then leverage a direct measure of managers' hoarding actions using the firm's HR data to show that managers who face stronger hoarding incentives are more likely to take hoarding actions.

	\vspace{-0.3cm}
	\subsection{Manager-Reported Hoarding Incentives and Actions}
	\label{sec:manager_survey}
	\vspace{-0.2cm}
	Responses to the manager survey described in Section \ref{sec:manager_svy_descr} reveal that managers face strong incentives to hoard talent instead of supporting workers in their career development outside the team. 

    To assess whether managers perceive a tension between developing talent and retaining employees, the manager survey asked respondents to rate their agreement with the following statement: ``Talent development entails a conflict of interest for leaders, because more developed employees are more likely to leave the team.'' Managers responded on a four-point scale ranging from ``I strongly agree'' to ``I totally do not agree.'' See Appendix Table \ref{tab:QF} for the raw response distribution for this statement.  The majority (55\%) of managers agree or strongly agree, thereby explicitly acknowledging that supporting workers' career development entails a conflict of interest because more developed workers are more likely to leave the team (Bar 1 of Figure \ref{fig:manager_incentives}).\footnote{Note that to the extent that social desirability bias affected manager responses, this figure likely represents an underestimate of the share of managers who perceive a conflict of interest.}

To better understand whether the perceived conflict of interest reflects a lack of rewards for managers' talent development efforts, the manager survey included two additional statements rated on the same four-point agreement scale.  First, to capture how talent development is valued relative to a core managerial responsibility—team performance—respondents were presented with the statement: ``The firm values a manager’s impact on employee career development at least as much as their impact on team performance.'' Second, respondents were asked to indicate agreement with the statement: ``A leader’s direct intervention (e.g., encouragement, increasing visibility) has a large impact on employees’ career development.'' 
This second question serves as a useful comparison, as it contrasts how managers perceive the rewards for talent development with their perceived ability to take meaningful action. See Appendix Table \ref{tab:QF} for raw response distributions.

Responses to these questions suggest that managers perceive effort expended on talent development as having high impact but yielding limited rewards. While 96\% of managers believe that their direct intervention has a large impact on their workers' career development, only 36\% of managers perceive their impact on workers' careers to be valued by the firm at least as much as their impact on immediate team performance (Bars 2 and 3 of Figure \ref{fig:manager_incentives}). 

To further probe the robustness of this finding, I draw on a separate set of survey questions that use a multiple-choice format rather than an agreement scale. These questions provide direct benchmarks of talent development relative to other core managerial responsibilities. To assess perceived impact, managers were asked:  ``Where do you think leaders can create the most impact for [Company]?''  Responses were recorded on a four-point scale ranging from ``Very high impact'' to ``Very low impact.'' To contrast this with perceived rewards, managers were also asked: “When it comes to leaders’ compensation and promotion prospects, how important do you think is a record of success in each of the following?'' This question used a similar four-point scale, ranging from ``Very important'' to ``Not important.'' Both questions asked managers to evaluate six core responsibilities that are considered central to middle management at the firm: strategic orientation, vision, operational efficiency, talent development, culture and morale, and innovation. See Appendix Tables \ref{tab:QA} and \ref{tab:QB} for raw response distributions.

The survey responses reveal a notable gap between perceived impact and perceived reward for talent development. While a large majority of managers (87\%) report that talent development is an area where they can create high impact for the firm, only 40\% indicate that a track record in developing employees is important for their own compensation and promotion prospects. This gap corroborates the finding that despite recognizing the potential benefits of investing in their employees’ careers, managers perceive a lack of incentives for doing so.\footnote{See Appendix Table \ref{table:survey_talenthoardingmanager} for survey quotations of managers describing this lack of incentives. For instance, one manager stated: ``Managers pursue their own goals and often prevent further development of workers, because they are not rewarded for developing talent.''} 

Comparing managers' attitudes about talent development and their other core responsibilities are similarly revealing. First, when comparing the perceived importance of different responsibilities for managers' compensation and promotion prospects, talent development emerges as the lowest-rated dimension (Panel B of Appendix Figure \ref{fig:benchmarking_QAB}). Second, the discrepancy between perceived impact (Panel A) and perceived reward (Panel B) is more pronounced for talent development than for any other responsibility, suggesting a particularly sharp misalignment between what managers believe benefits the firm and what is recognized in their own advancement.  

In the appendix, I present additional results that benchmark responses across these different survey questions. For example, when comparing these results to the earlier agreement-based questions (specifically, the share of managers who agree that talent development is valued by the firm and that their actions meaningfully influence workers' careers), the patterns are strikingly similar. See Appendix Section \ref{sec:appendix_benchmark} for details on this comparison and a discussion of additional benchmarking exercises. Together, these survey findings support the interpretation that talent development is viewed as under-incentivized by managers, despite being recognized as impactful.

	How widespread is talent hoarding? I use two complementary survey questions to assess the magnitude of managerial talent hoarding at the firm. First, my preferred measure uses a survey question that was designed to directly reflect the definition of talent hoarding from the conceptual framework: managers deliberately reducing the likelihood that a worker leaves the team. The question asked, ``How often might leaders find themselves in situations where they need to dissuade a team member from exploring opportunities in another department due to immediate team needs or performance goals?'' Responses were measured on a four-point scale ranging from ``Never'' to ``Very often.'' See Appendix Table \ref{tab:QG} for the raw response distribution.  I find that 75\% of managers respond that this situation occurs at least sometimes, indicating that the broad majority of managers explicitly recognize the existence of talent hoarding. 

    My second measure probes the issue more indirectly by focusing on foregone investment in employee development rather than actively dissuading worker mobility. Managers were asked, ``What are the reasons that may prevent leaders from investing time and effort towards their employees’ career development?'' Respondents could select multiple response options among the following: leaders' lack of knowledge, limited resources, risk of losing talent, need to prioritize short-term targets, and lack of interest on the part of the employee. See Appendix Table \ref{tab:QD} for the raw response distribution. I find that 45\% of managers cite the risk of losing workers as a key reason for not investing in employees. Moreover, 66\% of managers cite the need to prioritize short-term targets over long-term employee development, which accords with the responses to the first question and underscores the importance of team performance in managerial incentives. Collectively, these responses corroborate the finding that talent hoarding is a common reaction to prevailing managerial incentives.
	\subsection{Measuring  Hoarding Actions in the Data}
	\label{sec:th_measure}
	
	Measuring managers' talent hoarding actions in the data is difficult because it typically occurs through interpersonal interactions. The survey evidence suggests that one common way in which managers limit workers' opportunities to leave the team is by suppressing worker visibility.  To capture high-stakes talent hoarding actions in the data, I  identify managers' systematic suppression of worker visibility based on a measure of worker visibility that I collect from the firm's HR databases. Specifically, I measure  systematic differences between potential ratings and performance ratings.  Performance ratings are meant to provide task-specific feedback to workers about their past performance in their current position.  Potential ratings are designed to inform the firm about a worker's future potential for higher-level positions and thus are meant to identify workers who would be likely to perform well if they were promoted. Both ratings are conducted simultaneously by a worker's direct supervisor and are very similar to common worker assessments, such as the nine-box grid, that are used by many organizations across the world (\citealp{cappelli2014talent}). 

    
	An important distinction between performance and potential ratings is the extent of their visibility outside of a worker's team. Performance ratings are private signals to the worker and are not shared with other units in the firm. If a worker applies for a job in a different unit, that unit will not be able to access the worker's past performance ratings. The firm's HR department describes the privacy norms regarding performance ratings as similar to the strong taboo in Germany of talking about one's salary. In contrast, potential ratings are public signals of worker talent and are widely circulated within the firm. The HR department regularly circulates lists of high-potential workers, making them highly visible outside of worker's current team. 
	
	Intuitively, a manager who wants to hoard talent should give workers lower (public) potential ratings relative to their (private) performance ratings. A survey of the firm's employees suggests that this method of talent hoarding is commonplace.\footnote{For instance, one worker states, ``Supervisors suppress potential ratings because of fear that employees will leave their current position for a promotion.'' (Appendix Table \ref{table:survey_talenthoarding}).} For a given worker, the difference between performance and potential ratings may reflect worker-specific factors that are unrelated to talent hoarding; however, because managers have discretion in conducting these evaluations, comparing systematic differences between the ratings across workers can capture manager behavior.\footnote{Illustrating the importance of manager discretion in this setting, \cite{Benson2021potential} find in contemporaneous work that managers in a large retailer overrate men's potential compared to women's, possibly as a reaction to men's higher turnover risk.} 
	
	My measure of talent hoarding is defined by the difference between the actual and predicted potential ratings a manager assigns to their workers. Specifically, I take the residuals from the OLS estimation of the following regression of the potential rating given to worker $i$ in quarter $t$ by manager $m$ on their performance rating and other worker characteristics:
	\vspace{-0.3cm}
	\begin{align}\label{eqn:potential}
		\text{potential}_{imt} &=  \beta_1\text{performance}_{imt} + \beta_X X_{it}+\beta_t + \varepsilon_{imt} 
	\end{align}
	
	\noindent In Equation \ref{eqn:potential}, $X_{it}$ denotes the vector of controls described in Section \ref{sec:firmdata}, $\beta_t$ represents quarter fixed effects.  I compute the average difference (over workers and quarters) between predicted and actual potential ratings for each manager. This defines a manager-level measure of talent hoarding, reflecting a manager’s systematic tendency to assign public potential ratings that are higher or lower than expected given worker performance and characteristics based on all workers in their team during my sample period.

   Figure \ref{fig:talenthoarding_distribution} shows the distribution of the mean deviation between predicted and actual potential ratings. The data for this figure is at the manager level (i.e., containing one observation for each of the over 6,000 unique managers in my sample). The distribution is centered close to zero (median = –0.006, SD = 0.212) and reveals substantial variation. Many managers assign potential ratings roughly in line with predictions, but there is meaningful mass in both tails. Managers in the left tail appear to inflate worker visibility by assigning higher-than-expected potential ratings. In contrast, managers in the right tail consistently assign lower-than-expected potential ratings, suppressing worker visibility in a manner consistent with strategic talent hoarding. 

    In my empirical analysis, I use both the continuous measure of a manager’s mean deviation between predicted and actual potential ratings, and discrete classifications based on terciles of this distribution. The continuous measure enables a graded interpretation: larger positive deviations correspond to greater suppression of worker visibility. To analyze meaningful variation in managerial behavior, I classify managers in the top tercile (mean deviation above 0.1036, shaded in orange) as having a higher tendency to suppress visibility. This classification allows me to focus on managers who are most likely engaged in systematic hoarding behavior, and thus also allows a direct comparison with the survey-based results that rely on binary measures of managerial hoarding actions. Robustness tests in Appendix Sections  \ref{sec:appendix_manger} and \ref{sec:appendix_robustness_rotation} confirm that results are consistent when using the continuous measure, a median split, or a more conservative quantile cutoff.

	Appendix Section \ref{sec:robust_measure} provides additional information on the construction and validity of this measure. I show that manager-level hoarding measure are stable over time and not simply driven by noise or measurement error. The talent hoarding measure is also not simply a function of managers' ability to assess talent, suggesting that differences between performance and potential ratings are unlikely to be a result of managers' involuntary mistakes.  In addition, the underrating of potential captured by this measure does not appear to represent managers' correct assessment of workers' future performance. Instead, managers' suppression of the public signals of worker talent is consistent with their incentive to reduce visibility outside of the team.
	
	This measure allows me to directly test empirical predictions for the circumstances under which managers should engage in more hoarding (Section \ref{sec:predictions_managers}) as well as to document how talent hoarding depresses workers' career development (Section \ref{sec:instrument}). 
	
	\subsection{Testing Empirical Predictions for Managerial Hoarding Actions}
	\label{sec:predictions_managers}
	To provide evidence that managers purposefully hoard talent based on managers' revealed actions, I use the direct measure of talent hoarding I construct to assess whether hoarding actions are more likely to occur when managers face greater hoarding incentives. Following the empirical predictions from my conceptual framework in Section \ref{sec:framework}, I relate my measure of managerial talent hoarding for each of the over 6,000 managers in my sample to the incentives these managers face. Specifically, I analyze whether managers are more likely to be classified as hoarding-prone based on the measure discussed in Section \ref{sec:th_measure} if they face stronger incentives to hoard. To conduct this test, I estimate a linear model for managers' hoarding propensity using an OLS regression on an indicator that managers face a strong incentive to hoard and a set of manager controls (i.e., gender, German citizenship, age, firm tenure, and education).
	
	Motivated by the conceptual framework, I construct three proxy measures that capture variation in managers' hoarding incentives. First, the framework predicts that managers should engage in more talent hoarding if the gains from hoarding are larger. To proxy for these gains, I use information from managers' variable compensation to measure the portion of their overall compensation that is related to team performance.  Second, the framework predicts that talent hoarding should be greater if a worker's departure has a larger impact on team performance. I proxy for this dimension using the size of the manager's team. Intuitively, smaller teams have fewer people who can compensate for a single worker's departure.  Third, the framework predicts more hoarding if managers incur lower costs of talent hoarding. I proxy for this dimension using variation in the visibility of worker talent. In functional areas where individual achievements are difficult to observe (e.g., corporate functions such as HR or marketing), it should be very easy for managers to hoard talent because the manager has a high degree of control over worker visibility. On the other hand, in areas such as engineering where many workers file patents, it should be more difficult for managers to hoard.\footnote{High-visibility functions are defined as engineering, project management, and quality management. Low-visibility functions are defined as administration, finance, safety and health, HR, marketing, it, logistics, and purchasing.} See Appendix Section     \ref{sec:hoarding_incentives_details} for additional details on how these proxies are constructed and distributed in my sample.

    Managers who face stronger incentives to hoard talent are substantially more likely to engage in hoarding behavior. Table~\ref{fig:hoarding_predictions} reports results from OLS regressions of a binary indicator for hoarding-prone managers based on my measure of suppressing potential ratings on the three  proxies for hoarding incentives. Column 1 shows that managers with a 1 percentage point higher share of performance-related compensation (i.e., stronger financial incentives for hoarding) are 0.19 percentage points more likely to hoard talent ($p$ = 0.000). This effect corresponds to a 13 percentage point difference in predicted hoarding rates between the 90th and 10th percentiles of the financial incentive distribution. Column 2 uses team size as a proxy for hoarding incentives, and finds that managers with smaller teams are significantly more likely to hoard: a one-person increase in team size is associated with a 1.3 percentage point reduction in hoarding behavior ($p$ = 0.000), implying a 13 percentage point difference between the 90th and 10th percentiles of team size. Column 3 shows that managers in low-visibility functional areas are 4.0 percentage points more likely to hoard ($p$ = 0.002), relative to peers in high-visibility areas.

These results confirm that talent hoarding is substantially more common among managers who face stronger incentives to retain talent within their teams, in line with the predictions of my conceptual framework. Importantly, these findings are robust to using alternative specifications of the incentive measures, such as binary indicators for a high share of pay based on team performance or an indicator for leading a small team, as well as to alternative versions of the outcome variable, including the continuous measure of hoarding behavior (Appendix Tables~ \ref{tab:incentives_reg} and~\ref{tab:hoarding_predictions_outcome}).

Higher levels of talent hoarding among managers with stronger incentives are also evident when analyzing a complementary measure of talent hoarding based on worker assignments to high-visibility trainings. At the firm, managers can assign their workers to in-person trainings, which are organized by a centralized and well-recognized department at the firm. These trainings are prestigious and typically increase worker visibility outside their team. For example, lists of training participants are widely circulated within the firm, and many workers list trainings on their CVs. I use the frequency with which managers restrict access to these training opportunities relative to what would be expected given worker characteristics to construct a measure of talent hoarding. The construction of this measure is analogous to that of the baseline measure using potential ratings. See Appendix Section \ref{sec:appendix_manager_training} for more details.
	
The effects of hoarding incentives on restricting access to high-visibility trainings are similar in both direction and economic magnitude to those observed using the potential ratings–based measure (Appendix Table~\ref{tab:hoarding_predictions_training}). Managers whose performance-related compensation is 1 percentage point higher are 0.20 percentage points more likely to hoard talent ($p$ = 0.000). Similarly, a one-person increase in team size results in a 1.4 percentage point reduction in hoarding ($p$ = 0.000). Lastly, managers in low-visibility functional areas are 2.98 percentage points more likely to hoard ($p$ = 0.021). These patterns are highly robust to using alternative measures of both outcome and regressors (Appendix Tables \ref{tab:incentives_training_reg} and \ref{tab:hoarding_predictions_training_rob}).

In addition to the administrative analysis, which captures high-stakes realized outcomes, I use data from the manager survey to examine whether those facing stronger incentives are more likely to report talent hoarding. The advantage of the survey data is that the outcome measures are based on questions that explicitly elicit hoarding behavior, providing more direct insights into how managers perceive hoarding actions. I construct the incentive measures in the survey using a similar approach as in the administrative data (see Appendix Section~\ref{sec:robust_incentives_survey} for details), and analyze the two separate survey questions related to hoarding behavior described in Section \ref{sec:manager_survey}.

Table~\ref{table:survey_hoarding_predictions} shows that the effects of hoarding incentives on perceptions of talent hoarding in the survey are similar in both direction and economic magnitude to those observed using administrative measures. Columns 1 to 3 use my primary question, which asks respondents how often managers feel the need to dissuade team members from exploring opportunities in other departments (i.e., perceived hoarding). See Appendix Table \ref{tab:QG} for the raw response distribution and question wording. A 1 percentage point increase in the share of performance-related compensation leads to a 0.27 percentage point increase in perceived hoarding ($p$ = 0.019). Similarly, a one-person increase in team size results in a 0.4 percentage point decrease in perceived hoarding ($p$ = 0.026), and managers in low-visibility functional areas are 6.7 percentage points more likely to perceive hoarding ($p$ = 0.277).

Columns 4 to 6 use an alternative measure of perceived hoarding based on a question asking whether the risk of losing talent prevents managers from investing time and effort in their employees’ career development (see Appendix Table   \ref{tab:QD}  for the raw response distribution and  question wording).  The results are similar: a 1 percentage point increase in performance compensation yields a 0.32 percentage point increase in perceived hoarding ($p$ = 0.038); a one-person increase in team size reduces perceived hoarding by 0.65 percentage points ($p$ = 0.003); and managers in low-visibility areas are 14.4 percentage points more likely to perceive hoarding ($p$ = 0.010).

These patterns are consistent across both survey questions and align closely with those in the administrative data, reinforcing the interpretation that stronger managerial incentives are associated with greater hoarding behavior. These findings are also robust to using alternative specifications of the incentive variables (Appendix Table \ref{table:survey_hoarding_predictions_robust}).  Placebo tests further confirm that these relationships are specific to talent hoarding and do not simply capture other organizational barriers (Appendix Table~\ref{table:survey_placebo}).

	Taken together, these results provide evidence using several complementary measures of manager actions that managers are more likely to engage in talent hoarding if they face greater hoarding incentives. Interpreted through the lens of my conceptual framework, these results confirm that more misaligned incentives lead to higher levels of talent hoarding. 
	\vspace{-0.3cm}
	\section{Evidence from Workers: Effects of Manager Rotations}
	\label{sec:instrument}
	\vspace{-0.3cm}
	This section provides evidence that talent hoarding has meaningful impacts on worker outcomes. I use manager rotations as a source of quasi-random variation in workers' exposure to talent hoarding. When a manager learns that they will move to a new position on a different team, they no longer have an incentive to hoard workers on their current team. For workers whose manager will soon rotate, this creates a temporary window of time during which they will not be subject to talent hoarding. I test for the impacts of talent hoarding using workers' internal job applications, which represent high-stakes actions that are critical for workers' careers within the firm. This quasi-experimental analysis provides evidence that complements the earlier analysis of managers. 
While the manager-level analysis isolates specific managerial actions—such as suppressing worker visibility—it captures only one particular hoarding action, namely, the suppression of worker visibility outside of the team. In contrast, the worker-level analysis based on manager rotations encompasses a broader set of hoarding actions, including less easily-measured actions such as discouraging applications or exerting implicit pressure.
    Section \ref{sec:design_rotation} introduces the rotation design. To support the interpretation of rotation effects as being driven by talent hoarding, Section \ref{sec:mechanism} tests predictions from the conceptual framework. 
		\vspace{-0.3cm}
	\subsection{Rotation Design}
	\label{sec:design_rotation}
	\vspace{-0.2cm}
	Manager rotations represent typical features of managers' career progression and are not restricted to specific managers or teams. As in many other firms, managers switch positions within the firm to fulfill requirements for future promotions (\citealp{CullenPerezTruglia}).\footnote{I use the term manager rotation in line with previous research to refer to the event, in which a manager leaves their team for a different position within the firm.} For instance, in order to become an executive, the firm requires managers to collect experience in different parts of the firm. In general, internal job switches are routine and encouraged by the firm. Consistent with the institutional environment, I empirically document that manager rotations appear orthogonal to pre-existing worker characteristics by comparing workers who are exposed to a manager rotation to those that are not (Panel A of Appendix Table \ref{table:iv_validity_fs_ind}). I find that both sets of workers are observationally similar with respect to their demographics and their past performance.  Panel B of Table \ref{table:iv_validity_fs_ind} finds a similar balance with respect to manager characteristics, such as demographics, performance, and pre-event team-level outcomes. Appendix Figure \ref{fig:rotation_absent} uses an event study design to document the absence of pre-trends in key team-level outcomes, such as absenteeism, further corroborating that manager rotations are not restricted to teams facing specific circumstances. 
	
    My main analysis uses 1,359 manager rotations that result from a worker's direct supervisor leaving their team to make an internal job transition within the firm.\footnote{Appendix Section \ref{sec:appendix_robustness} presents a robustness test using manager exits, which yields similar patterns.} During the four-year study period, 20\% of managers rotate at least once. Rotations do not occur on a regular schedule, and workers cannot easily predict when managers will leave the team; Appendix Figure \ref{fig:rotation_timeposition} documents large variation in the time that managers spend in a position before rotating. The firm reports that managers typically learn about their new position two to three quarters before they rotate, at which point they inform their teams about the rotation, following a strict policy at the firm that requires them to announce their transitions as soon as possible.\footnote{Survey responses, indicating that the news of internal applications spreads quickly by internal gossip, further corroborate the immediacy of these announcements since it is readily apparent when someone has violated this policy.} 
	
	Because managers know they will not be affected by workers leaving the team once they find out about their departure, manager rotations should temporarily reduce hoarding incentives, and thus increase worker applications. I illustrate the dynamic effects of manager rotations using a quarterly event study of applications around the quarter in which a manager rotates. I estimate a specification with worker and quarter fixed effects, binning event time dummy variables at $t=-8$ and $t=4$, and clustering standard errors at the worker and rotation level. Panel A of Figure \ref{fig:msw_dynamics} presents quarterly event study coefficients and demonstrates that manager rotations result in a distinct transitory increase in worker applications. Event time $t=0$ denotes the quarter in which a manager rotates. Application rates increase up to three quarters before the manager rotation takes place, which is when managers start to inform their teams about their departure. However, before  $t=-3$, when managers do not know yet about their job rotation, trends in applications are flat. In the quarter of rotation,  applications increase by 2.3 percentage points, almost doubling workers' baseline application rate of 2.7\%. As the manager's replacement settles in, application rates return to baseline levels after one quarter.\footnote{Note that applications taking a quarter to return to baseline is consistent with the fact that in a given quarter, only few job openings are available in the internal labor market, implying that not all workers who decide to apply once they find out about the manager rotation will immediately find a suitable job opening.} 
	
	Since manager rotations appear to have the largest impact on worker applications in the quarter of the rotation, the remainder of the analysis focuses on worker behavior in that quarter. I analyze the effect of manager rotations on worker applications in the same quarter by estimating a linear model for workers' internal application choices using an OLS regression of the following form:
	\vspace{-0.2cm}
	\begin{align}
		\text{Applied}_{it} = \delta_1 \text{Rotation}_{it} + \delta_X X_{it} + \delta_t + u_{it}  \label{eqn:fs} 
	\end{align} \\
	\vspace{-1.5cm}
	
	\noindent $\text{Applied}_{it}$ and $\text{Rotation}_{it}$ are indicators that worker $i$ in quarter $t$ applies for an internal job opening and experiences a manager rotation, respectively. $X_{it}$ includes a broad set of worker and position controls.\footnote{My baseline set of controls, described in Section \ref{sec:firmdata}, include female, age, German citizenship, educational qualifications, marital status, family status, parental leave, firm tenure,  division, function, location, full-time, hours, and number of direct reports. Appendix Table \ref{table:robustness_controls} documents that my results are robust to including different controls.} $\delta_t$ represents quarter fixed effects.  Robust standard errors are two-way clustered at the rotation and worker level.
	
	Manager rotations have large positive impacts on worker applications in the quarter of rotation. Column 1 of Table \ref{table:results_all}, Panel A presents OLS estimates for the effect of manager rotations on worker applications, based on Equation \ref{eqn:fs}. When a manager rotates, applications increase by 2.3 percentage points, representing a 78\% increase compared to the baseline application rate of 2.9\%. The application effect is not a result of managers taking their subordinates with them to the new team or of workers replacing managers in their old position: 97\% of applications at the firm are for positions outside of the worker's current team and not to the manager's new team. In addition, the observed impact does also not depend on the characteristics of the outgoing or incoming managers (Appendix Table \ref{table:incoming}). If manager rotations reflect the impact of talent hoarding, these findings imply that managerial talent hoarding deters a meaningful share of applicants from exploring job opportunities outside the team. The following subsection presents evidence in favor of this interpretation.\footnote{Appendix Section~\ref{sec:confounder} further supports this interpretation by presenting a series of robustness checks that suggest that key alternative explanations, such as correlated shocks, team-specific milestones, worker-manager loyalty, or role-model effects, are unlikely to be main drivers in this setting.} Interpreted in this light, this finding accords with workers' survey responses discussed in Section \ref{sec:background}, in which 41\% of workers indicate that they are afraid to apply to internal positions due to talent hoarding. 
	\vspace{-0.3cm}
	\subsection{Does Talent Hoarding Drive Rotation Effects?}
	\label{sec:mechanism}
	\vspace{-0.2cm}
	This section provides evidence that talent hoarding is a key mechanism underlying the impacts of manager rotations. The conceptual framework predicts that different types of workers face different degrees of talent hoarding, depending on factors like productivity and manager costs of hoarding. I test these predictions by examining how the effect of manager rotations varies along several of these dimensions. In each case, the analysis includes a continuous interaction between the rotation indicator and a standardized version of the respective proxy for stronger hoarding incentives. This approach allows me to estimate how the effect of manager rotation on application behavior varies with each dimension. The baseline effect of a manager rotation is shown in blue in Panel A of Figure~\ref{fig:mechanism_departurecost}, while the estimated change in this effect associated with a one-standard-deviation shift in the characteristic is shown in orange in the respective panels.

	\vspace{2mm} 
	\textit{Worker Quality.}---The conceptual framework in Section \ref{sec:framework} predicts that managers are more likely to hoard high-productivity workers. Therefore, when managers rotate and talent hoarding temporarily subsides, application effects should be larger for the high-productivity workers who experienced more talent hoarding than their low-productivity peers. To test this prediction, I construct a measure of worker quality as the predicted value from a regression of applicants' internal hiring probabilities on observable characteristics such as education, tenure, and function. The predicted value of this regression provides an index of worker quality for all workers, weighting worker characteristics by their importance for hiring prospects within the firm.\footnote{This definition of worker quality has the advantage that it reflects the values that the firm places on worker qualifications. Robustness exercises in Appendix Section \ref{sec:appendix_robustness_rotation} find similar patterns when using alternative measures such as educational qualifications or past performance.} 

    Panel~A  shows that the average effect of a manager rotation on application rates is 2.3 percentage points. Panel~B of Figure~\ref{fig:mechanism_departurecost} shows that a one-standard-deviation increase in worker quality is associated with a 1.2 percentage point larger increase in applications following a manager rotation ($p$ = 0.000). The positive and statistically significant interaction confirms that higher-quality workers are much more responsive to manager rotations, consistent with the idea that they face stronger hoarding incentives ex ante.
    	\vspace{2mm}
        
	\textit{Departure Costs.}---Another prediction is that managers hoard talent more when the costs of worker departures are larger. I test this prediction using two measures of departure costs. The first measure is the number of other workers in the team, since a larger team contains more workers that can compensate for a teammate's departure.  Panel C of Figure~\ref{fig:mechanism_departurecost} shows that the effect of manager rotations on applications is significantly larger in smaller teams. The figure plots the interaction of manager rotations with a standardized (negated) measure of team size, so that positive coefficients indicate stronger effects for smaller teams. By including detailed worker and position characteristics, this analysis compares workers with similar qualifications in similar positions, but in teams of different sizes. A one-standard-deviation decrease in team size is associated with a 0.7 percentage point larger increase in applications following a manager rotation ($p$ = 0.008). Measuring departure costs using the average number of days required to fill a worker's vacated position yields similar results. Panel~D of Figure~\ref{fig:mechanism_departurecost} shows that a one-standard-deviation increase in replacement duration is associated with a 0.54 percentage point increase in the application effect following a rotation ($p$ = 0.056). These results support the prediction that workers in roles with higher departure costs are more subject to hoarding behavior.

	\vspace{2mm}
	\textit{Talent Hoarding Propensity.}---The framework also predicts that managers who are more prone to hoarding talent should have larger effects on application behavior when they rotate. I test this prediction using my measure of managerial hoarding propensity, based on the average deviation between a manager’s private performance ratings and the publicly visible potential ratings they assign to their workers (see Section	\ref{sec:th_measure} for details).

\begin{figure}[!htb]
		\caption{Differential Effects of Hoarding-Prone Managers by Worker Quality}
		\begin{minipage}[b]{0.8\linewidth}
			\centering	
			\includegraphics[scale=1.0]{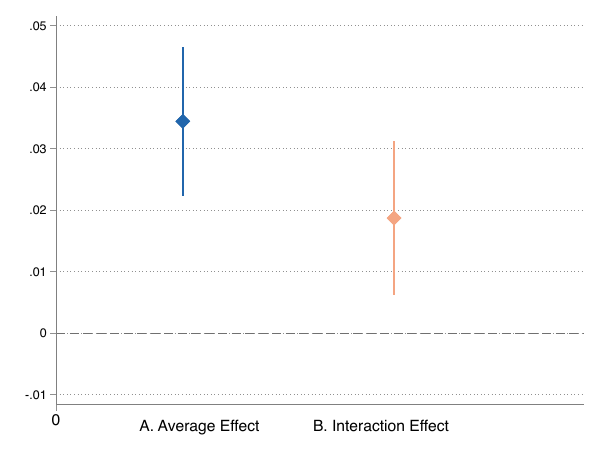}
		\end{minipage}	\\
		\begin{minipage}[b]{0.9\linewidth}
			\footnotesize \textit{Notes}:   This figure illustrates how rotation effects of hoarding-prone managers differ by worker quality. I focus on workers whose managers are in the  top tercile (visibility-suppressing) of the hoarding measure. The measure is based on the average deviation between a manager’s actual and predicted potential ratings.  I estimate two separate regressions: one that includes the main effect of manager rotation (plotted in blue) and one that adds an interaction between manager rotation and the standardized worker quality index (plotted in orange). The interaction coefficient reflects how the effect of a manager rotation changes with a one-standard-deviation increase in worker quality. Robust standard errors are clustered at the worker and rotation level. 95\%-level confidence intervals are displayed. Controls: Female, age, German citizenship, educational qualifications, marital status, family status, parental leave, firm tenure, division, function, location, full-time, hours, number of direct reports, and quarter fixed effects. N=3xx,xxx.  
		\end{minipage}	
		\label{fig:mechanism_heteroquality}
	\end{figure}

    The impacts of manager rotations are strongly correlated with direct measures of managers' propensities to hoard talent. Panel E of Figure~\ref{fig:mechanism_departurecost} shows that workers experience significantly larger application effects when a manager with a higher propensity to hoard talent rotates. I interact the manager rotation indicator with a standardized version of the hoarding measure. A one-standard-deviation increase in hoarding propensity is associated with a 0.61 percentage point larger increase in applications ($p$ = 0.033), confirming that workers respond more strongly when a hoarding-prone manager leaves. The framework also predicts that higher-quality workers are hoarded more intensively by hoarding-prone managers. Figure~\ref{fig:mechanism_heteroquality} further tests this prediction by focusing on workers whose departing manager falls in the top tercile of the hoarding distribution and then examines how the rotation effect varies with the continuous measure of worker quality. Among workers whose manager had a high propensity to hoard, applications increase by 3.6 percentage points on average following a rotation (baseline effect). A one-standard-deviation increase in worker quality is associated with an additional 1.7 percentage point increase in applications ($p$ = 0.005), demonstrating that the suppressive effect of hoarding-prone managers is concentrated among high-quality workers.

	\vspace{2mm}
	\textit{Risk of Retaliation}---In survey results (Section \ref{sec:survey}), workers report fear of manager retaliation as a reason they refrain from applying. If workers are more likely to refrain from applying to positions that carry a greater risk of retaliation, applications to these positions should exhibit particularly large increases around rotations. I test this prediction by comparing workers' applications to job openings that are closer in proximity to their current position (i.e., within versus outside their current division, functional area, and location). Intuitively, the risk of retaliation is larger if managers are more likely to find out about worker applications. Appendix Table \ref{table:mechanism_distance} shows that manager rotations have much larger and statistically distinguishable effects on applications in close proximity to workers' current position with respect to all three dimensions, confirming the prediction.\footnote{Manager rotations could lead to an opposite prediction on proximity if they were to operate through the visibility channel instead of retaliation. However, while the suppression of visibility is a common form of talent hoarding, manager rotations do not immediately increase visibility since worker evaluations only occur one to two times a year.} 
	
	\vspace{2mm}
	\textit{External Transitions.}---While managers may frequently learn about workers' applications to internal positions and possibly retaliate or interfere with those applications, features of the institutional setting make this less likely to be the case for applications to positions outside the firm.\footnote{In Germany, it is very uncommon that potential employers contact the current firm to inquire about the candidate. Except for a few specific occupations, such as academics, the use of reference letters is also very uncommon, implying that relative to internal transitions, managers have both limited ability to learn about or interfere with external applications.} Under talent hoarding, the impact of manager rotations should thus be larger for internal applications than for external applications outside the firm.  Because I do not observe external applications, I test this prediction by comparing the effect of manager rotations on internal job transitions within the firm and external job transitions out of the firm. Panel B of Figure \ref{fig:msw_dynamics} shows that manager rotations only increase worker transitions within the firm and not outside of the firm, even though both types of transitions trend identically in quarters prior to the rotation. Columns 1 and 2 of Appendix Table \ref{table:mechanism_internalexternal_reg} confirm this finding. Manager rotations increase internal transitions by 1.1 percentage points but have negligible effects on external transitions (0.11 percentage points) compared to the identical baseline rate of 0.8\% ($p$ of difference = 0.000).
	
	\vspace{-0.3cm}
	\subsection{Discussion and Interpretation}
	\vspace{-0.2cm}
	The preceding results provide empirical evidence that the worker application responses to manager rotations are to a large part driven by managerial talent hoarding.
   Appendix Section~\ref{sec:confounder} complements this interpretation by providing evidence against several plausible alternative mechanisms---including bad-news shocks, milestone completions, loyalty, match quality, and role-model effects---as the primary drivers of the observed effects. Although it is not possible to exhaustively rule out all potential alternative mechanisms, the observed patterns impose strong restrictions on the manner in which such mechanisms would have to affect worker behavior, which provides evidence against a quantitatively meaningful role for many plausible alternative mechanisms. Nonetheless, it remains possible that the observed responses to rotations may also capture factors that are unrelated to talent hoarding. Yet, to the extent that talent hoarding is a key driver of responses to manager rotations, analyzing impacts of rotations is informative about the impacts of hoarding on workers' careers. This interpretation is supported by the analysis focusing on managers in Section \ref{sec:managers}. The following section analyzes impacts on workers' subsequent career outcomes.
	\vspace{-0.4cm}
	\section{Consequences of Talent Hoarding} 
	\label{sec:results_misallocation} 
	\vspace{-0.3cm}
	The preceding results indicate that talent hoarding effectively deters workers from applying for job opportunities outside the team. This finding naturally begs the question of whether deterred applications would have been successful, and whether talent hoarding has negative consequences for the firm overall. To answer these questions, this section measures impacts on the realized allocation of talent. In principle, talent hoarding can have negative consequences for the firm through many different channels, such as depressing human capital investment. Motivated by prior work studying the allocation of talent (\citealp{holmstrom1979}, \citealp{rosen1982}, \citealp{lazear2015value}), I focus on providing suggestive evidence that managerial hoarding affects how talent is allocated in the firm.
	\vspace{-0.3cm}
	\subsection{IV Approach}
	\vspace{-0.2cm}
	To test whether deterred applicants would have been successful, I use manager rotation as an instrument for worker applications to estimate the success probability of marginal applicants, who would not have applied in absence of a manager rotation. Equation \ref{eqn:rf} represents the reduced-form model for the effect of manager rotation on workers' hiring outcomes:
	\vspace{-0.2cm}
	\begin{align}
		\text{Hired}_{it} = \theta_1 \text{Rotation}_{it} + \theta_X X_{it} + \theta_t  + \epsilon_{it}   \label{eqn:rf}
	\end{align}
	\noindent In Equation \ref{eqn:rf}, $X_{it}$ denotes the vector of controls described in Section \ref{sec:firmdata}, $\theta_t$ represents quarter fixed effects.  $\text{Hired}_{it}$ is a binary indicator, which is always zero if workers do not apply. Standard errors are two-way clustered at the rotation and worker level. 
	The IV-estimator divides the reduced-form effect of manager rotation, $\theta_1$, by the first-stage effect on application choice, $\delta_1 $:
	\begin{align}
		\beta_{IV} = \frac{\theta_1}{\delta_1 }  \label{eqn:iv}
	\end{align}
	I estimate $\beta_{IV}$ by two-stage least squares, which can be interpreted as the local average treatment effect (LATE), defined as the effect of applications on hiring outcomes for workers induced to apply by manager rotations. Because workers can only get hired if they apply and there are no always takers, LATE equals the treatment effect on the treated (TOT).  To interpret $\beta_{IV}$ as LATE, four assumptions are required to hold: relevance, independence, exclusion, and monotonicity (\citealp{angrist1995}, \citealp{angrist1996}). I now provide evidence in support of these assumptions.
	
	\vspace{2mm}
	\textit{Relevance.}--- Rotations have large and significant effects on worker applications (Section \ref{sec:instrument}).
	
	\vspace{2mm}
	\textit{Independence.}--- Manager and worker characteristics are balanced and neither worker applications nor other key team-level outcomes, such as absenteeism or bonus pay, are characterized by pre-trends, suggesting that rotations are as good as randomly assigned (Section \ref{sec:instrument}).
	
	\vspace{2mm}
	\textit{Exclusion.}---The exclusion restriction requires that manager rotation does not affect hiring outcomes other than through workers' decisions to apply. This assumption would be violated if departing managers intercede in workers' hiring outcomes, for instance by taking workers with them to their new team or replacing themselves with workers. However, this occurs in less than 3\% of applications, suggesting that departing managers generally do not make subsequent hiring decisions for their subordinates. Alternatively, departing managers may influence hiring outcomes by reaching out to hiring managers. I test for this possibility by comparing applications where the posted job is located in the same area as workers' current job, making it more likely that the departing manager interacts with the hiring manager. However, I do not find that relative hiring rates are higher for job openings to which managers have closer ties (Panel A of Appendix Table \ref{table:iv_validity_excl}). 
	
	Another violation could occur if manager rotation increases worker qualifications, making them more likely to get hired. If this channel is quantitatively important, one would expect to see larger hiring effects for workers under managers with higher quality or who have been exposed to the manager for longer. I measure manager quality using past means of three outcomes:  team turnover, team absenteeism, and managers' performance rating. Panel B of Appendix Table \ref{table:iv_validity_excl} documents that the marginal hiring probability is similar for high-quality vs low-quality managers. Panel C of Appendix Table \ref{table:iv_validity_excl} shows that marginal hiring probabilities are comparable across exposure length, even for workers who have been with the manager for only one quarter, suggesting that such a channel is unlikely to violate the exclusion restriction.
	\vspace{2mm}
	
	\textit{Monotonicity.}--- My results do not appear to be biased by the existence of a large defier population. In my setting, defiers are individuals who would have applied in the absence of a manager rotation, but whose unobserved propensity to apply is reduced by the instrument.  Following \cite{arnold2018racial} and \cite{bhuller2020incarceration}, I show that the first-stage relationship between applications and manager rotations remains positive for 16 key subgroups of workers (Panel A of Appendix Table \ref{table:iv_mono}). To further test for the presence of defiers, I use two proxies for workers' unobserved application propensity: workers' own past application behavior and the leave-out team mean of past application rates. Panel B of Appendix Table \ref{table:iv_mono} shows that even workers, for whom the proxies predict that they are more likely to be defiers because of their higher application propensity in the past, do not experience lower or negative application effects when managers rotate.
	\vspace{-0.3cm}
	\subsection{Impacts on the Allocation of Talent}
	\vspace{-0.2cm}
 To document the extent to which managerial hoarding deters overall talent mobility in the firm, I analyze impacts on applications for any type of internal position. I begin by estimating whether deterred applicants would have been successful in getting hired. Column 2 of Table \ref{table:results_all}, Panel A reports 2SLS estimates of Equation \ref{eqn:iv}. I find that marginal applicants face a 49.1\% hiring probability. This large and positive hiring probability implies that a meaningful share of marginal applicants forgo high-stakes applications that would have been successful, indicating that the negative consequences of talent hoarding are non-negligible. This hiring probability of marginal applicants is also much higher than the average hiring likelihood at the firm, which is 27.6\%, suggesting that marginal applicants are positively selected based on the assessments of the firm's hiring managers. 
    
    A complier analysis based on \cite{abadie2003} corroborates that talent hoarding reduces applicant quality.\footnote{Under standard IV assumptions, complier characteristics can be estimated as $E[X_{it} \vert \text{Compliers}]$ for some characteristic $X_{it}$.  I calculate average complier characteristics and standard errors by performing 2SLS using the first-stage Equation \ref{eqn:fs} and a reduced-form equation replacing the outcome variable in Equation \ref{eqn:rf} with $X_{it}A_{it}$, where $X_{it}$ corresponds to a characteristic of individual $i$ and $A_{it}$ is a binary indicator for $i$ applying in quarter $t$.  I compute characteristics for always takers, who apply even in the presence of talent hoarding, by estimating an OLS regression of $X_{it}A_{it}(1-Z_{it})$ on $A_{it}(1-Z_{it})$, which allows me to estimate $E[X_{it} \vert \text{Always takers}]$.}  Panel A of Table \ref{table:complier} compares average characteristics across the entire employee population (Column 1), always takers who apply even in absence of manager rotations (Column 2), and marginal applicants who only apply if a manager rotates (Column 3). I find that marginal applicants are positively selected.  While 48.9\% of always takers hold a graduate degree, this is true for 59.4\% of marginal applicants. Similarly, 61.8\% of always takers received high performance ratings prior to applying, compared to 70.4\% of marginal applicants. These results are consistent with the predictions of the conceptual framework, which indicate that talent hoarding reduces the share of high-quality workers in the applicant pool, potentially limiting the ability of the firm to fill positions with suitable candidates.
	
	The fact that marginal applicants are of higher quality and that the firm would have liked to hire a high share of them suggests that these applicants would have been good fits for the new positions. To provide suggestive evidence that these applicants would have likely been productive in their new positions, I use data on employees' performance ratings, which are designed to provide task-specific feedback on whether a worker has accomplished her tasks in the past evaluation cycle. Since most workers in a team perform very similar tasks, performance ratings are particularly well-suited for drawing comparisons across workers within teams. I assess the possibility of forgone performance by estimating a 2SLS regression in which the outcome is defined as a worker landing a position \textit{and} performing better than the average performance of other workers in the new team one year later. Column 3 of Table \ref{table:results_all}, Panel A presents the results of this analysis and shows that 17.7\% of marginal applicants who apply to and land a new position also outperform their teammates. 

  Because of the important role that allocating workers to higher-level positions plays for the efficient allocation of talent in organizations, I also analyze applications to higher-level positions, using pay as a simple measure to identify these positions.\footnote{Higher-level positions are defined as those yielding a pay increase relative to workers' current position. Robustness tests in Appendix Section \ref{sec:appendix_robustness} find similar results using an alternative measure to identify higher-level positions.} Panel B of Table \ref{table:results_all} shows that 20.1\% of marginal applicants, who would not have applied in absence of a manager rotation, land a higher-level position and outperform their team, indicating that a non-negligible share of deterred applicants could have been productive in higher-level positions. This finding suggests that in the presence of talent hoarding, firms may be failing to fully realize productivity gains because workers are not assigned to positions in which they would be productive.
  
\begin{figure}[!htb]
		\caption{Worker-Reported Perceptions of Manager by Worker Gender}
		\centering
		\includegraphics[scale=1.0]{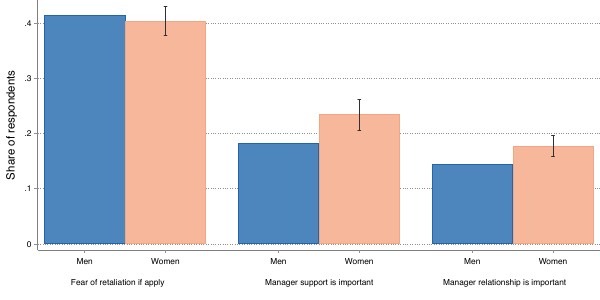}
		\label{fig:talenthoarding_surveygend}
		\vspace{0.3cm}
		\begin{minipage}[b]{0.9\linewidth}
			\footnotesize \textit{Notes}: This figure depicts workers' survey responses about their managers separately estimated by gender. Bars 1 and 2 represent the share of workers who report fearing retaliation if managers find out about internal applications. Bars 3 and 4 represent the share of workers who report manager support as critical for their career development. Bars 5 and 6 report the share of workers who rank a good relationship with their manager as  most important job feature. While men and women are similarly likely to report fearing retaliation, women are 26\% more likely to  value manager support and  22\% more likely to value a good relationship. See Appendix Section \ref{sec:app_survey} for more information on the employee survey and the exact question wording. Robust standard errors are used and 95\%-level confidence intervals are displayed. Controls: Age, German citizenship, educational qualifications, family status, firm tenure, function, full-time, and hours. N=1x,xxx. 
		\end{minipage}	
	\end{figure}

	In addition to creating efficiency costs, talent hoarding may impact between-group inequality if some workers are more affected than others. The survey evidence in Section \ref{sec:survey} indicates that talent hoarding  occurs through direct interpersonal interactions, suggesting that impacts may be larger for workers who depend more heavily on managerial support or are more sensitive to confrontation with their manager.  In line with previous work (\citealp{bertrand2011}), employees' survey responses highlight the possibility that talent hoarding may affect women more than men (Figure \ref{fig:talenthoarding_surveygend}). Women are 26\% more likely to mention the importance of manager support for their career development ($p$ = 0.000) and are 22\% more likely to rank a good relationship with their supervisor as the most important feature of their job ($p$ = 0.001). These differences suggest that women rely more on managers' career guidance and place more value on preserving a good relationship with their manager. 
	
	I conduct three exercises to assess the extent of gender differences. First, I conduct the complier analysis separately by gender in order to analyze whether talent hoarding affects women at a higher part of the quality distribution.  Panels B and C of Table \ref{table:complier} compare average characteristics across the entire employee population (Columns 4 and Column 7), always takers who apply even in absence of manager rotations (Columns 5 and 8), and marginal applicants who only apply if a manager rotates (Columns 6 and Column 9). For both men and women, marginal applicants are positively selected. However, the extent of this positive selection is more pronounced for women. While 61.2\%  of marginal female  applicants hold a graduate degree, this is only true for 36.3\%  of always takers. For men, 59.4\% of marginal applicants and 52.5\% of always takers hold a graduate degree. Similarly, 77.7\% of women  at the margin received high performance ratings relative to 61.5\% of always takers. Among men, 69.2\% of marginal applicants versus 62.0\% of always takers received high ratings. These patterns suggest that higher-quality women are affected by talent hoarding. 
	
    Second, I compare mobility effects by gender by estimating the hiring probabilities for men and women at the margin. Columns 3 and 4 of Table \ref{table:results_gender} report the 2SLS estimates of Equation \ref{eqn:iv}, which are separately estimated by gender and capture the probability of landing a higher-level position for marginal applicants. I find that both men and women experience positive and statistically significant marginal hiring probabilities of 41.2\% and 62.3\%, respectively. These hiring probabilities are much larger than those of always takers in my sample, which are 26.9\% for men and 28.6\% for women. Although the gender difference among marginal applicants is not statistically significant, the extent of positive selection within women is striking (i.e., marginal vs. average applicants), and suggests that talent hoarding may be more consequential for women's career outcomes.
	
\begin{figure}[!htb]
		\caption{Potential Outcomes by Application Status for Marginal Applicants}
		\vspace{2mm}
		\centering
		\includegraphics[scale=1.8]{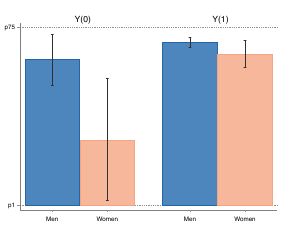}\\
		\label{fig:potentialoutcomes}
		\vspace{0.3cm}
		\begin{minipage}[b]{0.9\linewidth}
			\footnotesize \textit{Notes}: This figure shows potential outcomes with respect to log real annual earnings in quarter $t+4$, measured in percentiles. The left two bars represent the potential outcomes for marginal applicants had they not applied, labeled $Y(0)$. The right two bars represent the potential outcomes for marginal applicants had they applied, labeled $Y(1)$. Robust standard errors are clustered at the worker and rotation level and 95\%-level confidence intervals are displayed. Controls: Female, age, German citizenship, educational qualifications, marital status, family status, parental leave, firm tenure, division, function, location, full-time, hours, number of direct reports, and quarter fixed effects. N=3xx,xxx.   
		\end{minipage}	
	\end{figure}

	A third test for disparate impacts by gender compares the average potential outcomes with respect to compliers' earnings in the treated state (i.e., when marginal applicants apply because talent hoarding ceases) to those in the untreated state (i.e., when marginal applicants do not apply due to talent hoarding). This analysis leverages the potential outcomes framework that follows from the interpretation of $\beta_{IV}$ as the LATE for marginal applicants. The advantage of this approach is that it allows a comparison of the same set of individuals across two different potential outcomes, avoiding potential composition bias. Figure \ref{fig:potentialoutcomes} presents estimates of log annual earnings in quarter $t+4$ for marginal applicants in both potential outcomes states by gender. Outcomes are reported in terms of percentiles at the firm. Both men and women experience higher earnings if they choose to apply; however, the larger gains realized by women lead to a reduction in the pay gap of 86\%. These findings suggest that talent hoarding may exacerbate gender inequality in the firm, raising the possibility that talent hoarding has non-negligible consequences with respect to both efficiency and equity in the internal labor market. 
	\vspace{-0.3cm}
	\section{Conclusion}
	\label{sec:conclusion}
	\vspace{-0.3cm}
	This paper provides the first empirical evidence that talent hoarding is an important source of frictions in organizations. Using a unique combination of survey and administrative data from a large manufacturing firm, I show that managers face strong incentives to hoard talented workers in their team and act accordingly. My findings indicate that managerial talent hoarding is commonplace (explicitly reported by 75\% of managers)  and has non-negligible effects on the internal allocation of talent within the firm. While my results provide the first detailed insights on talent hoarding, additional evidence suggests that such talent hoarding behavior is endemic. In a survey of the top publicly listed companies in Germany, 83\% cite talent hoarding as a crucial friction in their organization (\citetalias{hkp}). Firm surveys in other countries, such as the United States, document that talent hoarding is not limited to German organizations (\citealp{i4cp}, \citealp{kornferry}).
	
    Because talent hoarding arises due to misaligned incentives, a natural solution would be to more closely align the incentives of managers with those of the firm. In theory, firms could reward managers based on their subordinates' future performance, irrespective of whether subordinates remain on the team. However, HR leaders typically state that realigning incentives through financial compensation is infeasible (\citetalias{hkp}). Similarly, when asked under which conditions managers would be more likely to support workers' career development, only a minority of managers cite financial incentives. Instead, 68\% of managers highlight the ease of replacing lost talent as key condition. In addition, policies that increase application rates---such as implementing regular application schedules and having other organizational agents, such as the HR department, directly invite workers to apply for positions---could reduce the scope for managers to engage in talent hoarding. The survey I conduct shows that for such policies to be effective, the firm must be able to deter managers from retaliating against workers, for instance by assuring full confidentiality for applicants. 

    Although the potential benefits of these policies for efficient talent allocation may be large, it is possible that such policies could discourage managers from investing in their workers' human capital. While a nuanced assessment of the costs and benefits associated with managerial talent hoarding incentives is outside the scope of this paper, survey responses suggest the existence of additional costs of talent hoarding that are separate from those relating to talent allocation. For example, surveyed workers who report being subject to talent hoarding are 30\% more likely to report having searched for external jobs, indicating that talent hoarding may create unwanted turnover of high-quality workers the firm would like to retain. In addition, the survey responses indicate that managers hoard talent by deterring workers from pursuing training programs and from participating in high-profile projects, limiting skill development. Together, these findings suggest that managerial talent hoarding has broad impacts on talent in organizations.

	\begin{singlespace}
		\bibliography{literature202202}
		\nocite{*}
	\end{singlespace}

	\clearpage
	\newpage
	\section{Figures and Tables}

	
	\begin{figure}[!htb]
		\caption{Self-Reported Misaligned Incentives Among Managers}
		\centering
		\vspace{0.5cm}
		\begin{minipage}[b]{1.0\linewidth}
			\centering	
			\includegraphics[scale=1.5]{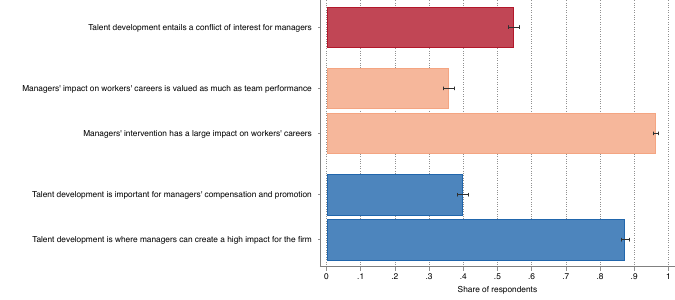}
		\end{minipage}	\\
		\begin{minipage}[b]{1.0\linewidth}
			\vspace{0.5cm}
			\footnotesize \textit{Notes}: This figure uses the manager survey to depict the incentives managers face with respect to supporting their workers' career development. 
Each bar reflects the share of managers who indicated a particular response option. The exact wording of each question, along with the full distribution of responses, is provided in the respective Appendix Tables indicated in parentheses.  The first bar shows the share of managers (55\%) who agree with the statement that developing workers entails a conflict of interest for managers, because more developed workers are more likely to leave the team (Appendix Table \ref{tab:QF}). The second bar shows the share of managers (36\%) who agree with the statement that managers' impact on developing workers is valued at least as much as their impact on team performance (Appendix Table \ref{tab:QF}). The third bar shows the share of managers (96\%) who agree with the statement that a manager's direct intervention has a large impact on workers' career development (Appendix Table \ref{tab:QF}). The fourth bar shows the share of managers (40\%) who indicated that a record of success with respect to talent development is important for managers' own compensation and promotion prospects (Appendix Table \ref{tab:QB}). The fifth bar shows the share of managers (87\%) who indicated that managers can create a lot of impact for the firm with respect to talent development (Appendix Table \ref{tab:QA}).  See Appendix Section \ref{sec:app_survey_manager} for more information on the manager survey. 95\%-level confidence intervals are displayed.  Controls: Female, age, German citizenship, firm tenure, experience, function. N=3,xxx. 
		\end{minipage}	
		\label{fig:manager_incentives}
	\end{figure}
	
	\begin{figure}[h]
		\thisfloatpagestyle{plainlower}
		\caption{Distribution of Manager-Level Deviations in Potential Ratings}
		\centering
		\includegraphics[scale=3.75]{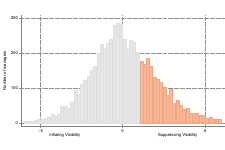}
		\label{fig:talenthoarding_distribution}
		\begin{minipage}[b]{0.9\linewidth}
			\footnotesize \textit{Notes}: This figure depicts the mean deviation by manager between predicted and actual potential ratings, which captures systematic discrepancies in worker visibility, and serves as a measure of managers' propensities to hoard talent. The y-axis reports the number of unique managers with a given deviation.  I assess managers' systematic underreporting of public potential ratings by comparing managers' actual potential rating to the predicted potential rating based on managers' own assessment of worker performance and worker characteristics. Positive values indicate managers who, on average, assign lower public potential ratings than predicted based on their own performance evaluations of workers, suggesting suppression of visibility. Negative values indicate managers who inflate visibility by assigning public ratings that exceed the predicted level.  For ease of interpretability, I label hoarding-prone managers as those in the upper tercile of the distribution (shaded in orange) because they have a high tendency to suppress visibility, while those in the two lower terciles (shaded in gray) are either likely to enhance visibility or provide potential ratings that are roughly aligned with expectations.  See Appendix Table \ref{table:measure_summ} for detailed summary statistics of the distribution.
            Controls: Female, age, German citizenship, educational qualifications, marital status, family status, parental leave, firm tenure, division, function, location, full-time, hours, number of direct reports, and quarter fixed effects. N=6,xxx. 
		\end{minipage}	
	\end{figure}

	\FloatBarrier
	\begin{figure}[!ht]
		\thisfloatpagestyle{plainlower}
		\caption{Effect of Manager Rotations on Applications and Job Transitions}
		\centering
		\begin{minipage}[b]{0.8\linewidth}
			\centering
			\caption*{Panel A. Application Probability}
			\includegraphics[scale=0.7]{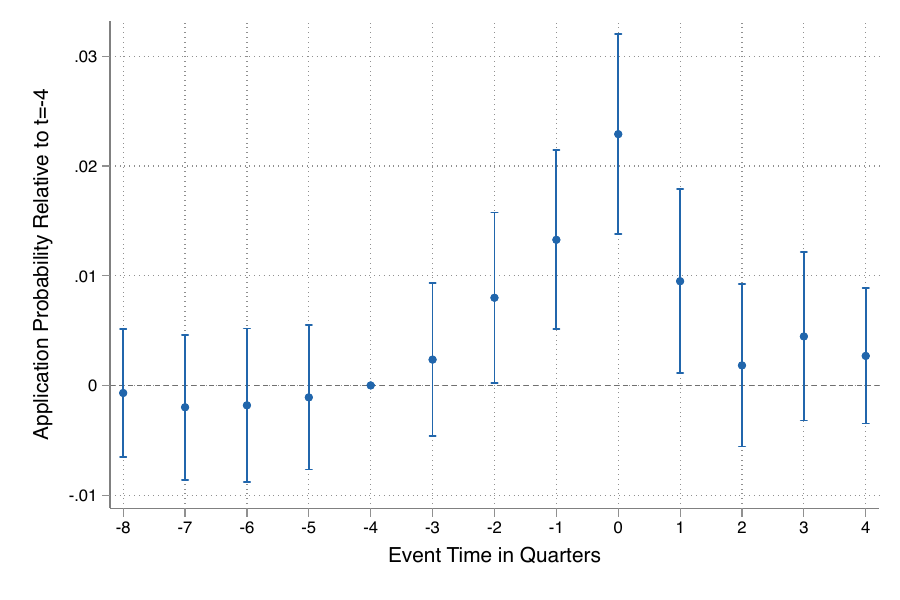}
		\end{minipage}	\\
		\begin{minipage}[b]{0.8\linewidth}
			\centering
			\caption*{Panel B. Cumulative Transition Probability}
			\includegraphics[scale=0.7]{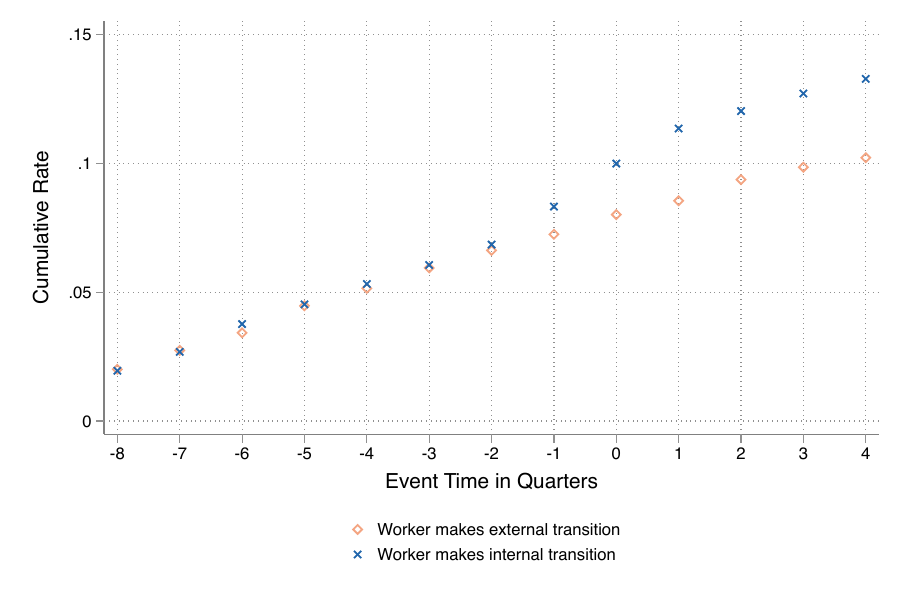}
		\end{minipage}	\\
		\label{fig:msw_dynamics}
		\begin{minipage}[b]{1.0\linewidth}
			\footnotesize \textit{Notes}: This figure depicts  internal applications (Panel A) and job transitions (Panel B) around a manager rotation.  Panel A presents estimates from an event study regression, in which the outcome is an indicator that the worker  applied for any internal position in a  quarter and event time is defined relative to the occurrence of a manager rotation. The specification includes worker and quarter fixed effects, but no other controls. I bin event time dummy variables at $t=-8$ and $t=4$. The mean application rate as of $t=-4$ is 0.027. The sample also includes those who have not experienced a manager rotation. Robust standard errors are clustered at the worker and rotation level and 95\%-level confidence
intervals are displayed. Panel B plots the cumulative share of workers who have exited the team via internal (i.e., within the firm) and external (i.e., out of the firm) transitions around a manager rotation. Workers are assigned to their team as of ten quarters before the team experiences a manager rotation. N=3xx,xxx.
		\end{minipage}	
	\end{figure}
	\FloatBarrier

	\FloatBarrier
	\begin{figure}[!ht]
		\thisfloatpagestyle{plainlower}
		\caption{Heterogeneity in Application Responses by Hoarding Incentive Proxies}
		\begin{minipage}[b]{0.8\linewidth}
			\centering
			\includegraphics[scale=1.0]{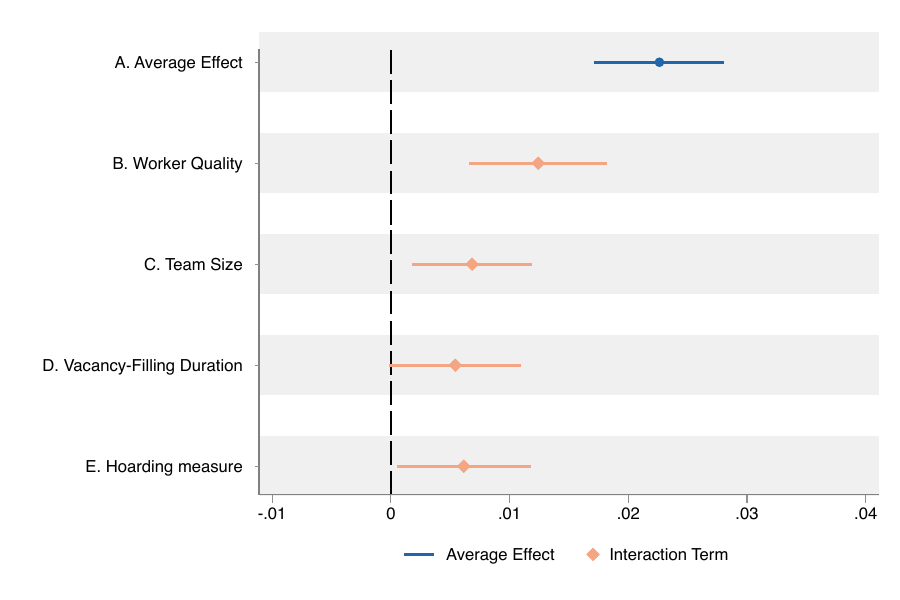}
		\end{minipage}	\\
		\label{fig:mechanism_departurecost}	
		\begin{minipage}[b]{1.0\linewidth}
			\footnotesize \textit{Notes}:    This figure examines how the effect of manager rotations on worker applications varies with proxies for talent hoarding incentives. The outcome is an indicator for whether a worker applied to an internal job in a given quarter. Each coefficient stems from a separate regression based on Equation \ref{eqn:fs}.  Panel A shows the baseline average effect of a manager rotation (in blue). 
            The orange coefficients show the effect  of the rotation interacted with a standardized (mean-zero, SD=1) version of the respective proxy for hoarding incentives.  Panel B interacts rotations with an index of predicted worker quality (constructed using the predicted value from an OLS regression of workers' internal hiring probability on worker characteristics).
Panel C uses a negated measure of team size so that higher values correspond to smaller teams. Panel D uses the average duration (in days) it takes to fill vacancies in the worker’s functional area.
Panel E uses a continuous measure of manager hoarding propensity, defined as the average deviation between workers' predicted and actual potential ratings. See Section \ref{sec:th_measure} for details on how this measure is constructed.  Appendix Figures 	\ref{fig:rob_workerquality}	and 	\ref{fig:rob_hoard}	 show robustness to alternative specifications.
Robust standard errors are clustered at the worker and rotation level. 95\%-level confidence intervals are shown. Controls include: gender, age, German citizenship, education, marital and family status, parental leave, firm tenure, division, function, location, full-time status, hours worked, number of direct reports, and quarter fixed effects. N=3xx,xxx.
		\end{minipage}	
	\end{figure}

	
	\FloatBarrier

	\clearpage
	\newpage
	
	\begin{table}[h]
		\caption{Summary Statistics of Analysis Sample}
		\begin{center}
			\scalebox{1.0}{{
\def\sym#1{\ifmmode^{#1}\else\(^{#1}\)\fi}
\begin{tabular}{l*{1}{cccc}}
\hline\hline
\multicolumn{2}{l}{\textbf{Panel A}: All employees} & & & \\
                                        &      Mean &    Std. deviation&       p25  & p75 \\
\hline
Female             &        0.21&        0.40&        0.00&        0.00\\
German citizen    &        0.90&        0.30&        1.00&        1.00\\
Age (years)          &       43.33&        9.97&       35.00&       51.25\\
Tenure at firm (years)&       13.23&        9.62&        5.00&       18.75\\
Schooling (years)  &       15.81&        2.73&       12.00&       18.00\\
Married             &        0.62&        0.49&        0.00&        1.00\\
Children                &        0.73&        0.44&        0.00&        1.00\\
Technical position &        0.63&        0.48&        0.00&        1.00\\
Full-time           &        0.92&        0.26&        1.00&        1.00\\
Weekly hours       &       41.18&        4.54&       40.75&       43.50\\
\hline
Observations    &\multicolumn{4}{c}{3xx,xxx}        \\
 \hline 
   \\[0.1mm]     
 \multicolumn{2}{l}{\textbf{Panel B}: Managers} & & & \\
                                        &      Mean &    Std. deviation&       p25  & p75 \\
\hline
Female              &        0.10&        0.31&        0.00&        0.00\\
German citizen      &        0.92&        0.26&        1.00&        1.00\\
Age (years)         &       47.75&        7.72&       42.25&       53.50\\
Tenure at firm (years)&       16.61&        9.01&        9.75&       22.50\\
Schooling (years)   &       16.68&        2.34&       16.00&       18.00\\
Married             &        0.78&        0.41&        1.00&        1.00\\
Children            &        0.88&        0.32&        1.00&        1.00\\
Technical position  &        0.59&        0.49&        0.00&        1.00\\
Full-time           &        0.97&        0.16&        1.00&        1.00\\
Weekly hours        &       42.83&        2.21&       43.50&       43.50\\
Number of direct reports&        8.61&        4.78&        5.00&       11.00\\
 \hline
Observations    &\multicolumn{4}{c}{7x,xxx}        \\
\hline\hline\\
\multicolumn{5}{p{0.8\linewidth}}{\footnotesize \textit{Notes}: This table reports summary statistics for  my quarterly analysis sample. Panel A describes all employees in my sample. Panel B provides summary statistics for the subsample of  managers, defined as those with responsibility over a team.   A technical position is defined as a job related to engineering, IT, quality management, or production. In total, my sample consists of over 300,000 employee-by-quarter observations from 2015 to 2018. }\\
\end{tabular}
}}
		\end{center}
		\label{table:sample_descriptive}
	\end{table}

	\begin{table}
		\caption{Heterogeneity in Managers' Suppression of Public Ratings by Incentives}
		\bigskip
		\begin{center}
			\scalebox{1.0}{{
\def\sym#1{\ifmmode^{#1}\else\(^{#1}\)\fi}
\begin{tabular}{l*{3}{c}}
	  \hline  \hline 
\underline{Outcome variable} & \multicolumn{3}{c}{Top tercile hoarding measure}\\ \cline{2-4}
	 
                                       &\multicolumn{1}{c}{(1)}           &\multicolumn{1}{c}{(2)}         &\multicolumn{1}{c}{(3)}        \\
\hline
 \\[0.1mm] 
Financial incentive (share)    &   0.1892   &-- & -- \\
                            &   (0.033)           & --   &  --        \\  
                             \\[0.1mm]     
Team size    &  --   & -0.0133 & -- \\ 
&   --         & (0.002)     & --        \\  
 \\[0.1mm]     
Talent visibility    &   --   & -- & 0.0397 \\
&   --        & --    &  (0.013)         \\  
                            \\[0.1mm]       
                             
\hline
Outcome Mean            &    0.3000               &    0.3000                    &     0.3000       \\
Predicted difference p90-p10 & 0.1302 & -0.1308 & 0.0397 \\
Observations        &       6,xxx    &    6,xxx        &   6,xxx        \\
 
\hline \hline					\multicolumn{4}{p{0.6\linewidth}}{\footnotesize \textit{Notes}: This table examines whether managerial talent hoarding, measured by the suppression of public ratings, is more prevalent among managers with stronger incentives to hoard talent.   I use three distinct measures of hoarding incentives. Column 1 captures financial incentives using the share of total compensation that is performance-related. Column 2 uses the number of team members overseen by the manager. Column 3 uses a binary indicator for low talent visibility, defined as working in a functional area with limited visibility of individual achievements. Each coefficient is estimated separately based on a linear model for managers’ hoarding propensity using an OLS regression of the indicator that managers are hoarding-prone, defined by their mean deviation between predicted and actual potential ratings being in the top tercile of the manager distribution, on one incentive measure at a time. The predicted difference  p90-p10 refers to the predicted difference in the outcome between the 90th and 10th percentiles of the continuous regressors (between values 1 and 0 for binary indicators), holding all other covariates at their mean values. See Section \ref{sec:th_measure} for details on the construction of this alternative hoarding measure. See Appendix Section \ref{sec:hoarding_incentives_details} for definitions of the incentive measures and Appendix Table \ref{tab:incentives_summ} for their distribution in the sample. See Appendix Tables   \ref{tab:incentives_reg} and  \ref{tab:hoarding_predictions_outcome} for complementary results that show robustness of my findings to alternative specifications of the regressors and the outcome variable. See Appendix Table  \ref{tab:hoarding_predictions_training}  for parallel results using managers’ withholding of access to high-visibility training opportunities as an alternative measure of hoarding behavior. Controls: Female, age, German citizenship, firm tenure, educational qualification. Robust standard errors in parentheses.}
 \end{tabular}}

}
		\end{center}
	 \label{fig:hoarding_predictions}
	\end{table}
	
	\begin{table}
		\caption{Heterogeneity in Manager-Reported Talent Hoarding by Incentives}
		\bigskip
		\begin{center}
			\scalebox{1.0}{{
\def\sym#1{\ifmmode^{#1}\else\(^{#1}\)\fi}
\begin{tabular}{lccccccc}
	  \hline  \hline 

\underline{Outcome variable:}	   &\multicolumn{3}{c}{Dissuade workers from leaving} &   &\multicolumn{3}{c}{Risk of losing talent}   \\ \cline{2-4} \cline{6-8}
                                       &\multicolumn{1}{c}{(1)}           &\multicolumn{1}{c}{(2)}         &\multicolumn{1}{c}{(3)} &  &\multicolumn{1}{c}{(4)}           &\multicolumn{1}{c}{(5)}         &\multicolumn{1}{c}{(6)}         \\
\hline
 \\[0.1mm] 
Financial incentive (share)     &    0.2723  &-- & -- &&  0.3226  & -- & --  \\
                            &    (0.116)  &-- & -- &&  (0.156)  & -- & --   \\      
                            \\[0.1mm]     
Team size     &   -- &  -0.0040 & -- && -- & -0.0065  & -- \\
                            & --           &  (0.002)  & -- && -- &  (0.002)  & -- \\
                            \\[0.1mm]  
Talent visibility      &    --  &  -- &0.0669 & & -- & -- & 0.1443\\
                            &   --        &  --   &(0.062)  && -- & -- & (0.056)    \\    [0.1mm]    
                          
\hline
Outcome Mean            &     0.7506           &       0.7506              &     0.7506 &        &      0.4461          &       0.4461        &      0.4461     \\
Predicted difference p90-p10 & 0.0343 & -0.0464 &0.0669 && 0.0406 & -0.0747 &0.0143 \\
Observations        &       3,xxx    &    3,xxx        &   3,xxx       & &       3,xxx    &    3,xxx        &   3,xxx \\

\hline \hline					\multicolumn{8}{p{\linewidth}}{\footnotesize \textit{Notes}: This table uses data from the manager survey to examine whether managers who face stronger incentives to hoard talent are more likely to report talent hoarding behaviors. Columns 1 to 3 measure talent hoarding based on whether respondents agree that managers feel the need to dissuade team members from exploring opportunities in other departments. The corresponding survey question reads: \textit{How often might leaders find themselves in situations where they need to dissuade a team member
from exploring opportunities in another department due to immediate team needs or performance
goals?} See Appendix Table \ref{tab:QG} for the distribution of responses. Columns 4 to 6 measure talent hoarding based on whether respondents agree that the risk of losing talent prevents managers from investing time and effort in workers’ career development. The corresponding survey question reads:  \textit{What are the reasons that may prevent leaders from investing time and effort towards their
employees’ career development?} See Appendix Table \ref{tab:QD} for the distribution of responses. 
 To proxy for managers’ incentives to hoard talent, I use three measures: 1) the share of total compensation that is performance-related, 2) the number of team members overseen by the manager,  3) a binary indicator for low talent visibility,  defined as working in a functional area with limited visibility of individual achievements.  Each coefficient is estimated separately based on a linear model for managers’ hoarding propensity using an OLS regression of the survey-based indicator that managers are hoarding-prone on one incentive measure at a time. The predicted difference  p90-p10 refers to the predicted difference in the outcome between the 90th and 10th percentiles of the continuous regressors (between values 1 and 0 for binary indicators), holding all other covariates at their mean values. See Appendix Section \ref{sec:robust_incentives_survey} for how these incentives are measured and Appendix Table  \ref{tab:incentives_summ_survey} for how these measures are distributed in my sample. Appendix Tables \ref{table:survey_hoarding_predictions_robust} and \ref{table:survey_placebo} present robustness checks.  See Appendix Section \ref{sec:app_survey_manager} for additional details on the survey instrument. Controls: Female, age, German citizenship, firm tenure, experience, function. Robust standard errors in parentheses. N=3,xxx.}
 \end{tabular}}

}
		\end{center}
		\label{table:survey_hoarding_predictions}
	\end{table}
	

	\begin{table}
		\caption{Testing for Random Assignment of Manager Rotations}
		\bigskip
		\begin{center}
			\scalebox{1.0}{{
\def\sym#1{\ifmmode^{#1}\else\(^{#1}\)\fi}
\begin{tabular}{l*{2}{c}}
	  \hline  \hline \\
   \multicolumn{2}{l}{\textbf{Panel A}: Workers} & \\
 &   No  Rotation&     Rotation  \\
 &\multicolumn{1}{c}{(1)}         &\multicolumn{1}{c}{(2)}      \\
\hline

German citizen      &        0.90&        0.88\\
Female              &        0.21&        0.20\\
Age (years)         &       43.34&       42.79\\
Tenure at firm (years)&       13.24&       12.73\\
Schooling (years)   &       15.81&       15.98\\
Married             &        0.62&        0.61\\
Children            &        0.73&        0.73\\
High performance rating&        0.63&        0.61\\
Team size           &        9.08&        8.99\\
Quarters worked with manager&        9.72&        9.31\\
Recorded absence    &        0.09&        0.10\\
\hline  \\[0.1mm] 
 \multicolumn{2}{l}{\textbf{Panel B}: Managers} & \\
 &   No  Rotation&     Rotation  \\
 &\multicolumn{1}{c}{(1)}         &\multicolumn{1}{c}{(2)}      \\
\hline

German citizen      &        0.93&        0.91\\
Female              &        0.10&        0.12\\
Age (years)         &       47.97&       46.31\\
Tenure at firm (years)&       16.72&       15.48\\
Schooling (years)   &       16.60&       16.81\\
Married             &        0.79&        0.79\\
Children            &        0.89&        0.89\\
High performance rating&        0.79&        0.75\\
Team size           &        6.75&        6.58\\
Team: Quarters worked with manager&        9.53&        9.40\\
Team: Share recorded absence in past 4Q&        0.11&        0.12\\
Team: Share applied in past 4Q&        0.03&        0.04\\
Team: Share turnover in past 4Q&        0.01&        0.01\\
\hline\hline 
 \multicolumn{3}{p{0.8\linewidth}}{\footnotesize \textit{Notes}: This table tests for random assignment of manager rotations. Panel A focuses on worker characteristics and compares those who do not experience their manager switching in a given quarter (Column 1) to those who do experience their manager switching in a given quarter (Column 2). The total number of quarterly observations is 3xx,xxx. Panel B focuses on manager and team characteristics and compares managers who do not switch in a given quarter (Column 1) to those who switch in a given quarter (Column 2). The total number of quarterly observations is 7x,xxx.}
 \end{tabular}}
}
		\end{center}
		\label{table:iv_validity_fs_ind}
	\end{table}

	\clearpage
	\newpage
	
	\FloatBarrier
	\begin{table}
		\caption{Effect of Manager Rotation on Worker Outcomes}
		\begin{center}
			\scalebox{1.0}{{
\def\sym#1{\ifmmode^{#1}\else\(^{#1}\)\fi}
	\begin{tabular}{lccc}

	  \hline  \hline   \\[0.1mm] 

     \multicolumn{4}{l}{\textbf{Panel A}: Any Position}  \\
   & Applied & Hired & Hired + Perform  \\ 
   & OLS & IV & IV \\
& (1) & (2) &(3)   \\
\hline
Manager Rotation      &     0.0226 &- &-       \\
                            &     (0.003)        & - &-    \\
                            
Applied      &    -  &   0.4913&      0.1770 \\
                              &    -    & (0.059)         &     (0.045)      \\
\hline
Outcome Mean            &      0.0290         &      0.0080         &      0.0037            \\
Observations        &       3xx,xxx          &  3xx,xxx   &  3xx,xxx      \\
\hline   \\[0.1mm] 

 \multicolumn{4}{l}{\textbf{Panel B}: Higher-Level Position} \\
 & Applied & Hired & Hired + Perform\\
 & OLS & IV & IV \\
 & (1) & (2) &(3)   \\
\hline
Manager Rotation    &      0.0106&           -           &            -          \\
                           &     (0.002)         &    -                  &          -            \\
                            
Applied    &       -               &      0.4623&      0.2012 \\
                              &          -             &     (0.078)         &     (0.064)         \\
\hline
Outcome Mean          &         0.0102         &      0.0028         &      0.0013         \\
Observations        &       3xx,xxx          &  3xx,xxx   &  3xx,xxx      \\
\hline \hline					\multicolumn{4}{p{0.6\linewidth}}{\footnotesize \textit{Notes}: This table reports the effects of manager rotations on workers' career outcomes. Each coefficient is based on a separate regression. Panel A focuses on any type of internal position. Panel B focuses on higher-level positions, defined as positions that would yield a  pay increase relative to workers' current pay. Column 1 reports the first-stage effect of manager rotation on applications based on Equation \ref{eqn:fs}. Column 2 reports estimates from a two-stages least squares regression on landing a position that instruments for applying with manager rotation based on Equation \ref{eqn:iv}, which represents the LATE. Column 3 estimates a similar two-stages least squares regression, but use an indicator for landing a position \textit{and} performing  better than the leave-out team average one year later as outcome variable. Robust standard errors are clustered at the worker and rotation level. Controls: Female, age, German citizenship, educational qualifications, marital status, family status, parental leave, firm tenure, division, function, location, full-time, hours, number of direct reports, and quarter fixed effects. }
 \end{tabular}}
}
		\end{center}
		\label{table:results_all}
	\end{table}
	

	\begin{table}
		\caption{Effect of Manager Rotation on Worker Outcomes by Gender}
		\begin{center}
			\scalebox{1.0}{{
\def\sym#1{\ifmmode^{#1}\else\(^{#1}\)\fi}
\begin{tabular}{l*{6}{c}}
	  \hline  \hline 
	&   \multicolumn{6}{c}{ Higher-Level Position}  \\ \cline{2-7}
	  &\multicolumn{2}{c}{Applied}   &\multicolumn{2}{c}{Hired} &  \multicolumn{2}{c}{Hired + Perform} \\
      &\multicolumn{2}{c}{OLS}  &\multicolumn{2}{c}{IV} &  \multicolumn{2}{c}{IV} \\
                    &         Men         &    Women     &            Men         &       Women  &          Men         &       Women       \\
                                       &\multicolumn{1}{c}{(1)}         &\multicolumn{1}{c}{(2)}     &\multicolumn{1}{c}{(3)}         &\multicolumn{1}{c}{(4)}          &\multicolumn{1}{c}{(5)}         &\multicolumn{1}{c}{(6)}     \\
\hline
 \\[0.1mm] 
Manager Rotation      &   0.0102&      0.0118  &-&- &- &- \\
                            &      (0.002)         &     (0.004)        & -&-    &- &-    \\       
                            \\[0.1mm]       
                            
  Applied    &     - &      -    &  0.4179&      0.6231&      0.1891  &      0.2364      \\     
                              &     -           &  -       &   (0.087)         &     (0.184)         &     (0.074)         &     (0.133)       \\       
                            \\[0.1mm]       
\hline
Outcome Mean              &     0.0094         &      0.0135         &      0.0025         &      0.0039         &      0.0012       &      0.0021        \\
Observations        &       3xx,xxx        &      8x,xxx           &   3xx,xxx     &       8x,xxx    & 3xx,xxx       &      8x,xxx    \\
\hline \hline					\multicolumn{7}{p{0.8\linewidth}}{\footnotesize \textit{Notes}:  This table reports the effects of manager rotations on workers' career progression by gender. Each coefficient is based on a separate regression. Columns 1 and 2 report first-stage effects of manager rotation on applications to higher-level positions, defined as positions that would yield a pay increase relative to workers' current pay, based on Equation \ref{eqn:fs}. Columns 3 and 4 report estimates from a two-stages least squares regression on landing a position that instruments for applying with manager rotation based on Equation \ref{eqn:iv}. Columns 5 and 6 estimate a similar two-stages least squares regression, but use an indicator for landing a position  \textit{and}  performing  better than the leave-out team average one year later as outcome variable. Robust standard errors are clustered at the worker and rotation level. Controls: Female, age, German citizenship, educational qualifications, marital status, family status, parental leave, firm tenure, division, function, location, full-time, hours, number of direct reports, and quarter fixed effects. }\\ \end{tabular}}}
		\end{center}
		\label{table:results_gender}
	\end{table}
	\FloatBarrier

    \begin{landscape}
        
	\begin{table}
		\caption{Characteristics of Marginal Applicants (in \%)}
		\begin{center}
			\scalebox{0.85}{{
\def\sym#1{\ifmmode^{#1}\else\(^{#1}\)\fi}
\begin{tabular}{lccc| ccc|ccc}
	  \hline  \hline 
   & \multicolumn{3}{c}{\textbf{Panel A}: All} & \multicolumn{3}{c}{\textbf{Panel B}: Men}  & \multicolumn{3}{c}{\textbf{Panel C}: Women}  \\
& All & Always& Marginal & All & Always& Marginal &All & Always & Marginal\\
&workers  & takers& applicants&workers  & takers& applicants & workers & takers & applicants\\
& (1) & (2) &(3) & (4) & (5) & (6) & (7) & (8) & (9) \\
\hline

German citizen          &    89.8&    88.3&    90.4       &    90.7&    89.2&    89.0  &    86.4&     85.2&    95.8\\
Age $>=$40 years          &    60.4&    45.3&    50.4 &    63.6&    49.7&    50.6  &    47.7&    30.0&    49.1\\
Married             &    61.7&    56.6&     53.6    &    64.5&    62.2&    58.6  &    50.5&    36.8&    36.6\\
Children            &    73.3&    69.7&    70.0     &    76.0&    74.6&    65.5  &    62.7&    52.4&    88.2\\
In position$<$2 years  &    36.4&    38.8&    40.4 &    36.1&    38.1&    40.9 &    37.3&    41.1&    38.1\\
In position 2-5 years  &    36.4&    39.9&    43.9 &    36.3&    39.2&    42.1 &    36.6&    42.0&    50.7\\
In position$>=$5 years &    27.0&    21.1&    16.0 &    27.3&    22.4&    17.3 &    25.6&    16.4&    12.0\\
Technical position  &    63.2&    57.3&    60.3  &     71.0&    65.3&    69.9 &    32.7&    29.2&    26.7\\
Full-time            &    92.5&    94.4&    97.7          &    97.4&    97.9&    101.3  &    73.3&    82.1&    85.5\\
Graduate degree     &     47.6&    48.9&    59.4    &    51.5&    52.5&    59.4 &    32.3&    36.3&    61.2\\
High performance    &    63.2&    61.8&    70.4    &    63.7&    62.0&    69.2 &    60.7&    61.5&    77.7\\ 

\hline \hline					\multicolumn{10}{p{0.9\linewidth}}{\footnotesize \textit{Notes}: This table illustrates results from a complier analysis, as described in Section \ref{sec:results_misallocation}. Panel A contains all employees. Panel B focuses only on men. Panel C focuses only on women.  Each number represents the share of individuals in a given group that exhibit the respective characteristic. The first column in each panel shows raw means. The second column in each panel represents always takers. The third column in each panel represents marginal applicants, who only apply if managers rotate and talent hoarding temporarily abates. Each number in the second and third column of each panel is based on a separate regression including controls and represents an adjusted mean (in \%). Controls: Female, age, German citizenship, educational qualifications, marital status, family status, parental leave, firm tenure, division, function, location, full-time, hours, number of direct reports, and quarter fixed effects. 
 N=3xx,xxx. }\\ \end{tabular}}

}
		\end{center}
		\label{table:complier}
	\end{table}

    \end{landscape}

	
	\renewcommand{\appendixpagename}{Online Appendix}
	\clearpage
	\newpage

	\appendix
    \appendixpage
\addappheadtotoc
	\setcounter{table}{0}
	\renewcommand{\thetable}{\thesection\arabic{table}}
	\setcounter{figure}{0}
	\renewcommand\thefigure{\thesection\arabic{figure}}

	\section{Appendix Figures}

	\FloatBarrier
	\begin{figure}[p]
		\thisfloatpagestyle{plainlower}
		\caption{Number of Manager Rotations by Managers' Length in Position}
		\centering
		\includegraphics[scale=0.7]{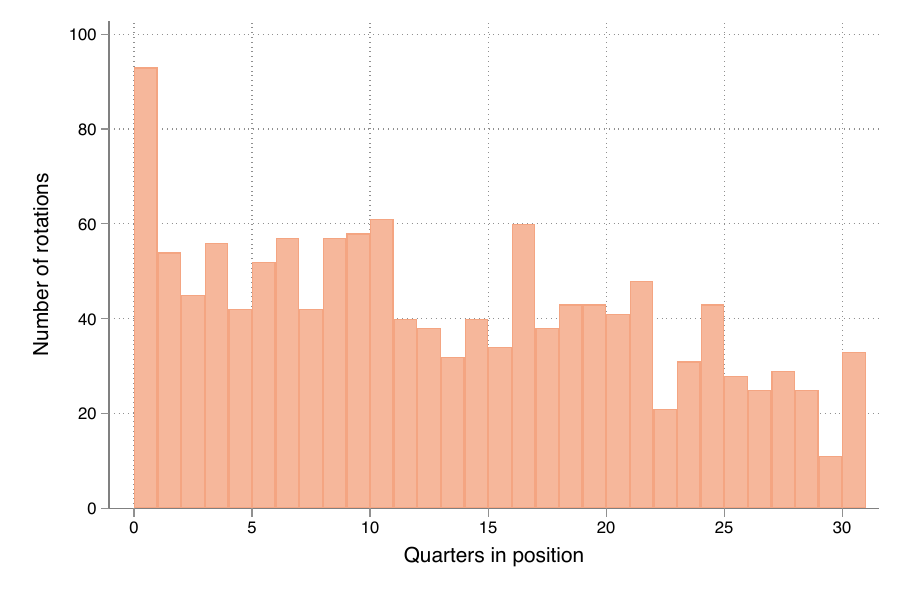}
		\label{fig:rotation_timeposition}
		\begin{minipage}[b]{0.9\linewidth}
			\footnotesize \textit{Notes}: This figure illustrates variation in the timing of manager rotations, as measured by the number of quarters a manager has been in their position at the time of rotation. The  number of observations is  1,359, representing the total number of all manager rotations I use for my analysis. 
		\end{minipage}	
	\end{figure}

	
		\FloatBarrier
	\begin{figure}[p]
		\thisfloatpagestyle{plainlower}
		\caption{Effect of Manager Rotation on Team-Level and Manager-Level Outcomes}
		\centering
		\begin{subfigure}{0.45\textwidth}
			\centering
			\caption*{Panel A. Team Absenteeism}
			\includegraphics[width=\linewidth]{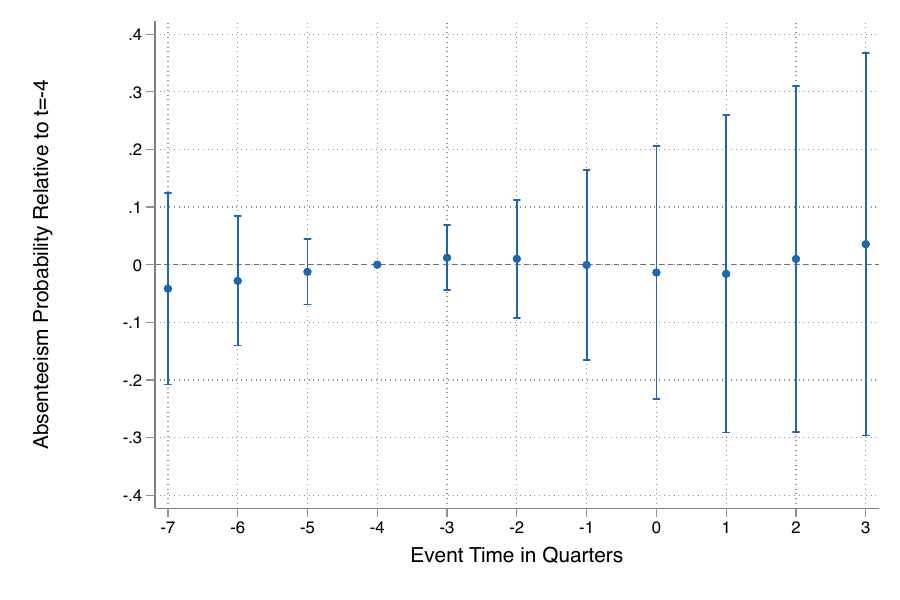}
		\end{subfigure}%
		\begin{subfigure}{0.45\textwidth}
			\centering
			\caption*{Panel B. Team Log(Bonus)}
			\includegraphics[width=\linewidth]{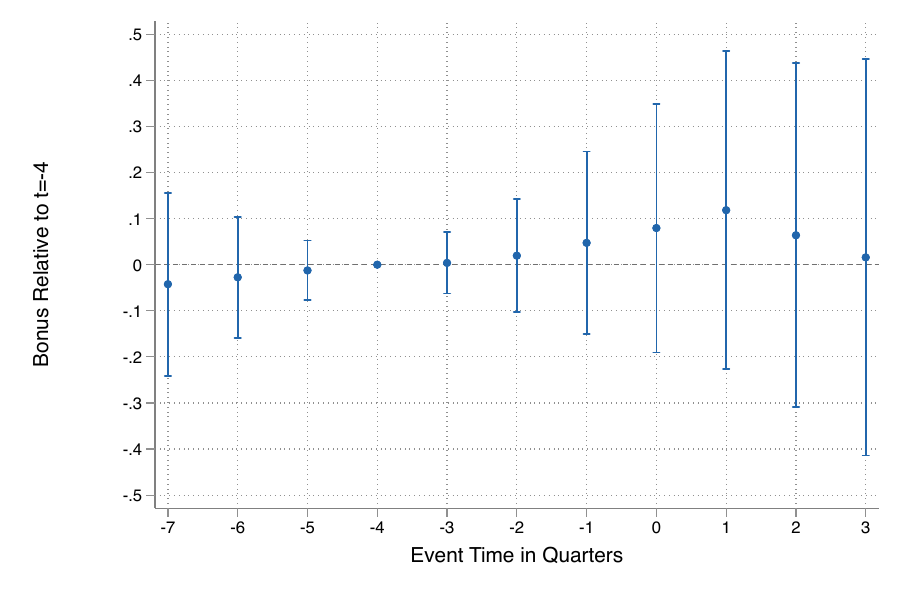}
		\end{subfigure}
		\begin{subfigure}{0.45\textwidth}
			\centering
			\caption*{Panel C. Manager Performance}
			\includegraphics[width=\linewidth]{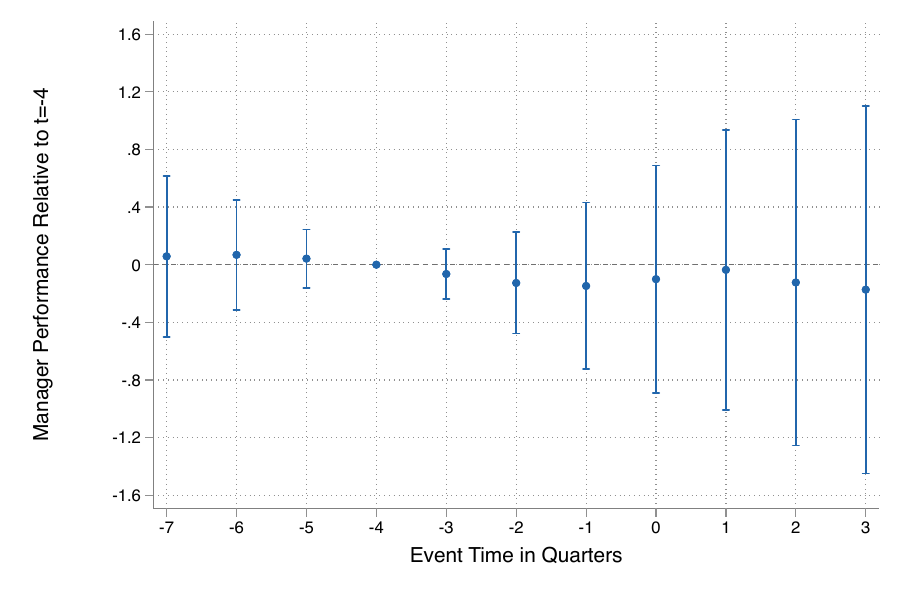}
		\end{subfigure}%
		\begin{subfigure}{0.45\textwidth}
			\centering
			\caption*{Panel D. Manager Log(Bonus)}
			\includegraphics[width=\linewidth]{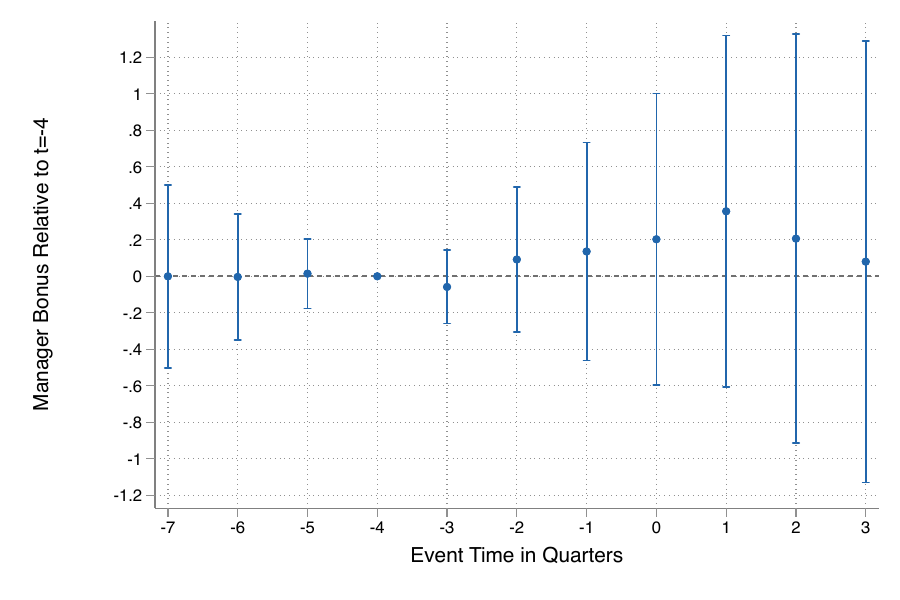}
		\end{subfigure}
		\label{fig:rotation_absent}
		\begin{minipage}[b]{0.9\linewidth}
			\footnotesize \textit{Notes}: This figure documents the absence of pre-trends in key team-level and manager-level outcomes before a manager rotation. Panel A shows team-level absenteeism rates. Panel B shows team-level log bonus payments. Panel C shows the manager's performance rating. Panel D shows the manager's log bonus payment. Event time is defined relative to the occurrence of a manager rotation. The specification includes team and quarter fixed effects, but no other controls. I  bin  event time dummy variables at $t=-8$ and $t=4$ and cluster standard errors at the team level. Robust standard errors are used and 95\%-level confidence intervals are displayed.  The mean outcome in Panels A (C) as of $t=-4$ is 0.09 (2.95). Outcome means for Panel B and D are unreported for confidentiality.  N=1,xxx. 
		\end{minipage}	
	\end{figure}
	\FloatBarrier

	\clearpage
	\newpage

	\section{Appendix Tables}
	
	\FloatBarrier
	\begin{table}[!ht]
		\caption{Workers' Survey Quotations}
		\begin{center}
			\scalebox{0.9}{{
\def\sym#1{\ifmmode^{#1}\else\(^{#1}\)\fi}
\begin{tabularx}{\textwidth}{|l|X|}
\hline\hline
 Form of talent hoarding & Quotations by workers       \\
\hline
   Informal underrating&``[My] former boss made me look bad at the potential new superiors behind my back in order to keep me on the team.''   \\
   \hline
Formal underrating&``Supervisors suppress potential ratings because of fear
that employees will leave their current position for a promotion.'' \\
& ``Out of fear that workers will leave the team, supervisors tend to underrate employees.''\\
& ``My supervisor would certainly find out if I applied to another position at the firm and that would have a negative impact on my assessment.'' \\
\hline
Suppressed development &``[Supervisors] keep employees in their positions by
preventing further development, rejecting training courses, and increasing
workloads to prevent capacity for new tasks and development measures.'' \\
& ``Career development is not supported by direct supervisors, instead it is actively blocked with the goal of keeping people in their current positions.'' \\
   \hline
Soft pressure& ``I decided not to apply internally ... because the message
communicated to me in the employee dialogue was that I can't leave the team within three years of joining.''\\
& ``My boss strongly hinted that in order for my success until now to be considered I need to follow through on my project until the very end otherwise my efforts will not be fully taken into account.'' \\
   \hline
Threats &``... my supervisor communicated very openly in a workshop that if one of his employees applied and was not hired, his career in his current department would also be at an end.'' \\
   \hline
Retaliation &  ``The position should fit exactly so that negative effects of applying regarding the current manager are worthwhile.'' \\

&``Fear of negative reactions from the supervisor: I have seen it from
many colleagues who openly stated that they were applying. These colleagues
then received no further training and no more interesting projects. The
supervisor had written them off. This made the months leading up to the final
change of job very difficult.'' \\ 
\hline\hline \multicolumn{2}{p{0.95\linewidth}}{\footnotesize \textit{Notes}: This table displays worker quotations regarding talent hoarding at the firm. The quotations are based on workers' free-text responses to questions on internal career development at the firm.}\\\end{tabularx}}
}
		\end{center}
		\label{table:survey_talenthoarding}
	\end{table}
	\FloatBarrier
	
	\FloatBarrier
	\begin{table}[!ht]
		\caption{Managers' Survey Quotations}
		\begin{center}
			\scalebox{0.9}{{
	\def\sym#1{\ifmmode^{#1}\else\(^{#1}\)\fi}
	\begin{tabularx}{\textwidth}{|l|X|}
		\hline\hline
Statement & Quotations by managers       \\
		\hline

  Misaligned incentives
& ``Regarding the development of your subordinates, there often is a conflict of interests for managers, since the employee then usually leaves the team and the position will not be approved to be refilled again.''\\
& ``Managers have no interest in developing talents because they have no direct benefit from it.''\\
 &``Managers pursue their own goals and often prevent further development of workers, because they are not rewarded for developing talent.''\\
&``Selfish managers are not willing to promote or recommend subordinates to other areas of the firm, even if that would add value to the firm.'' \\
&``[Middle] Managers can't improve the situation themselves since they depend on the upper management. That is why they block the development of their subordinates.'' \\
	\hline
Risk of losing talent
&``I observe that managers are not interested in letting  good workers leave their current position or develop them, otherwise they would have to fill the current position again.'' \\
&``Managers don't actively support workers in switching positions because the vacant position is usually not (immediately) filled again.'' \\
& ``Positions are cut if an employee takes on a new job or even dies, so managers are afraid that their team will shrink and they don't know how to do all the work with the rest of the team. Managers would be more open to talk to employees about developing them if they would know that they can fill the vacant position again.'' \\
	& ``Many managers are not necessarily interested in developing workers or helping them to get a better job within the firm, because they would lose a good worker. Switches to other areas at the firm are not encouraged, even if it would have been the right move for the worker.''   \\
	&``If you are good at what you are doing, it is very unlikely that you will be suggested for a higher-level position, especially if that position is in a different team.'' \\
	&``Supervisors have the policy to do whatever it takes to keep  people in their team, even if this means ignoring the development of the team members.'' \\

\hline\hline \multicolumn{2}{p{0.95\linewidth}}{\footnotesize \textit{Notes}: This table displays manager quotations regarding talent hoarding at the firm. The quotations are based on managers' free-text responses to questions on internal career development at the firm. }\\\end{tabularx}}
}
		\end{center}
		\label{table:survey_talenthoardingmanager}
	\end{table}
	\FloatBarrier

	\begin{table}[h]
		\caption{Comparison of Analysis Sample to Representative Survey of German Workforce}
		\begin{center}
			\scalebox{0.9}{{
\def\sym#1{\ifmmode^{#1}\else\(^{#1}\)\fi}
\begin{tabular}{l*{3}{c}}
\hline\hline
                    &\multicolumn{1}{c}{\shortstack{Analysis sample}}&\multicolumn{1}{c}{\shortstack{Large manufacturers}}&\multicolumn{1}{c}{\shortstack{German workforce}}\\
                                        &\multicolumn{1}{c}{(1)}&\multicolumn{1}{c}{(2)}&\multicolumn{1}{c}{(3)}\\
\hline
Female              &        0.21&        0.21&        0.45\\
German citizen     &        0.89&        0.85&        0.88\\
Age  (years)               &        43.3&        44.1&        43.9\\
Married             &        0.61&        0.59&        0.54\\
Children                &       0.80&        0.62&        0.62\\
Education (years)     &          15.8&        11.7&        11.8\\
Firm Tenure  (years)&        13.3&        16.8&        13.2\\
Weekly hours        &          41.1&        40.5&        38.1\\
Team leadership&        0.18&        0.25&        0.29\\
\hline
Observations        &       1xx,xxx&        1,472&       13,791\\
\hline\hline \multicolumn{4}{p{0.85\linewidth}}{\footnotesize \textit{Notes}: This table compares the employees in my sample to a representative survey of German workers (BIBB). Because the most recent version of the BIBB was conducted in 2018, I restrict my quarterly sample to 2018. Column 1 reports summary statistics for my analysis sample. Column 2 uses data from the BiBB survey and restricts to employees in manufacturing firms with at least 250 employees in West Germany. Column 3 uses data from the BiBB survey for all labor force participants in West Germany. Columns 2 and 3 are weighted to achieve representativeness of the German population. Team leadership is defined based on whether someone supervises direct reports.}\\ \end{tabular}}

}
		\end{center}
		\label{table:bibb_2018}
	\end{table}


    	\begin{table}[!h]
		\caption{Heterogeneity in Manager Hoarding  by Incentives Using an Alternative Measure Based on Suppression of Training Opportunities}
		\bigskip
		\begin{center}
			\scalebox{0.80}{{
	\def\sym#1{\ifmmode^{#1}\else\(^{#1}\)\fi}
	\begin{tabular}{l*{3}{c}}
		\hline  \hline 
		
		\underline{Outcome variable} &\multicolumn{3}{c}{Top tercile training measure}        \\ \cline{2-4}
		&\multicolumn{1}{c}{(1)}           &\multicolumn{1}{c}{(2)}         &\multicolumn{1}{c}{(3)}        \\
		\hline
		\\[0.1mm] 
		Financial incentive (share)    &  0.2007   &-- & -- \\
		&    (0.033)       & --   &  --        \\  
		\\[0.1mm]     
		Team size    &  --   &   -0.0143 & -- \\ 
		&   --         &  (0.001)      & --        \\  
		\\[0.1mm]     
		Visibility    &   --   & -- &  0.0298  \\
		&   --        & --    & (0.013)        \\  
		\\[0.1mm]       
		
		\hline
		Outcome Mean            &     0.3001              &    0.3001                  &      0.3001        \\
	Predicted difference p90-p10 & 0.1380 & -0.1403 & 0.0298 \\
		Observations        &       6,xxx    &    6,xxx        &   6,xxx        \\
		
		\hline \hline					\multicolumn{4}{p{0.65\linewidth}}{\footnotesize \textit{Notes}:  This table uses an alternative measure of managers' hoarding actions based on managers’ withholding of access to high-visibility training opportunities to examine whether managerial talent hoarding is more prevalent among managers with stronger incentives to hoard talent.  I use three distinct measures of hoarding incentives. Column 1 captures financial incentives using the share of total compensation that is performance-related. Column 2 uses the number of team members overseen by the manager. Column 3 includes a binary indicator for low talent visibility, defined as working in a functional area where individual achievements are less visible. Each coefficient is estimated separately based on a linear model for managers’ hoarding propensity using an OLS regression of the indicator that managers are hoarding-prone, defined as being in the top tercile of the distribution of mean deviations between predicted and actual training assignments, on one incentive measure at a time. The predicted difference  p90-p10 refers to the predicted difference in the outcome between the 90th and 10th percentiles of the continuous regressors (between values 1 and 0 for binary indicators), holding all other covariates at their mean values. See Appendix Section \ref{sec:appendix_manager_training} for details on the construction of this alternative hoarding measure. See Appendix Section \ref{sec:hoarding_incentives_details} for definitions of the incentive measures and Appendix Table \ref{tab:incentives_summ} for their distribution in the sample. Appendix Tables \ref{tab:incentives_training_reg} and \ref{tab:hoarding_predictions_training_rob} present complementary robustness results using alternative specifications of the regressors and outcome variable.  See Table \ref{fig:hoarding_predictions}  for my main results using managers’ suppression of potential ratings as baseline measure of hoarding behavior.Controls: Female, age, German citizenship, firm tenure, educational qualification. Robust standard errors in parentheses.}
\end{tabular}}

}
		\end{center}
	 \label{tab:hoarding_predictions_training}
	\end{table}

	
	\clearpage
	\newpage

	\begin{table}
		\caption{Effect of Manager Rotation by Transition Type and Incumbent Status}
		\begin{center}
			\scalebox{0.86}{{
\def\sym#1{\ifmmode^{#1}\else\(^{#1}\)\fi}
\begin{tabular}{l*{4}{c}}
\hline\hline
          &\multicolumn{2}{c}{Job transition}       &\multicolumn{2}{c}{Internal application}   \\
                    &    Internal         &    External           &   Incumbent       &     Newly hired      \\
                                        &\multicolumn{1}{c}{(1)}         &\multicolumn{1}{c}{(2)}      &\multicolumn{1}{c}{(3)}         &\multicolumn{1}{c}{(4)}       \\
\hline
Manager Rotation    &      0.0111&      0.0011       &      0.0190&      0.0275\\
                    &     (0.002)         &     (0.001)       &   (0.004)   &      (0.004)\\
\hline
Outcome Mean        &       0.008         &       0.008          &       0.027         &       0.030        \\
P-value of t-test         &\multicolumn{2}{c}{0.0000}               &\multicolumn{2}{c}{0.1297}      \\
Observations        &      3xx,xxx         &      3xx,xxx         &      1xx,xxx        &      1xx,xxx         \\
\hline\hline \multicolumn{5}{p{0.75\linewidth}}{\footnotesize \textit{Notes}: This table provides evidence in support of talent hoarding as underlying mechanism. Columns 1 and 2 illustrate the effect of manager rotations on workers' job transitions within the firm (Column 1) and out of the firm (Column 2). Consistent with talent hoarding, only internal transitions are affected by rotations. Columns 3 and 4 document the application effects of manager rotations for workers who have been in the team before the manager arrived (Column 3) and workers who in the past were hired by the rotating manager (Column 4). I do not find that rotations have larger effects for workers who the manager was able to select herself, which is what one would expect under manager-worker-specific match effects. Robust standard errors are clustered at the worker and rotation level. Controls: Female, age, German citizenship, educational qualifications, marital status, family status, parental leave, firm tenure, division, function, location, full-time, hours, number of direct reports, and quarter fixed effects. }\\
\end{tabular}
}
}
		\end{center}
		\label{table:mechanism_internalexternal_reg}
	\end{table}
	

	\begin{table}
		\caption{Effect of Manager Rotation by Destination of Application}
		\begin{center}
			\scalebox{0.86}{{
\def\sym#1{\ifmmode^{#1}\else\(^{#1}\)\fi}
\begin{tabular}{l*{6}{c}}
	\hline\hline
&\multicolumn{2}{c}{Division}  &\multicolumn{2}{c}{Functional area}  &\multicolumn{2}{c}{Location}  \\
		&\multicolumn{1}{c}{Same} &\multicolumn{1}{c}{Different} 	 &\multicolumn{1}{c}{Same} &\multicolumn{1}{c}{Different}	&\multicolumn{1}{c}{Same}	 &\multicolumn{1}{c}{Different} \\
		&\multicolumn{1}{c}{(1)}         &\multicolumn{1}{c}{(2)}         &\multicolumn{1}{c}{(3)}         &\multicolumn{1}{c}{(4)}         &\multicolumn{1}{c}{(5)}         &\multicolumn{1}{c}{(6)}         \\
	\hline
	Manager Rotation    &     0.01587& 0.00709&   0.01437&     0.00813& 0.01768   &     0.00437     \\
	&     (0.002)         &     (0.002)         &     (0.002)         &     (0.002)         &     (0.002)         &     (0.001)         \\
	\hline
	Outcome Mean        &        0.01438       &         0.01434    &     0.01575          &           0.01310  &       0.01777     &          0.00963      \\
P-value of t-test         &\multicolumn{2}{c}{0.0013}               &\multicolumn{2}{c}{0.0214} &\multicolumn{2}{c}{0.0000}       \\
Observations        &       3xx,xxx         &       3xx,xxx             &      3xx,xxx    &  3xx,xxx    &       3xx,xxx         &       3xx,xxx       \\
\hline\hline \multicolumn{7}{p{0.9\linewidth}}{\footnotesize \textit{Notes}: This table illustrates heterogeneity in the effect of manager rotations on worker applications by the destination of applications. Under talent hoarding, workers should disproportionally hold off on applications about which managers are likely to find out (e.g., because of proximity to the hiring manager). I assess three dimension of proximity compared to worker's current position: same versus different division (Columns 1 and 2), same versus different functional area (Columns 3 and 4), and same versus different location (Columns 5 and 6). Each coefficient stems from a separate regression based on Equation \ref{eqn:fs}.  Robust standard errors are clustered at the worker and rotation level. Controls: Female, age, German citizenship, educational qualifications, marital status, family status, parental leave, firm tenure, division, function, location, full-time, hours, number of direct reports, and quarter fixed effects. }\\
\end{tabular}
}
}
		\end{center}
		\label{table:mechanism_distance}
	\end{table}
	

	\begin{landscape}
	    
	\begin{table}
		\caption{Effect of Manager Rotation by Manager Characteristics}
		\begin{center}
			\scalebox{0.9}{{
\def\sym#1{\ifmmode^{#1}\else\(^{#1}\)\fi}
\begin{tabular}{l*{8}{c}}
		\hline\hline   \\[0.1mm] 
		 \multicolumn{9}{l}{\textbf{Panel A}: Outgoing Manager}   \\	
	\hline
	&\multicolumn{2}{c}{Manager gender}&\multicolumn{2}{c}{Manager vs worker gender}&\multicolumn{2}{c}{Manager experience} &\multicolumn{2}{c}{Team absenteeism}\\
	&\multicolumn{1}{c}{Male} &\multicolumn{1}{c}{Female}                                      &\multicolumn{1}{c}{Same} &\multicolumn{1}{c}{Opposite}                    &\multicolumn{1}{c}{Low} &\multicolumn{1}{c}{High} &\multicolumn{1}{c}{Low} &\multicolumn{1}{c}{High} \\
	&\multicolumn{1}{c}{(1)}&\multicolumn{1}{c}{(2)}&\multicolumn{1}{c}{(3)}&\multicolumn{1}{c}{(4)}&\multicolumn{1}{c}{(5)}&\multicolumn{1}{c}{(6)} &\multicolumn{1}{c}{(7)}&\multicolumn{1}{c}{(8)}\\
	\hline
	Manager Rotation&      0.02112&                  0.03554       &   0.02105       &        0.02900    & 0.02014  &       0.02691   &  0.02637&      0.02098    \\
	&     (0.003)        &   (0.009)                          &        (0.003)                    &         (0.006)              &            (0.005)            &        (0.006)              &    (0.006) &     (0.003)  \\
	\hline
	Outcome Mean        &       0.029         &       0.029        &       0.029        &       0.029         &       0.029         &       0.029     & 0.029 &     0.029 \\
	P-value of t-test   &\multicolumn{2}{c}{0.1092}    &\multicolumn{2}{c}{0.2538}     &\multicolumn{2}{c}{0.2894}   &\multicolumn{2}{c}{0.3691}  \\
	Observations               &       3xx,xxx          &       3xx,xxx          &       3xx,xxx          &      3xx,xxx         &       3xx,xxx          &       3xx,xxx    &     3xx,xxx          &       3xx,xxx   \\
	\hline \\
			 \multicolumn{9}{l}{\textbf{Panel B}: Incoming Manager}   \\	
	\hline
	&\multicolumn{2}{c}{Manager gender}&\multicolumn{2}{c}{Manager vs worker gender}&\multicolumn{2}{c}{Manager tenure} &\multicolumn{2}{c}{Functional area}\\
	&\multicolumn{1}{c}{Male} &\multicolumn{1}{c}{Female}                                      &\multicolumn{1}{c}{Same} &\multicolumn{1}{c}{Opposite}                    &\multicolumn{1}{c}{Low} &\multicolumn{1}{c}{High} &\multicolumn{1}{c}{Same} &\multicolumn{1}{c}{Different} \\
	&\multicolumn{1}{c}{(1)}&\multicolumn{1}{c}{(2)}&\multicolumn{1}{c}{(3)}&\multicolumn{1}{c}{(4)}&\multicolumn{1}{c}{(5)}&\multicolumn{1}{c}{(6)} &\multicolumn{1}{c}{(7)}&\multicolumn{1}{c}{(8)}\\
	\hline
	Manager Rotation&       0.02151&                  0.02532       &0.02138     &        0.02396  &        0.02511      &       0.02171  &  0.02025 &     0.02374   \\
	&     (0.003)        &   (0.008)                          &        (0.003)                    &         (0.006)              &            (0.005)            &        (0.006)              &  (0.003) &   (0.005)  \\
	\hline
	Outcome Mean        &       0.029         &       0.029        &       0.029        &       0.029         &       0.029         &       0.029     & 0.029 &     0.029 \\
	P-value of t-test   &\multicolumn{2}{c}{0.6399}    &\multicolumn{2}{c}{0.7248}     &\multicolumn{2}{c}{0.6242}   &\multicolumn{2}{c}{0.5324}  \\
	Observations               &       3xx,xxx          &       3xx,xxx          &       3xx,xxx          &      3xx,xxx         &       3xx,xxx          &       3xx,xxx       &      3xx,xxx          &       3xx,xxx  \\
	\hline \\
\hline\hline \multicolumn{9}{p{0.9\linewidth}}{\footnotesize \textit{Notes}: This table illustrates heterogeneity in the effect of manager  rotations on worker applications by the characteristics of the outgoing manager (Panel A) and the incoming manager (Panel B). Each coefficient stems from a separate regression based on Equation \ref{eqn:fs}, where the rotation event is restricted to managers with a given characteristic. Panel A compares application effects by the outgoing manager's gender (Columns 1 and 2), manager vs worker gender (Columns 3 and 4), bottom vs top quartile of manager experience (Columns 5 and 6), and the past absenteeism rate of the manager's team split at the sample median  (Columns 7 and 8). Panel B compares application effects by the incoming manager's gender (Columns 1 and 2), manager vs worker gender (Columns 3 and 4), bottom vs top quartile of manager tenure at the firm (Columns 5 and 6), and whether the incoming manager was in the same vs different functional area before (Columns 7 and 8). In both panels, I find no statistical difference in the effect between each pair-wise comparison as indicated by the p-value of the corresponding t-test. Robust standard errors are clustered at the worker and rotation level. Controls: Female, age, German citizenship, educational qualifications, marital status, family status, parental leave, firm tenure, division, function, location, full-time, hours, number of direct reports, and quarter fixed effects. }\\ \end{tabular}}
}
		\end{center}
	\label{table:incoming}
	\end{table}
	\end{landscape}


	\begin{table}
		\caption{Testing for Instrument Exclusion}
		\begin{center}
			\scalebox{0.8}{{
\def\sym#1{\ifmmode^{#1}\else\(^{#1}\)\fi}
\begin{tabular}{l*{6}{c}}
			   \hline  \hline \\
		 \multicolumn{7}{l}{\textbf{Panel A}: IV Estimates on Hiring by Rotating Manager's  Ties to Destination}  \\
\hline
		   &\multicolumn{6}{c}{Rotating manager has worked in the job opening's ...}   \\
		   &\multicolumn{2}{c}{Division}   &\multicolumn{2}{c}{Functional area}   &\multicolumn{2}{c}{Location} \\
		 &\multicolumn{1}{c}{Different}&\multicolumn{1}{c}{Same} 		 &\multicolumn{1}{c}{Different}&\multicolumn{1}{c}{Same} 		 &\multicolumn{1}{c}{Different}&\multicolumn{1}{c}{Same} \\
		  &\multicolumn{1}{c}{(1)}&\multicolumn{1}{c}{(2)}&\multicolumn{1}{c}{(3)}&\multicolumn{1}{c}{(4)}&\multicolumn{1}{c}{(5)}&\multicolumn{1}{c}{(6)}\\
		  \hline
		  Applied   &      0.44231&            0.50917          &  0.31052                &               0.57442    &             0.59534       &       0.44680        \\
		  &     (0.116)           &    (0.070)               &       (0.089)                &                      (0.076)    &               (0.168)           &          (0.062)        \\
		  \hline
		  Outcome Mean        &       0.003         &       0.005         &       0.002      &       0.005         &       0.002        &       0.006       \\
		  Observations        &      3xx,xxx         &      3xx,xxx         &     3xx,xxx        &      3xx,xxx        &      3xx,xxx       &     3xx,xxx          \\
		    \hline \\
	 \multicolumn{7}{l}{\textbf{Panel B}: IV Estimates  on Hiring by Rotating Manager's Quality}  \\
  \hline
  &\multicolumn{2}{c}{Turnover}   &\multicolumn{2}{c}{Absenteeism}  &\multicolumn{2}{c}{Performance}  \\
                                &\multicolumn{1}{c}{Low} &\multicolumn{1}{c}{High}                    &\multicolumn{1}{c}{Low} &\multicolumn{1}{c}{High}                    &\multicolumn{1}{c}{Low} &\multicolumn{1}{c}{High} \\
                     &\multicolumn{1}{c}{(1)}&\multicolumn{1}{c}{(2)}&\multicolumn{1}{c}{(3)}&\multicolumn{1}{c}{(4)}&\multicolumn{1}{c}{(5)}&\multicolumn{1}{c}{(6)}\\
\hline
Applied             &      0.48749&    0.53386&      0.55033&    0.53776&     0.57052&     0.60847\\
                    &     (0.068)         &      (0.118)       &     (0.104)         &       (0.134)            &       (0.128)        &      (0.126)          \\
\hline
Outcome Mean        &       0.008         &       0.008             &       0.008         &       0.008         &       0.008         &       0.008         \\
  Observations        &      3xx,xxx         &      3xx,xxx         &     3	xx,xxx        &      3xx,xxx        &     3xx,xxx       &     3xx,xxx          \\
\hline \\
	 \multicolumn{7}{l}{\textbf{Panel C}: IV Estimates on Hiring by Rotating Manager's Exposure to Workers }  \\
	 \hline
	 &  \multicolumn{6}{c}{Exposure length in quarters} \\
	           &       $\leq$1         &         2-3         &         4-6         &         7-8         &        9-10         &       11-14            \\
	  &\multicolumn{1}{c}{(1)}&\multicolumn{1}{c}{(2)}&\multicolumn{1}{c}{(3)}&\multicolumn{1}{c}{(4)}&\multicolumn{1}{c}{(5)}&\multicolumn{1}{c}{(6)}\\
	  \hline
	  Applied               &     0.47322&     0.31729 &     0.52088&     0.45054&     0.46512 &     0.44879\\
                    &     (0.220)         &     (0.136)         &     (0.190)         &     (0.142)         &     (0.207)         &     (0.156)         \\
\hline  \\
	  \hline
	  Outcome Mean        &       0.009         &       0.008         &       0.009         &       0.009         &       0.008         &       0.008             \\
	  Observations        &       3x,xxx         &       5x,xxx         &      5x,xxx         &      2x,xxx         &       2x,xxx         &       3x,xxx             \\
	  \hline \\
	 \hline\hline
	  \multicolumn{7}{p{0.9\linewidth}}{\footnotesize \textit{Notes}: This table finds no evidence for a violation of the exclusion restriction regarding manager rotation as instrument for worker applications. Each panel presents two-stages least squares estimates of applying on getting hired based on Equation \ref{eqn:iv}. \underline{Panel A} estimates hiring outcomes for applications to which rotating managers have varying degrees of formal ties. Each column uses a different split of applications based on whether the application is to a different area of the firm, where area is defined as division (Columns 1 and 2), functional area (Columns 3 and 4), and location (Columns 5 and 6).  \underline{Panel B} estimates hiring outcomes by rotating managers' quality using past means for three outcomes:   team turnover (Columns 1 and 2), team absenteeism (Columns 3 and 4), and manager's performance rating (Columns 5 and 6). Each column uses a different manager type to define the rotation event based on whether the team-level outcome is in the bottom or top quartile of the team distribution. \underline{Panel C} estimates hiring outcomes by managers' length of exposure to workers. Each column uses a different split of workers based on the number of quarters the worker has been with the manager. Robust standard errors are clustered at the rotation and worker level. Controls: Female, age, German citizenship, educational qualifications, marital status, family status, parental leave, firm tenure, division, function, location, full-time, hours, number of direct reports, and quarter fixed effects. }\\ \end{tabular}}
}
		\end{center}
		\label{table:iv_validity_excl}
	\end{table}


	\begin{table}
		\caption{Testing for Instrument Monotonicity}
		\begin{center}
			\scalebox{0.8}{{
\def\sym#1{\ifmmode^{#1}\else\(^{#1}\)\fi}
\begin{tabular}{l*{4}{c}}
	\hline\hline\\
	 \multicolumn{5}{l}{\textbf{Panel A}: Application Effects of Manager Rotations by Subgroup}  \\
\hline
                     \multicolumn{1}{c}{Sample} & Observations & Baseline application rate & First-stage effect & Standard error \\
                      \multicolumn{1}{c}{(1)}&\multicolumn{1}{c}{(2)}&\multicolumn{1}{c}{(3)}&\multicolumn{1}{c}{(4)}&\multicolumn{1}{c}{(5)}\\
                    \hline
                    Age $\leq$ 40 years & 1xx,xxx&  0.04011  &  0.02334& (0.005)  \\
                    Age $>$ 40 years &  2xx,xxx&   0.02149 &  0.02083&  (0.003) \\
                    Tenure $\leq$ 5 years & 1xx,xxx &  0.03572 &  0.02918& (0.006) \\
                    Tenure $>$ 5 years & 2xx,xxx&   0.02669 & 0.02019& (0.003) \\
                    Schooling $\leq$ 13 years &1xx,xxx  &  0.02100   &  0.01161& (0.004)\\
                    Schooling $>$ 13 years &  2xx,xxx& 0.03122& 0.02575& (0.004) \\
                    Married & 2xx,xxx& 0.02694  &   0.02003& (0.003)\\
                    Not married & 1xx,xxx&0.03351  & 0.02635 &  (0.005) \\
                    Parent & 2xx,xxx& 0.02794&  0.02199& (0.003)   \\
                    Non-parent & 1xx,xxx&  0.03351  & 0.02391&    (0.006)   \\
                    German citizen & 3xx,xxx& 0.02844& 0.02289&   (0.003)  \\
                   Non-German citizen & 3x,xxx&  0.03395 &   0.01927 & (0.010)    \\
      Team leadership& 7x,xxx& 0.02457 & 0.01699&  (0.006) \\
                    No team leadership  & 3xx,xxx& 0.03001&  0.02337& (0.003)   \\
                    High performance& 2xx,xxx& 0.02931 &  0.02785& (0.004) \\
                    Low performance & 1xx,xxx&   0.03066   &  0.01719& (0.004) \\           
\hline \\
 \multicolumn{5}{l}{\textbf{Panel B}: Application Effects of Manager Rotations by Predicted Application Propensity}  \\
\hline
&    Never     &     Applied        &Low-application          &High-application          \\
& applied  & before & team & team \\
&\multicolumn{1}{c}{(1)}         &\multicolumn{1}{c}{(2)}           &\multicolumn{1}{c}{(3)}         &\multicolumn{1}{c}{(4)}       \\
\hline
Manager Rotation   &         0.01792  &        0.04240  &      0.01813&      0.02506 \\
&     (0.002)   &    (0.012)          &     (0.003)         &     (0.007)         \\
\hline
Outcome Mean      &       0.019    &       0.113            &       0.022        &       0.062         \\
Observations        &      3xx,xxx      &       4x,xxx               &      2xx,xxx        &       8x,xxx        \\
\hline \\
\hline \hline 
\multicolumn{5}{p{1.2\linewidth}}{\footnotesize \textit{Notes}: This table finds no evidence for a violation of the monotonicity assumption regarding manager rotation as instrument for worker applications. \underline{Panel A} presents first-stage effects of manager rotation on worker applications for several subpopulations of interest, as indicated by Column 1. Estimation is based on Equation \ref{eqn:fs} and conducted separately in each subpopulation. Column 2 contains the number of observations, Column 3 presents baseline application rates, Column 4 provides first-stage effects, and Column 5 contains robust standard errors, clustered at the rotation and worker level. I find that manager rotation has a positive and statistically significant first-stage effect (Column 4) for each subpopulation.  \underline{Panel B} presents first-stage effects of manager rotation on applying by workers' predicted application propensity. I use two approaches to predict workers' unobserved application propensity. Columns 1 and 2 split the sample by workers' own past application activity. Columns 3 and 4 split the sample by whether teams' leave-out application rates in the past were high or low. I find a significant and positive first-stage effect even for the subsets of workers who previously had a high propensity to apply (Columns 2 and 4).  Robust standard errors are clustered at the rotation and worker level. Controls: Female, age, German citizenship, educational qualifications, marital status, family status, parental leave, firm tenure, division, function, location, full-time, hours, number of direct reports, and quarter fixed effects.}\\ \end{tabular}}
}
		\end{center}
		\label{table:iv_mono}
	\end{table}

	\clearpage 
	\section{Theoretical Appendix}
	\label{sec:appendix_framework}
	
	In this section, I provide the formal derivations for predictions 4 and 5 referenced in Section \ref{sec:framework}. \\
	
	\noindent \textbf{Prediction 4.}  If $\beta_1 < \beta_2$ $\implies$ Pr[$i$ applies$\vert \beta= \beta_1$]>Pr[$i$ applies$\vert \beta= \beta_2$]  \\ This prediction implies  that greater levels of talent hoarding reduce the number of workers who apply for a  promotion. 
	
	Workers apply if  $q(\alpha_i,\beta) b \geq c + \varepsilon_i$, where $\varepsilon_i \sim \Psi$ captures worker-specific heterogeneity. A worker's probability to apply can be expressed as Pr[$i$ applies$ \vert \beta_m$] =
	$\Psi$$\big($q($\alpha_i$,$\beta_m$)$b - c$$\big)$.  Because workers' promotion probability is decreasing in talent hoarding ($\frac{\partial q}{\partial \beta}<0$), 
	if $\beta_1 < \beta_2$: 
	\begin{align*}
		q(\alpha_i,\beta_1)  &> q(\alpha_i,\beta_2) \\
		\Psi\big(q(\alpha_i,\beta_1)b - c\big) &> \Psi\big(q(\alpha_i,\beta_2)b - c\big) \\
		\implies \text{Pr}[i \text{ applies}\vert \beta= \beta_1] &> \text{Pr}[i \text{ applies}\vert \beta= \beta_2]
	\end{align*}
	
	\noindent \textbf{Prediction 5.} 
	If $\alpha_1 < \alpha_2$ and $\beta_1 < \beta_2$ $\implies$ $\frac{\text{Pr}[i \text{ applies} \vert \alpha_2, \beta_1]}{\text{Pr}[i \text{ applies} \vert \alpha_1, \beta_1]} > \frac{\text{Pr}[i \text{ applies} \vert \alpha_2, \beta_2]}{\text{Pr}[i \text{ applies} \vert \alpha_1, \beta_2]}$ \\
	This prediction implies that greater levels of talent hoarding change the composition of applicants, causing a lower share of workers with high productivity in the applicant pool. 	
	
	Let $r(\alpha_i, \beta_m)$ be $\text{Pr}[i \text{ applies} \vert \alpha_i, \beta_m]$=  $\Psi$$\big($q($\alpha_i$,$\beta_m$)$b - c$$\big)$. We have assumed that $\frac{\partial ^2 q}{\partial\beta\partial\alpha}<0, 
	\frac{\partial q}{\partial\beta}<0, \frac{\partial q}{\partial\alpha}>0 $. We want to show that $ \frac{\partial}{\partial \beta} \frac{r(\alpha_1,\beta)}{r(\alpha_2,\beta)} < 0$ for $\alpha_2 > \alpha_1$. 
	
	\begin{align*}
		\frac{\partial}{\partial \beta} \frac{r(\alpha_2,\beta)}{r(\alpha_1,\beta)} &=  \frac{\partial}{\partial \beta} \frac{\Psi(q(\alpha_2, \beta))}{\Psi(q(\alpha_1, \beta))} \\
		& = \frac{\Psi(q(\alpha_2, \beta)) \psi(q(\alpha_1, \beta)) \frac{\partial q(\alpha_1, \beta)}{\partial \beta} - \Psi(q(\alpha_1, \beta))  \psi(q(\alpha_2, \beta)) \frac{\partial q(\alpha_2, \beta)}{\partial \beta} }{\big[\Psi(q(\alpha_1, \beta)\big]^2} 
	\end{align*}
	Omitting the denominator since $\big[\Psi(q(\alpha_1, \beta)\big]^2 >0$ leaves to show that
	\begin{align*}
		\Psi(q(\alpha_1, \beta))  \psi(q(\alpha_2, \beta)) \frac{\partial q(\alpha_2, \beta)}{\partial \beta} &<\Psi(q(\alpha_2, \beta)) \psi(q(\alpha_1, \beta)) \frac{\partial q(\alpha_1, \beta)}{\partial \beta} 
	\end{align*}
	Rearranging leads to following expression
	\begin{align*}
		\underbrace{\frac{\frac{\partial q(\alpha_2, \beta)}{\partial \beta} }{\frac{\partial q(\alpha_1, \beta)}{\partial \beta} }}_{<0} & < \underbrace{\frac{\Psi(q(\alpha_2, \beta)) \psi(q(\alpha_1, \beta)) }{\Psi(q(\alpha_1, \beta)) \psi(q(\alpha_2, \beta)) }}_{>0}
	\end{align*}
	
	Since the left-hand side of the equation is below zero and the right-hand side of the equation is above zero, it holds that  $\frac{\partial}{\partial \beta} \frac{r(\alpha_1,\beta)}{r(\alpha_2,\beta)} < 0$ for $\alpha_2 > \alpha_1.$

	\clearpage 
	\section{Data Appendix}
	\label{sec:appendix_data}
		\setcounter{table}{0}
	\renewcommand{\thetable}{\thesection\arabic{table}}
	\setcounter{figure}{0}
	\renewcommand\thefigure{\thesection\arabic{figure}}
This appendix provides additional details on the data sources used in the analysis. It begins by describing the construction of the administrative dataset, which links personnel records, talent management data, and internal application data provided by the firm. It then outlines the implementation of the employee and manager surveys, including survey design, response rates, and benchmarking exercises to aid interpretation of the survey responses.
	
	\subsection{Administrative Data Used in the Analysis}
	\label{sec:appendix_data_construction}

    This paper is based on a novel dataset combining personnel records and application data from a large manufacturing firm. To comply with the firm's data protection protocols, all data processing was conducted within a secure data environment managed by the firm, where the dataset was assembled and anonymized.

The dataset integrates information from three internal data sources: (1) personnel records, (2) talent management data, and (3) job application data. These sources were linked using a unique employee identifier.

\textbf{Personnel Records.} The personnel records include detailed demographic and employment information. The following variables were extracted: employment status, date of birth, gender, marital and family status, parental leave status, date of entry into the firm, highest educational degree attained, citizenship, position, functional area, division, work location, contractual working hours, number of direct reports, reporting distance to the CEO, managerial autonomy, base pay, bonus payments, absenteeism, exit date (if applicable), retirement date, manager ID, and team ID. To create a quarterly panel dataset, the data were collapsed to the employee-by-quarter level, keeping the first observation of an employee in a given quarter.

\textbf{Talent Management Data.} From the firm’s internal talent management systems, I obtained information on individual performance and potential ratings, nominations to talent development programs, and worker assignments to trainings.

\textbf{Job Application Data.} The application tracking system provides data on all internal applications submitted by employees. Extracted variables include application date and status, application outcomes (e.g., rejection, interview, offer), vacancy ID, hiring manager ID, hierarchy level of the vacancy, as well as the division, functional area, and location associated with each vacancy.

The three data sources were linked using a unique employee id within the firm's secure data environment.
	\subsection{Employee Survey}
	\label{sec:app_survey}

	\vspace{0.3cm}
    All employees in my sample were invited to share their perspectives on the internal labor market at the firm. The survey yielded over 15,000 responses, corresponding to a response rate of approximately 50\%. A summary of the survey content and key findings is presented in Section~\ref{sec:employee_svy_descr}. This appendix section provides additional information on the survey’s implementation and instrument.
    
    \subsubsection{Survey Implementation}
    \label{sec:app_survey_implementation}

\begin{figure}[h]
		\thisfloatpagestyle{plainlower}
		\caption{Invitation to Participate in Employee Survey}
		\centering
		\includegraphics[scale=0.5]{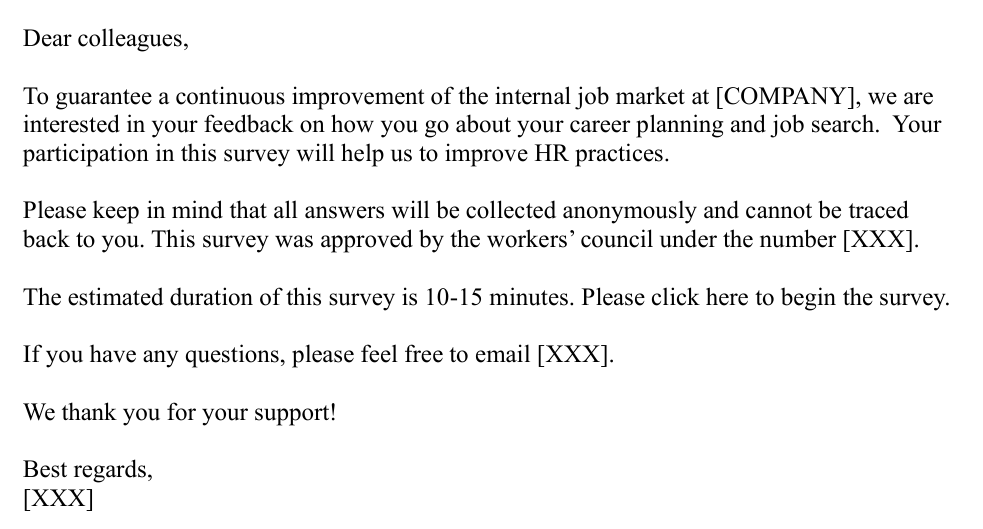}
		\label{fig:invitation_employee}
		\begin{minipage}[b]{0.9\linewidth}
			\footnotesize \textit{Notes}: This figure presents the wording of the survey invitation sent to employees. 
		\end{minipage}	
	\end{figure}

    The survey was conducted using the firm’s established procedures for internal employee surveys. It was formally approved by the firm’s data protection officer and the workers’ council. Invitations to participate were distributed via email by the Human Resources department. Appendix Figure~\ref{fig:invitation_employee} shows the full text of the invitation email.

The survey was administered through the firm’s secure online platform. Upon accessing the survey link, respondents were prompted to provide informed consent for their responses to be recorded and analyzed. The median completion time was 13 minutes. To ensure data reliability, I restrict the analysis to respondents who spent at least five minutes completing the survey. Appendix Table 	\ref{table:survey_X} compares respondents to the overall population and finds that both groups are very similar in terms of their characteristics. 

	\begin{table}[!ht]
		\caption{Comparison of Analysis Sample to Respondents of Employee Survey}
		\begin{center}
			\scalebox{1.0}{{
\def\sym#1{\ifmmode^{#1}\else\(^{#1}\)\fi}
\begin{tabular}{l*{2}{c}}
\hline\hline
                    &\multicolumn{1}{c}{Sample}
                                        &\multicolumn{1}{c}{Survey}\\
                                                            &\multicolumn{1}{c}{(1)}&\multicolumn{1}{c}{(2)}\\
\hline
Female              &        0.21&        0.24\\
German              &        0.90&        0.94\\
Age $<$30  years         &        0.10&        0.12\\
Age 30-39    years         &        0.30&        0.32\\
Age 40-49   years          &        0.31&        0.26\\
Age $>=$50   years         &        0.30&        0.28\\
Tenure $<=$ 2  years       &        0.10&        0.14\\
Tenure 3-5    years        &        0.12&        0.18\\
Tenure 6-9   years         &        0.12&        0.17\\
Tenure $>=$10   years      &        0.56&        0.51\\
Engineering         &        0.48&        0.41\\
Finance             &        0.08&        0.05\\
Marketing and Sales &        0.11&        0.07\\
\hline
Observations        &     3x,xxx  &       1x,xxx\\
\hline\hline \multicolumn{3}{p{0.5\linewidth}}{\footnotesize \textit{Notes}: This table compares average characteristics of employees in my analysis sample (Column 1) to employees who responded to the employee survey (Column 2). The sample of survey respondents is restricted to only contain responses who took at least five minutes to complete the survey.}\\
\end{tabular}
}
}
		\end{center}
		\label{table:survey_X}
	\end{table}

\clearpage
\newpage
    \subsubsection{Survey Instrument}
	\label{sec:app_survey_instrument}
	
	\noindent \textbf{A.} Please rate following six statements. ``Actively applying for positions at [Company] ...''  \{I strongly agree, I agree, Undecided, I do not agree, I totally do not
	agree\}  
	
	\textbf{A.1} ``... increases future promotion chances.'' 
	
	\textbf{A.2} ``... does not matter since jobs are only posted proforma.'' 
	
	\textbf{A.3} ``... would cause negative consequences by my current supervisor.''   
	
	\textbf{A.4} ``... is seen as disloyal to my current team.''  
	
	\textbf{A.5} ``... is appropriate once employees are unsatisfied with their job.'' 
	
	\textbf{A.6} ``... should only be done after checking in with one's direct supervisor.'' 
	\vspace{0.3cm}
	
	\noindent \textbf{B.} Which job characteristics are most important to you? Please select the two most important characteristics from the following list. \{Potential for training,
	Potential for promotion, Pay, Flexible hours, Location, Meaningful tasks, Familiar tasks, Challenging tasks, Good relationship with colleagues,
	Good relationship with supervisor\} 
	
	\vspace{0.3cm}
	\noindent \textbf{C.} Did someone recommend an open position to you in the past 12 months?
	
	\textbf{C.1} Yes
	
	\textbf{C.2} No
	
	\vspace{0.3cm}
	\noindent \textbf{D.} You indicated that someone recommended one or more open positions to you. Who recommended these positions to you? Please select all that apply.
	
	\textbf{D.1} Colleagues within your team
	
	\textbf{D.2} Colleagues outside of your team
	
	\textbf{D.3} Your supervisor
	
	\textbf{D.4} The supervisor responsible for the open position
	
	\textbf{D.5} Your HR responsible
	
	\textbf{D.6} Family or friends
	
	\vspace{0.3cm}
	
	\noindent \textbf{E.} At the end of this survey, we are interested in your personal opinion about current challenges and potential improvements with respect to careers at [Company]. 
	
	\textbf{E.1} What were the reasons why you decided in the past not to apply for internal job openings at [Company]? \{free-text response\}
	
	\textbf{E.2} What are the main challenges that you have encountered in your career development at [Company]? \{free-text response\}
	
	\textbf{E.3} What are some of the ways that [Company] could be helpful to you as you are planning your career? \{free-text response\}
	\vspace{0.3cm}

	\clearpage
	\newpage
	
	\subsection{Manager Survey}
	\label{sec:app_survey_manager}
	To provide direct evidence on managers’ incentives and talent hoarding behavior, I conducted a survey among managers in my sample. The survey achieved a response rate of 62\%, resulting in over 3,000 completed responses. A summary of the survey design and key findings is provided in Section~\ref{sec:manager_svy_descr}. This appendix section offers additional details on the survey implementation and questionnaire. It also presents the raw response distributions for the main survey questions used in the analysis and includes several benchmarking exercises to aid interpretation.
	\vspace{0.3cm}

        \subsubsection{Survey Implementation}
    \label{sec:app_survey_manager_implementation}

\begin{figure}[h]
		\thisfloatpagestyle{plainlower}
		\caption{Invitation to Participate in Manager Survey}
		\centering
		\includegraphics[scale=0.5]{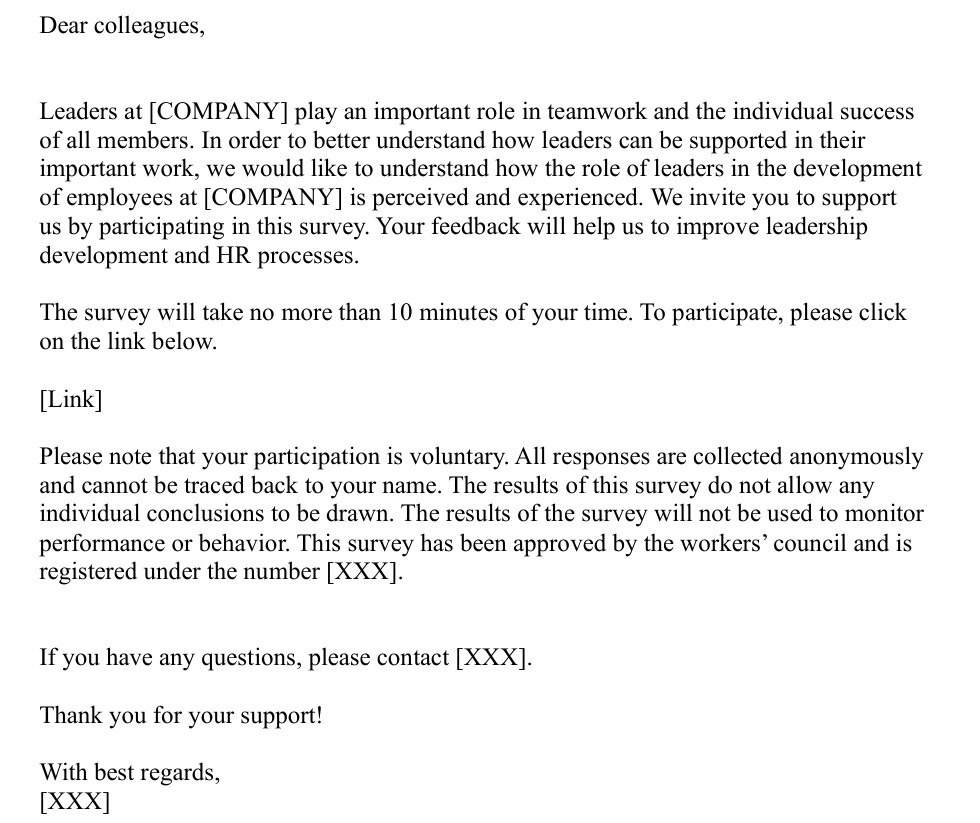}
		\label{fig:invitation_manager}
		\begin{minipage}[b]{0.9\linewidth}
			\footnotesize \textit{Notes}: This figure presents the wording of the survey invitation sent to managers. 
		\end{minipage}	
	\end{figure}

The survey was implemented using the firm’s standard procedures for internal employee surveys. Prior to launch, it received formal approval from both the firm’s data protection officer and the workers’ council. Managers were invited to participate via e-mail sent by the Human Resources department. Appendix Figure~\ref{fig:invitation_manager} displays the wording of the invitation email.

The survey was administered through the firm’s secure online survey platform. Upon accessing the survey link, respondents were prompted to provide informed consent for their responses to be recorded and analyzed. The median completion time was 14 minutes. To ensure data quality, the analysis sample is restricted to respondents who spent at least five minutes completing the survey. Appendix Table 		\ref{table:surveymanager_X} shows that respondents are similar to the overall population in terms of their characteristics. 

Several features suggest that the managers' interest and trust in the survey was high. While 62\% of managers responded to the survey overall, 61\% responded before a reminder was sent. Appendix Table \ref{table:surveymanager_reminderX} compares the demographics of managers who responded before a reminder was sent (Column 1) to those who responded afterwards (Column 2). I find that the characteristics of both groups of managers are very similar. In unreported results, I also find that the likelihood to respond before the reminder was sent is not correlated with managers' propensity to report talent hoarding using either of my two survey-based measures. 

    	\begin{table}[!ht]
		\caption{Comparison of Managers in Analysis Sample to Respondents of Manager Survey}
		\begin{center}
			\scalebox{1.0}{{
\def\sym#1{\ifmmode^{#1}\else\(^{#1}\)\fi}
\begin{tabular}{l*{2}{c}}
\hline\hline
                    &\multicolumn{1}{c}{Sample}
                                        &\multicolumn{1}{c}{Survey}\\
                                                            &\multicolumn{1}{c}{(1)}&\multicolumn{1}{c}{(2)}\\
\hline
Female              &        0.10&        0.16\\
German              &        0.92&        0.89\\
Age $<$30  years             &        0.01&        0.01\\
Age 30-39 years           &        0.17&        0.22\\
Age 40-49   years        &        0.40&        0.34\\
Age $>=$50 years         &        0.42&        0.43\\
Tenure $<=$ 2  years     &        0.03&        0.02\\
Tenure 3-5  years        &        0.06&        0.04\\
Tenure 6-9 years        &        0.11&        0.16\\
Tenure $>=$10  years     &        0.75&        0.78\\
Engineering         &        0.40&        0.33\\
Finance      &        0.08&        0.05\\
Marketing and Sales &        0.12&        0.07\\
\hline
Observations        &     7x,xxx  &       3,xxx\\
\hline\hline \multicolumn{3}{p{0.5\linewidth}}{\footnotesize \textit{Notes}: This table compares average characteristics of managers in my analysis sample (Column 1) to the managers who responded to the manager survey (Column 2). The sample of survey respondents is restricted to only contain responses who took at least five minutes to complete the survey.}\\
\end{tabular}
}
}
		\end{center}
		\label{table:surveymanager_X}
	\end{table}

    	\begin{table}[!ht]
		\caption{Summary Statistics of Managers Who Responded Before and After a Reminder was Sent}
		\begin{center}
			\scalebox{1.0}{{
\def\sym#1{\ifmmode^{#1}\else\(^{#1}\)\fi}
\begin{tabular}{l*{2}{c}}

\hline\hline
                    &\multicolumn{1}{c}{Before Reminder}
                                        &\multicolumn{1}{c}{After Reminder}\\

                                                            &\multicolumn{1}{c}{(1)}&\multicolumn{1}{c}{(2)}\\
\hline
Female              &        0.17&        0.15\\
German              &        0.89&        0.87\\
Age <$$30 years           &        0.02&        0.01\\
Age 30-39  years         &        0.23&        0.22\\
Age 40-49  years         &        0.34&        0.33\\
Age $>=$50  years        &        0.41&        0.44\\
Tenure $<=$ 2  years      &        0.02&        0.02\\
Tenure 3-5    years       &        0.04&        0.03\\
Tenure 6-9  years         &        0.16&        0.16\\
Tenure $>=$10  years      &        0.77&        0.79\\
Engineering         &        0.33&        0.33\\
Finance       &        0.05&        0.06\\
Marketing and Sales &        0.07&        0.07\\
\hline
Observations        &     2,xxx  &       1,xxx\\
\hline\hline \multicolumn{3}{p{0.7\linewidth}}{\footnotesize \textit{Notes}: This table compares average characteristics of managers who responded to the manager survey  before a reminder was sent (Column 1) and those who responded after a reminder was sent (Column 2). }\\
\end{tabular}
}
}
		\end{center}
		\label{table:surveymanager_reminderX}
	\end{table}
\clearpage
\newpage
    \subsubsection{Survey Instrument}
	\label{sec:app_survey_manager_instrument}

	\noindent \textbf{A.} Where do you think leaders can create the most impact for [Company]? Please rate the impact that leaders can create.  \{Very high impact, high impact, low impact, very low impact\}  
	
	\textbf{A.1} Strategic orientation
	
	\textbf{A.2} Vision 
	
	\textbf{A.3} Operational efficiency   
	
	\textbf{A.4} Talent development
	
	\textbf{A.5} Culture and morale
	
	\textbf{A.6} Innovation
	\vspace{0.3cm}

	\noindent \textbf{B.} When it comes to leaders’ compensation and promotion prospects, how important do you think is a record of success in each of the following?  Please rate the importance.  \{Very important, important, somewhat important, not important\}  
	
	\textbf{B.1} Strategic orientation
	
	\textbf{B.2} Vision 
	
	\textbf{B.3} Operational efficiency   
	
	\textbf{B.4} Talent development
	
	\textbf{B.5} Culture and morale
	
	\textbf{B.6} Innovation
	\vspace{0.3cm}

	\noindent \textbf{C.} What are the most effective ways that leaders can support their employees’ career development? Please select all that apply.

	\textbf{C.1} Training: Offer and recommend relevant trainings that can help employees enhance their skills and knowledge
	
	\textbf{C.2} Open dialogue: Establish a regular dialogue about career goals  
	
	\textbf{C.3} Career planning: Introduce a career planning process that provides clear guidance for employees to take next steps in their career   
	
	\textbf{C.4} Visibility: Ensure that decision-makers outside the team are aware of talented employees
	
	\textbf{C.5} Encouragement: Encourage employees to pursue new job opportunities throughout the organization
	
	\textbf{C.6} Information about job opportunities: Notify employees of job vacancies in other areas of the organization
	\vspace{0.3cm}
	
	
	\noindent \textbf{D.} What are the reasons that may prevent leaders from investing time and effort towards their employees’ career development? Please select all that apply. 
	
	\textbf{D.1} Leaders’ lack of knowledge about career development opportunities, trainings etc.
	
	\textbf{D.2} Limited resources to invest in employee development 
	
	\textbf{D.3} Risk of losing talent, because developing employees makes them more attractive to others
	
	\textbf{D.4} Need to prioritize short-term targets over long-term employee development
	
	\textbf{D.5} Lack of interest on the part of employees
	\vspace{0.3cm}

	\noindent \textbf{E.} Under which circumstances would leaders be more likely to support employees’ career development?  Please select all that apply. 
	
	\textbf{E.1} If the employee would develop within the team
	
	\textbf{E.2} If replacing employees who leave the team would be easier
	
	\textbf{E.3} If leaders who develop talent were to receive more visibility and recognition
	
	\textbf{E.4} If leaders who develop talent were more likely to be promoted 
	
	\textbf{E.5} If leaders who develop talent would receive financial incentives
	
	\vspace{0.3cm}
	
	\noindent \textbf{F.} Please rate your agreement with the following statements. \{I strongly agree, I agree, I do not agree, I totally do not
	agree\}  
	
	\textbf{F.1} ``A leader’s direct intervention (e.g., encouragement, increasing visibility) has a large impact on employees’ career development.'' 
	
	\textbf{F.2} ``Leaders’ impact on employees’ long-term career development is valued at least as much as their impact on team performance.'' 
	
	\textbf{F.3} ``Talent development entails a conflict of interest for leaders, because more developed employees are more likely to leave the team.'' 
	\vspace{0.3cm}
	
	\noindent \textbf{G.} How often might leaders find themselves in situations where they need to dissuade a team member from exploring opportunities in another department due to immediate team needs or performance goals?
	
	\textbf{G.1} Never
	
	\textbf{G.2} Sometimes
	
	\textbf{G.3} Often
	
	\textbf{G.4} Very often
	
\vspace{0.3cm}
	
	\noindent \textbf{H.}   At the end of the survey, we would like to ask for your expert assessment.  Based on your experience, which specific actions or strategies would be most effective in assisting leaders in their role? \{Free-text response\}

  \vspace{0.3cm}
	
	\noindent   Finally, we would like to ask for some background information about you, for statistical purposes.

\vspace{0.3cm}
	\noindent \textbf{I.}  Are you female?
    
\textbf{I.1} 	Yes

\textbf{I.2} 	No

\vspace{0.3cm}
	\noindent \textbf{J.}  How old are you?
    
\textbf{J.1} 	Below 30

\textbf{J.2}	30-39

\textbf{J.3}	40-49

\textbf{J.4} 50 or above

\vspace{0.3cm}
	\noindent \textbf{K.}  How many years have you been at [Company]? 
    
\textbf{K.1} 	Less than 1

\textbf{K.2} 	1-2

\textbf{K.3} 	3-5

\textbf{K.4} 	6-9

\textbf{K.5} 	10 or more

\vspace{0.3cm}
	\noindent \textbf{L.}  How many years of leadership responsibility at [Company] do you have?
    
\textbf{L.1} 	Less than 1

\textbf{L.2} 	1-2 

\textbf{L.3} 3-5

\textbf{L.4} 	6-9

\textbf{L.5} 	10 or more

\vspace{0.3cm}
	\noindent \textbf{M.} How big is the team that you currently lead? 
    
\textbf{M.1} 	3 or less people

\textbf{M.2}	4-5 people

\textbf{M.3}	6-9 people

\textbf{M.4}	10+ people

\vspace{0.3cm}
	\noindent \textbf{N.} What best describes your functional area? \{Dropdown\}

\vspace{0.3cm}
	\noindent \textbf{O.} 
    Which leadership level best describes your current position? \{Dropdown\}

\newpage
   \subsubsection{Response Distributions for Key Survey Questions}
	\label{sec:app_survey_manager_raw}
This section reports the full response distributions for the key survey questions analyzed in the main text. The tables below present the raw frequencies across answer categories for each item.

\begin{table}[h]
    \centering
    \caption{Perceived Impact of Leaders (\textbf{Question A)}}
    \begin{tabular}{lcccc}
        \hline \hline
        \textbf{Area of Impact} & \textbf{Very High} & \textbf{High} & \textbf{Low} & \textbf{Very Low} \\
        \hline
        \textbf{A.1} Strategic orientation & 0.20 &0.30 &0.36 & 0.14 \\
        \textbf{A.2} Vision &0.17 & 0.28 &0.41& 0.14\\
        \textbf{A.3} Operational efficiency & 0.34 & 0.50& 0.14 & 0.02 \\
        \textbf{A.4} Talent development & 0.40 & 0.47& 0.11 & 0.02 \\
        \textbf{A.5} Culture and morale & 0.55& 0.38 & 0.06 & 0.01 \\
        \textbf{A.6} Innovation & 0.09 &0.48 & 0.37& 0.06 \\
        \hline
        \multicolumn{5}{l}{Observations 3,xxx} \\
        \hline \hline
\multicolumn{5}{p{0.7\linewidth}}{\footnotesize \textit{Notes}: This table presents the share of managers who selected the respective answer choice in response to Question A: \textit{Where do you think leaders can create the most impact for \{Company\}? Please rate the impact that leaders can create.} Each row represents one dimension the respective dimension of interest. Each column presents one of four response options: \textit{"Very high impact"}, \textit{"High impact"}, \textit{"Low impact"}, and \textit{"Very low impact"}. Because respondents could only choose one response option for each dimension, the shares in each row sum to one.}\\
    \end{tabular}
    \label{tab:QA}
\end{table}

\begin{table}[h]
    \centering
    \caption{Record of Success is Important (\textbf{Question B)}}
    \begin{tabular}{lcccc}
        \hline
        \textbf{Area of Impact} & \textbf{Very High} & \textbf{High} & \textbf{Low} & \textbf{Very Low} \\
        \hline
        \textbf{B.1} Strategic orientation & 0.29& 0.49 & 0.18 & 0.04 \\
        \textbf{B.2} Vision & 0.19 & 0.46 & 0.28& 0.07 \\
        \textbf{B.3} Operational efficiency & 0.52 & 0.38 & 0.08 & 0.02 \\
        \textbf{B.4} Talent development 
       & 0.16 & 0.24 & 0.45& 0.15 \\
        \textbf{B.5} Culture and morale & 0.52 & 0.33 & 0.11 & 0.04 \\
        \textbf{B.6} Innovation & 0.18 & 0.51 & 0.26 & 0.05 \\
        \hline
        \multicolumn{5}{l}{Observations 3,xxx} \\
        \hline  \hline
        \multicolumn{5}{p{0.7\linewidth}}{\footnotesize \textit{Notes}: This table presents the share of managers who selected the respective answer choice in response to Question B: \textit{When it comes to leaders’ compensation and promotion prospects, how important do you think is a record of success in each of the following?  Please rate the importance.} Each row represents one dimension the respective dimension of interest. Each column presents one of four response options: \textit{"Very important"}, \textit{"Important"}, \textit{"Somewhat important"}, and \textit{"Not important"}. Because respondents could only choose one response option for each dimension, the shares in each row sum to one.}
    \end{tabular}
    \label{tab:QB}
\end{table}

\begin{table}[h]
    \centering
    \renewcommand{\arraystretch}{1.3} 
    \setlength{\tabcolsep}{12pt} 
    \caption{Ways Leaders Can Support Employee Career Development (\textbf{Question C})}
    \begin{tabular}{p{13cm}c}
        \hline
        \textbf{Support Strategy} & \textbf{Selected} \\
        \hline
        \textbf{C.1} Training: Offer and recommend relevant trainings to enhance skills & 0.69 \\
        \textbf{C.2} Open dialogue: Establish regular discussions about career goals & 0.83 \\
        \textbf{C.3} Career planning: Provide structured career guidance &  0.57 \\
        \textbf{C.4} Visibility: Ensure decision-makers recognize talented employees & 0.84 \\
        \textbf{C.5} Encouragement: Motivate employees to explore new opportunities &0.62 \\
        \textbf{C.6} Information: Notify employees of job vacancies within the company &  0.32\\
        \hline
       \multicolumn{2}{l}{Observations 3,xxx} \\
        \hline  \hline
                \multicolumn{2}{p{0.9\linewidth}}{\footnotesize \textit{Notes}: This table presents the share of managers who selected the respective answer  choice in response to Question C: \textit{What are the most effective ways that leaders can support their employees’ career development? Please select all that apply.} Because respondents could choose multiple answers, the shares do not sum to one. }\\
    \end{tabular}
    \label{tab:QC}
\end{table}

\begin{table}[h]
    \centering
    \renewcommand{\arraystretch}{1.3} 
    \setlength{\tabcolsep}{12pt} 
    \caption{Barriers to Leaders Investing in Employee Career Development (\textbf{Question D})}
    \begin{tabular}{p{13cm}c}
        \hline
        \textbf{Barrier} & \textbf{Selected} \\
        \hline
        \textbf{D.1} Lack of knowledge about career development opportunities &  0.54 \\
        \textbf{D.2} Limited resources to invest in employee development & 0.75 \\
        \textbf{D.3} Risk of losing talent if employees become more attractive to others & 0.45\\
        \textbf{D.4} Need to prioritize short-term targets over long-term development &0.66 \\
        \textbf{D.5} Lack of interest on the part of employees & 0.41 \\
        \hline
      \multicolumn{2}{l}{Observations 3,xxx} \\
        \hline  \hline
         \multicolumn{2}{p{0.9\linewidth}}{\footnotesize \textit{Notes}: This table presents the share of managers who selected the respective answer  choice in response to Question D:  \textit{What are the reasons that may prevent leaders from investing time and effort towards their employees’ career development? Please select all that apply.} Because respondents could choose multiple answers, the shares do not sum to one.}\\
    \end{tabular}
    \label{tab:QD}
\end{table}

\begin{table}[h]
    \centering
    \renewcommand{\arraystretch}{1.3} 
    \setlength{\tabcolsep}{12pt} 
    \caption{Circumstances Encouraging Leaders to Support Employee Career Development (\textbf{Question E})}
    \begin{tabular}{p{13cm}c}
        \hline
        \textbf{Circumstance} & \textbf{Selected} \\
        \hline
        \textbf{E.1} If the employee would develop within the team & 0.42 \\
        \textbf{E.2} If replacing employees who leave the team would be easier & 0.68 \\
        \textbf{E.3} If leaders who develop talent were to receive more visibility and recognition & 0.52 \\
        \textbf{E.4} If leaders who develop talent were more likely to be promoted & 0.22 \\
        \textbf{E.5} If leaders who develop talent would receive financial incentives & 0.21 \\
        \hline
   \multicolumn{2}{l}{Observations 3,xxx} \\
        \hline  \hline
     \multicolumn{2}{p{0.9\linewidth}}{\footnotesize \textit{Notes}: This table presents the share of managers who selected the respective answer  choice in response to Question E: \textit{Under which circumstances would leaders be more likely to support employees’ career development?  Please select all that apply.} Because respondents could choose multiple answers, the shares do not sum to one.}\\
    \end{tabular}
    \label{tab:QE}
\end{table}

\begin{table}[h]
    \centering
    \renewcommand{\arraystretch}{1.3} 
    \setlength{\tabcolsep}{5pt} 
    \caption{Agreement with Statements on Leadership and Career Development (\textbf{Question F})}
    \begin{tabular}{lcccc}
        \hline
        \textbf{Statement} & \textbf{Strongly Agree}  & \textbf{Agree} &  \textbf{Disagree} & \textbf{Strongly Disagree} \\
        \hline
        \textbf{F.1} Impact is large & 0.46& 0.50 & 0.03 & 0.01 \\
        \textbf{F.2} Impact is valued & 0.06 &0.30& 0.53 & 0.11 \\
        \textbf{F.3} Conflict of interest & 0.10 & 0.45& 0.38 & 0.07 \\
        \hline
    \multicolumn{5}{l}{Observations 3,xxx} \\
        \hline \hline
   \multicolumn{5}{p{0.85\linewidth}}{\footnotesize \textit{Notes}: This table presents the share of managers who selected the respective answer choice in response to Question F: \textit{Please rate your agreement with the following statements.} Each row represents one of the following three statements. F.1: \textit{"A leader's direct intervention (e.g., encouragement, increasing visibility."} F.2: \textit{"Leaders' impact on employees' long-term career development is valued at least as much as their impact on team performance."} F.3: \textit{"Talent development entails a conflict of interest for leaders, because more developed employees are more likely to leave the team"}. Each column presents one of four response options: \textit{"I strongly agree"}, \textit{"I agree"}, \textit{"I do not agree"}, and \textit{"I totally do not agree"}. Because respondents could only choose one response option for each statement, the shares in each row sum to one.}\\ 
    \end{tabular}
    \label{tab:QF}
\end{table}

\begin{table}[h]
    \centering
    \renewcommand{\arraystretch}{1.3} 
    \setlength{\tabcolsep}{8pt} 
    \caption{Frequency of Leaders Dissuading Team Members from Exploring Other Opportunities (\textbf{Question G})}
    \begin{tabular}{p{8cm}p{3cm}}
        \hline
        \textbf{Response} & \textbf{Selected} \\
        \hline
        \textbf{G.1} Never & 0.25 \\
        \textbf{G.2} Sometimes & 0.62 \\
        \textbf{G.3} Often &0.11 \\
        \textbf{G.4} Very Often & 0.02 \\
        \hline
  \multicolumn{2}{l}{Observations 3,xxx} \\
        \hline  \hline
 \multicolumn{2}{p{0.7\linewidth}}{\footnotesize \textit{Notes}: This table presents the share of managers who selected the respective answer choice in response to Question G: \textit{Where do you think leaders can create the most impact for \{Company\}? Please rate the impact that leaders can create.} Because respondents could only choose one response option for each statement, the shares in each row sum to one.}\\
    \end{tabular}
    \label{tab:QG}
\end{table}

\clearpage
\newpage

  \subsubsection{Benchmarking of Survey Findings}
\label{sec:appendix_benchmark}

Responses to the manager survey suggest that talent development is perceived as a relatively low-priority responsibility within the firm. To contextualize and assess the robustness of this finding, this section presents three benchmarking exercises that help interpret the strength and validity of these perceptions.

\vspace{2mm}
\noindent \textit{Comparison Across Managerial Responsibilities.} The first exercise demonstrates that talent development stands out relative to other manager responsibilities. One potential concern is that the reported low valuation of talent development reflects a general tendency among respondents to assign low importance to all managerial tasks. If so, the fact that 40\% of managers consider talent development important for their career prospects (Question B) might actually signal relatively high valuation. To assess this, I compare responses across six core managerial responsibilities evaluated in Questions A and B of the survey: strategic orientation, vision, operational efficiency, talent development, culture and morale, and innovation (see Appendix Tables~\ref{tab:QA} and~\ref{tab:QB} for response distributions).\footnote{Question A asked managers, ``Where do you think leaders can create the most impact for [Company]?'', while Question B asked ``When it comes to leaders’ compensation and promotion prospects, how important do you think is a record of success?''.}

Panel B of Appendix Figure~\ref{fig:benchmarking_QAB} shows that only 40\% of managers view a record of success in talent development as ``very important'' or  ``important'' for their own promotion and compensation prospects, the lowest of all responsibilities. The next lowest category, ``vision,'' was selected by 66\% of managers. In contrast, 87\% of managers report that they have a ``very high'' or ``high'' impact on talent development (Panel A), producing a 47 percentage point gap between perceived impact and perceived reward. This gap is larger than for any other responsibility. In comparison, the analogous gaps are considerably smaller or even of opposite sign for other responsibilities: operational efficiency (-6 p.p.), team morale (8 p.p.), innovation (-11 p.p.), strategic orientation (-28 p.p.), and vision (-20 p.p.). Talent development thus stands out as the domain where perceived personal impact is strong but perceived organizational valuation is weak.

\vspace{2mm}
\noindent\textit{Comparison Across Question Formats.} 
The second exercise probes robustness across questions to address the concern that results are driven by the wording of certain questions. In addition to asking managers about how much specific aspects of their role are rewarded (Question B), I also included a robustness question that elicits the agreement with the statement ``Leaders’ impact on employees’ long-term career development is valued at least as much as their impact on team performance'' (Question F.2). This question was designed to offer respondents a simpler comparison by focusing on only two aspects of manager responsibilities. Note that team performance is a natural comparison because it is typically the most salient aspect of middle managers' responsibilities. 

\begin{figure}[h]
		\thisfloatpagestyle{plainlower}
		\caption{Perceived Valuation of Talent Development Across Question Formats}
		\centering
		\includegraphics[scale=2.6]{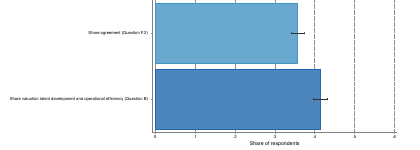}
		\label{fig:benchmarking_QBF}
		\begin{minipage}[b]{0.9\linewidth}
			\footnotesize \textit{Notes}: This figure draws on two separate questions from the manager survey to assess how managers perceive the firm's valuation of their efforts toward talent development. The first bar reports the share of managers who agreed with the statement in Question F.2: ``Leaders’ impact on employees’ long-term career development is valued at least as much as their impact on team performance.'' See Appendix Table \ref{tab:QF} for the full response distribution. The second bar is based on Question B and shows the share of managers who rated talent development as highly valued among those who also rated operational efficiency as highly valued, providing a benchmark for comparison. See Appendix Table \ref{tab:QB} for the full response distribution. Although the questions differ in format and framing, both yield a consistent insight: only a minority of managers perceive talent development to be strongly valued by the firm.  See Appendix Section \ref{sec:app_survey_manager} for more information on the manager survey. 95\%-level confidence intervals are displayed. N=3,xxx.
		\end{minipage}	
	\end{figure}

Only 36\% of respondents agreed with this statement (Appendix Table~\ref{tab:QF}), a share nearly identical to the 40\% who viewed talent development as important in Question B. Among those who believed operational efficiency (as a proxy for team performance) is highly valued, only 41\% said the same about talent development (Appendix Figure~\ref{fig:benchmarking_QBF}).\footnote{Since team performance is not explicitly included in Question B, I use operational efficiency as a proxy. While not identical, the two concepts are closely related, particularly in the context of the manufacturing firm I study, where the performance of many teams is evaluated based on the timely and efficient development of products.}  This consistency across survey formats reinforces the conclusion that managers perceive talent development as undervalued relative to other responsibilities.

To further probe robustness, I also compare the share of managers who agree with the statement ``A leader’s direct intervention (e.g., encouragement, increasing visibility) has a large impact on employees’ career development'' (Question F.1) to the share of managers who chose ``Very high impact'' or  ``High impact'' when rating their impact on talent development in Question A (see Appendix Table \ref{tab:QA} for the raw response distribution). The agreement share for the statement in F.1 is 96\% (Bar 3 of Figure \ref{fig:manager_incentives}), similarly high to the 87\% agreement share in Question A (Bar 5 of Figure \ref{fig:manager_incentives}). This comparison further corroborates the validity of the survey instrument.

\vspace{2mm}
\noindent\textit{Social Desirability Bias and Comparison Across Statements.} A third potential concern is that my estimates are contaminated by response biases, in particular social desirability bias and survey pessimism. However, three factors suggest this is unlikely to meaningfully distort the interpretation of results.

First, managers were not asked about their own values, but rather to assess the firm’s emphasis on different managerial tasks. There is no strong reason to believe that respondents would systematically underreport the firm's valuation of talent development. If anything, one might expect the opposite. Moreover, the survey was framed in neutral language and introduced as anonymous, reducing incentives to tailor responses to social expectations (see Section~\ref{sec:manager_svy_descr}).

Second, comparison across agreement-style questions suggests that managers do not avoid agreement generally. For instance, 96\% of respondents agreed that a leader’s intervention can shape career outcomes (F.1), and 55\% agreed with a more sensitive statement in F.3: ``Talent development entails a conflict of interest for leaders, because more developed employees are more likely to leave the team.'' The fact that a majority agreed with a potentially self-incriminating statement further suggests that social desirability bias is not dominating the pattern of responses.

Third, to the extent that social desirability bias is present, it likely leads to an understatement of talent hoarding behaviors or perceived organizational misalignment. Thus, any remaining bias would make the observed results conservative rather than overstated.

 	\begin{figure}[h]
		\thisfloatpagestyle{plainlower}
		\caption{Perceived Impact and Valuation Across Managerial Responsibilities}
		\centering
		\includegraphics[scale=2.9]{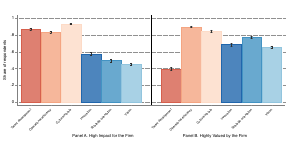}
		\label{fig:benchmarking_QAB}
		\begin{minipage}[b]{0.9\linewidth}
			\footnotesize \textit{Notes}: This figure summarizes managers’ perceptions of how their efforts with respect to talent development are rewarded versus how much impact they believe they can have on the firm, based on responses from the manager survey. Panel A draws on Question A of the survey (see Appendix Table~\ref{tab:QA} for the full question wording and response distribution). Each bar shows the share of managers who report that leaders can have a very high or high impact on the firm with respect to a given managerial responsibility. Panel B draws on Question B of the survey (see Appendix Table~\ref{tab:QB}). Each bar shows the share of managers who report that a record of success in the respective responsibility is considered very important or important for their own compensation and promotion prospects.
 See Appendix Section \ref{sec:app_survey_manager} for more information on the manager survey. 95\%-level confidence intervals are displayed. N=3,xxx. 
		\end{minipage}	
	\end{figure}

	\clearpage
	\newpage
	\section{Supplementary Results and Robustness}
	\label{sec:appendix_robustness}
	\setcounter{table}{0}
	\renewcommand{\thetable}{\thesection\arabic{table}}
	\setcounter{figure}{0}
	\renewcommand\thefigure{\thesection\arabic{figure}}

This appendix provides supplementary analyses and robustness checks that support the empirical findings presented in the main text. It is organized in three parts.  First, I provide additional detail and validation for the managerial talent hoarding measure, which is based on deviations between private performance ratings and public potential ratings.  Second, I present a series of robustness checks for the manager-level analysis discussed in Section~\ref{sec:managers}. These include alternative measures of hoarding incentives, hoarding outcomes, and placebo tests using unrelated organizational barriers. Third, I examine the robustness of the worker-level results discussed in Section~\ref{sec:instrument}, including sensitivity analyses of the rotation design and a discussion of alternative mechanisms that could plausibly explain the observed patterns.

\subsection{Construction and Validity of the Talent Hoarding Measure}
\label{sec:robust_measure}
This appendix section provides further detail and validation for the measure of managerial talent hoarding introduced in Section \ref{sec:managers}. As discussed in the main text, the measure builds on the firm’s internal performance and potential ratings, leveraging the institutional difference that performance ratings are meant to be private, while potential ratings are shared broadly and are meant to signal a worker’s readiness for promotion.

\begin{figure}[!htb]
			\caption{Manager-Reported Ways to Impact Workers' Careers}
		\centering	\includegraphics[scale=3.0]{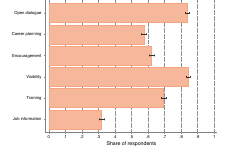}
		\label{fig:manager_talentdevelopment}
		\vspace{0.3cm}
		\begin{minipage}[b]{1.0\linewidth}
			\footnotesize \textit{Notes}: This figure uses responses from the manager survey to depict manager-reported incentives and actions with respect to talent development. Panel A shows the share of managers who indicated that a given condition would make managers more likely to support workers' career development. Each bar shows the share of managers who chose a given action as an effective way through which managers can impact their workers' career development. Managers were allowed to choose multiple answers.  See Appendix Table \ref{tab:QC} for the
full question wording and response distribution. See Appendix Section \ref{sec:app_survey_manager} for more information on the manager survey and the exact question wording.  95\%-level confidence intervals are displayed.   Controls: Female, age, German citizenship, firm tenure, function. N=3,xxx.
		\end{minipage}	
	\end{figure}

\textit{Motivation.---} Identifying managerial talent hoarding is challenging because hoarding often occurs through interpersonal interactions and is not directly observable in administrative data.  The empirical strategy in this paper leverages measures of worker visibility to infer talent hoarding based on the systematic suppression of publicly observable signals of worker talent. This approach is motivated by survey evidence from both workers and managers. Workers identify suppression of visibility as a key channel through which managers engage in hoarding behavior (see Section~\ref{sec:survey}). Consistent with this notion, 84\% of managers report that making workers visible outside the team is among the most important actions they can take to influence career advancement (Appendix Figure~\ref{fig:manager_talentdevelopment}).\footnote{Other response options included establishing an open dialogue with workers about their career development (84\%), allowing workers to participate in trainings (69\%), encouraging employees to pursue other job opportunities (62\%), providing guidance for career planning (58\%), and notifying workers about internal job opportunities outside the team (32\%). Since there is no objective benchmark for how to interpret these percentages, this question provides qualitative evidence that supports the conceptual relevance of visibility-related actions.} Survey responses also suggest that such managerial actions are highly consequential: workers who report being subject to talent hoarding in the form of retaliation are 5.1 percentage points (56\%) less likely to receive a job recommendation from their manager ($p$=0.000), and 4.6 percentage points (13\%) less likely to receive a recommendation from anyone in the firm ($p$=0.000). These findings motivate the use of visibility-related assessments as a meaningful way to detect strategic hoarding behavior.

\textit{Constructing Manager-Level Deviations.---} Building on the institutional difference in visibility between performance and potential ratings, I construct a manager-level measure of talent hoarding by estimating a model of expected potential ratings based on observed performance ratings and worker characteristics:
	
	\vspace{-0.3cm}
	\begin{align}\label{eqn:potential}
		\text{potential}_{imt} &=  \beta_1\text{performance}_{imt} + \beta_X X_{it}+\beta_t + \varepsilon_{imt} 
	\end{align}
where $\text{potential}_{imt}$ is the potential rating given by manager $m$ to worker $i$ in quarter $t$, $\text{performance}_{imt}$ is the corresponding performance rating, $X_{it}$ includes worker demographics and job controls, and $\beta_t$ captures quarter fixed effects. The residuals from this regression capture deviations between actual and predicted potential ratings. A residual below zero indicates that the manager assigns higher potential ratings than expected given the worker’s performance and characteristics, suggesting that the manager may be inflating visibility. A residual above zero implies that the manager systematically assigns lower-than-expected potential ratings, consistent with suppressing visibility and engaging in talent hoarding behavior. I compute the average residual at the manager level across all workers and quarters, which provides a time-invariant measure of each manager’s tendency to systematically underrate or overrate worker potential. The median manager is observed during 75\% of my sample period. For those, who are present during the entire sample period, the median number of worker ratings that is used for identifying hoarding is over 70, enhancing its reliability and reducing noise. See Panel A of Appendix Figure \ref{fig:relationships} for summary statistics of how performance and potential ratings are distributed in my sample. 

\begin{figure}[!ht]
		\thisfloatpagestyle{plainlower}
		\caption{Robustness to Alternative Measures of Manager Hoarding Propensity}
		\begin{minipage}[b]{0.8\linewidth}
			\centering
			\includegraphics[scale=1.0]{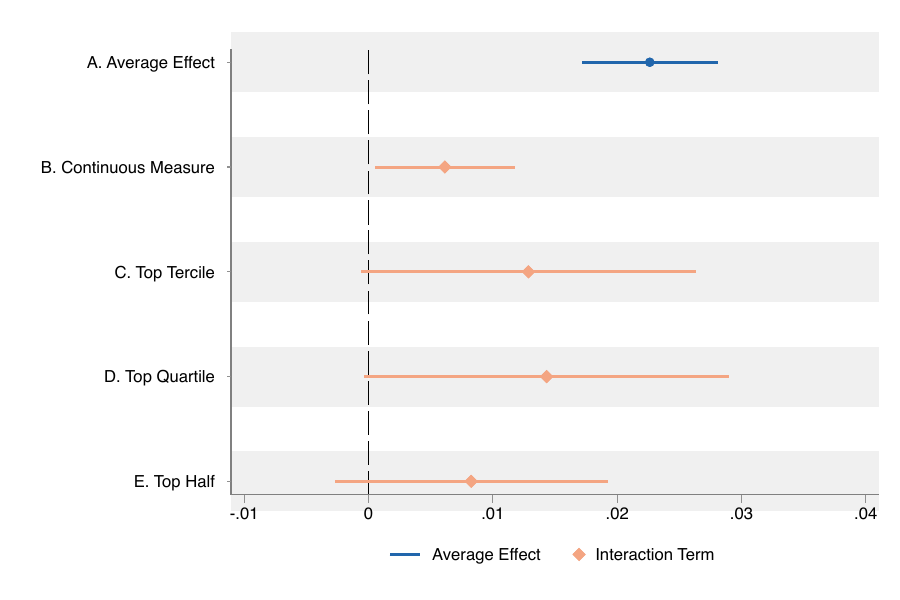}
		\end{minipage}	\\
		\label{fig:rob_hoard}	
		\begin{minipage}[b]{1.0\linewidth}
			\footnotesize \textit{Notes}:  This figure examines whether the effect of manager rotations on worker applications varies consistently across alternative specifications of the manager hoarding propensity measure. The outcome is an indicator for whether a worker applied to an internal job in a given quarter. Each coefficient stems from a separate regression based on Equation~\ref{eqn:fs}. Panel A shows the baseline average effect of a manager rotation (in blue). The orange coefficients show the interaction of the rotation effect with different versions of the manager hoarding propensity measure. Panel B uses the continuous version of the hoarding measure, defined as the average deviation between predicted and actual potential ratings. Panels C to E show results when discretizing this variable. Panel C uses an indicator for whether the manager is in the top tercile of the distribution, Panel D uses the top quartile, and Panel E uses the top half. In each case, the orange point represents the interaction of the rotation indicator with the respective indicator for being hoarding prone. Robust standard errors are clustered at the worker and rotation level. 95\%-level confidence intervals are shown. Controls include: gender, age, German citizenship, education, marital and family status, parental leave, firm tenure, division, function, location, full-time status, hours worked, number of direct reports, and quarter fixed effects. N = 3xx,xxx.
		\end{minipage}	
	\end{figure}

\textit{Interpretation as Measure of Talent Hoarding.---}  Figure~\ref{fig:talenthoarding_distribution} shows the distribution of the manager-level residual means. The distribution is centered around zero (mean = 0.018, median = –0.006) with substantial variation (SD = 0.212). This pattern indicates that while many managers do not systematically deviate from predicted values, a sizable subset consistently underrates or overrates worker potential. To analyze meaningful variation in managerial behavior, I classify managers in the top tercile of the hoarding measure (mean deviation above 0.1036) as hoarding-prone based on the visibility suppression this measure detects.  I use this definition in the main analysis and verify that results are not sensitive to the specific cutoff choice. Appendix Table \ref{tab:hoarding_predictions_outcome}, Appendix Table  
\ref{tab:hoarding_predictions_training_rob}, and Appendix Figure
\ref{fig:rob_hoard}	 confirm that findings are robust when using the continuous hoarding measure, a median split, or an alternative quantile-based definition.

Appendix Table 	\ref{table:measure_summ} complements the information provided by Figure~\ref{fig:talenthoarding_distribution} and summarizes the distribution of the hoarding measures, showing the mean, median, and standard deviation for managers who fall into the bottom, middle, and top tercile of the distribution. These summary statistics further underscore that while most managers do not suppress talent visibility, a substantial share does systematically shade down worker talent in public ratings.

Managers with stronger hoarding tendencies are significantly less likely to publicly recognize top-performing workers. The estimates from the linear probability model of actual potential ratings on performance ratings and worker characteristics suggests that a manager with a hoarding measure of zero has a 66.5\% likelihood of assigning high potential to a worker who received the top performance rating. In contrast, for a manager whose hoarding measure is one standard deviation above the mean and who is thus hoarding-prone, this likelihood is 22.5 percentage points lower. This difference illustrates the substantial difference in potential ratings that is captured by the constructed hoarding measure. Panel B of Appendix Figure \ref{fig:relationships} corroborates this notion and shows that across all performance ratings managers at higher deciles of the hoarding measure are less likely to rate their workers as having high potential. 

\textit{Validity of the Measure.---} To validate the measure, I assess its stability and predictive value.  First, I show that managers’ mean deviation is reasonably stable over time. When using earlier years to estimate a manager’s mean deviation, its correlation with the manager’s deviation based on later years is strongly positive (0.58). This finding suggests the measure captures persistent managerial behavior rather than temporary deviations or measurement error.

Several additional tests suggest that my measure indeed captures managerial hoarding actions. First, the underrating of potential that the measure captures does not appear to result from managers’ involuntary mistakes. I find no statistically detectable difference in managers’ hoarding probability based on the extent of experience they have in leading (and rating) a team. The underrating also does not appear to result from managers’ accurate assessment of worker potential. When managers with high propensities to hoard talent rotate, underrated workers not only experience increases in applications and hiring, but are also likely to perform well in subsequent positions, demonstrating that the low potential rating was inaccurate. In addition, while low potential ratings could in theory stem from the fact that managers have an incentive to hire low-potential workers to avoid the possibility of losing talent, this is not confirmed in the data. I find that talent hoarding effects occur both for incumbent workers and workers who are newly hired by a rotating manager (Columns 3 and 4 of Appendix Table 	\ref{table:mechanism_internalexternal_reg}) . 


A further placebo test examines whether the hoarding measure has predictive power in settings where talent hoarding should be less pronounced. This test helps rule out the concern that the hoarding measure simply reflects broader unobserved differences in managerial style or quality. If the hoarding measure were just a proxy for such general managerial traits, we would expect to see effects on a wide range of worker outcomes, including external job exits. Under the interpretation that managers suppress potential ratings to limit internal visibility and deter job applications, however, we would expect this behavior to affect workers’ internal, but not external career transitions, which are generally invisible to current managers and less subject to interference. Appendix Table \ref{table:measure_robustness} supports this: manager rotations have strong effects on internal applications and job transitions for managers with a high propensity to hoard talent (Columns 1 and 3), but no discernible effect on external job transitions out of the firm (Column 5). This placebo result strengthens the interpretation of the hoarding measure as capturing targeted managerial efforts to retain talent within the team, rather than reflecting unrelated confounding factors.

Taken together, this appendix section supports the validity of the talent hoarding measure by demonstrating that it is stable, systematically varies across managers, and is not confounded by noise or managerial skill. 

\FloatBarrier
\begin{figure}[!ht]
    \thisfloatpagestyle{plainlower}
    \caption{Correlations Between Potential, Performance, and Hoarding Measure}
    \centering
    \begin{minipage}[t]{0.49\linewidth}
        \centering
        \caption*{Panel A. Potential by Performance Rating}
        \includegraphics[width=\linewidth]{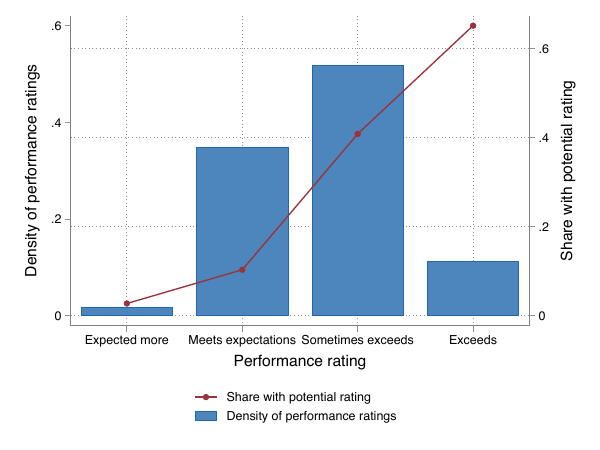}
    \end{minipage}
    \hfill
    \begin{minipage}[t]{0.49\linewidth}
        \centering
        \caption*{Panel B. Potential by Hoarding Decile}
        \includegraphics[width=\linewidth]{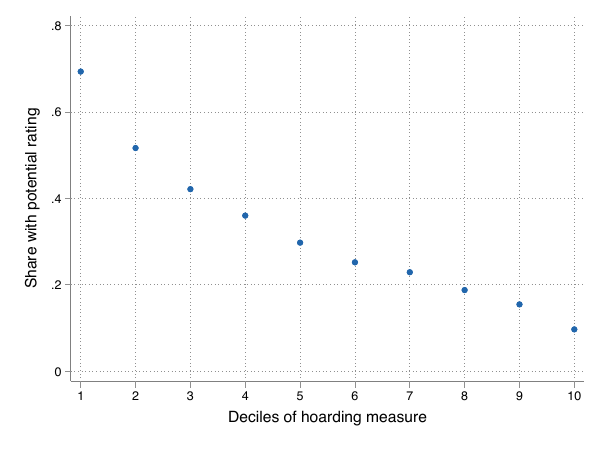}
    \end{minipage}
    \vspace{1em}
    
    \begin{minipage}[b]{\linewidth}
        \footnotesize \textit{Notes}: This figure presents the distributions of and correlations between performance ratings, potential ratings, and the hoarding measure based on the deviation between actual and predicted potential ratings. Panel A presents the density of performance ratings (left axis) and the share of workers with a potential rating for a given performance rating. Note that potential rating is a binary variable. Panel B presents the share of workers with a potential rating by decile of their manager's hoarding measure. N=3xx,xxx.
    \end{minipage}
    
    \label{fig:relationships}
\end{figure}

\begin{table}[!ht]
		\caption{Summary Statistics for Manager-Level Talent Hoarding Measure }
		\begin{center}
			\scalebox{0.8}{{
\def\sym#1{\ifmmode^{#1}\else\(^{#1}\)\fi}
	\begin{tabular}{lccc}
\hline  \hline 
& Mean & Median & Std. Dev. \\
\hline 
Bottom Tercile &-.2217&   -.1941&    .1049  \\
Middle Tercile & -.0016&   -.0061&    .0567 \\
Top Tercile &   .2578&    .2188&     .1363 \\
\hline 
Full Sample & .0102&   -.0061&    .2115 \\
\hline\hline \multicolumn{4}{p{0.45\linewidth}}{\footnotesize \textit{Notes}: This table summarizes the distribution of the manager-level talent hoarding measure, defined as the average residual between actual and predicted potential ratings from Equation~\ref{eqn:potential}. Managers are grouped into terciles based on this measure: the bottom tercile (inflating visibility) includes managers with residuals below –0.099, the middle tercile includes residuals between –0.099 and 0.1036, and the top tercile (suppressing visibility) includes managers with residuals above 0.1036. The table reports the median, mean, and standard deviation of the hoarding score in each group. See Figure~\ref{fig:talenthoarding_distribution} for the full distribution and Appendix Section~\ref{sec:robust_measure} for additional details. Sample includes N = 6,xxx managers.}
\\
\end{tabular}
}

}
		\end{center}
		\label{table:measure_summ}
	\end{table}

\begin{table}[!ht]
		\caption{Placebo Test: Manager Hoarding Measure and Career Outcomes}
		\begin{center}
			\scalebox{0.8}{{
\def\sym#1{\ifmmode^{#1}\else\(^{#1}\)\fi}
\begin{tabular}{l*{6}{c}}
	\hline\hline
	&\multicolumn{2}{c}{Panel A}  &\multicolumn{2}{c}{Panel B}  &\multicolumn{2}{c}{Panel C}  \\
&\multicolumn{2}{c}{Internal Applications}  &\multicolumn{2}{c}{Internal Transition}  &\multicolumn{2}{c}{External Transitions}  \\
		&\multicolumn{1}{c}{High} &\multicolumn{1}{c}{Low} 	 &\multicolumn{1}{c}{High} &\multicolumn{1}{c}{Low}	&\multicolumn{1}{c}{High}	 &\multicolumn{1}{c}{Low} \\
		&\multicolumn{1}{c}{(1)}         &\multicolumn{1}{c}{(2)}         &\multicolumn{1}{c}{(3)}         &\multicolumn{1}{c}{(4)}         &\multicolumn{1}{c}{(5)}         &\multicolumn{1}{c}{(6)}         \\
	\hline
	Manager Rotation    & 0.0344  &  0.0144 & 0.0166  & 0.0101   & -0.0014   &    0.0020      \\
	&     (0.006)          &    (0.005)        &    (0.004)      &      (0.003)            &     (0.002)        &     (0.002)           \\
	\hline
	Outcome Mean        &        0.029       &       0.029   &     0.008          &           0.008  &       0.008   &          0.008    \\

Observations        &       3xx,xxx         &       3xx,xxx             &      3xx,xxx    &  3xx,xxx    &       3xx,xxx         &       3xx,xxx       \\
\hline\hline \multicolumn{7}{p{0.9\linewidth}}{\footnotesize \textit{Notes}: This table presents a placebo test evaluating whether the talent hoarding measure predicts outcomes in settings where hoarding behavior should not matter. Each coefficient stems from a separate regression based on Equation \ref{eqn:fs}. Each panel compares rotations of managers with high versus low propensity to hoard, defined based on the top and bottom terciles of the manager-level hoarding score distribution. The outcomes of interest are internal applications (Panel A), internal job transitions within the firm (Panel B), and external job transitions out of the firm (Panel C). Robust standard errors are clustered at the worker and rotation level. Controls: Female, age, German citizenship, educational qualifications, marital status, family status, parental leave, firm tenure, division, function, location, full-time, hours, number of direct reports, and quarter fixed effects.  }\\
\end{tabular}
}
}
		\end{center}
		\label{table:measure_robustness}
	\end{table}

\clearpage
\newpage

\subsection{Robustness of Manager-Level Results} 
\label{sec:appendix_manger}

Section \ref{sec:predictions_managers} analyzes whether talent hoarding behaviors---measured using both administrative data and the manager survey---are more prevalent among managers with stronger incentives to hoard talent. This appendix section provides additional detail on the construction of key variables and presents a series of robustness checks to validate the main findings.

\subsubsection{Measures of Hoarding Incentives in the Administrative Data} 
\label{sec:hoarding_incentives_details}

This subsection provides additional details on the construction of the three baseline measures of managers’ incentives to hoard talent. It also presents robustness checks based on alternative versions of these measures.

As outlined in the conceptual framework in Section \ref{sec:framework}, hoarding behavior is expected to be more likely when:
(1) managers have stronger financial incentives tied to team performance,
(2) the cost of losing team members is higher (e.g., in smaller teams), and
(3) the visibility of individual talent is lower. I proxy for variation in these dimensions using the following measures:

\textbf{Financial Incentives.} I construct two measures of managers’ financial incentives. The primary measure, used in the main analysis, is the share of total compensation that is performance-related, based on the firm’s internal compensation records. This continuous variable reflects the relative strength of incentive alignment. As a robustness check, I also use the logarithm of the total amount of performance-related pay, which captures the absolute magnitude of the incentive. Additionally, I apply a categorical measure based on the firm's internal leadership classification system. This system categorizes managerial positions into those with high versus low shares of compensation tied to team performance, and is used by the firm for internal purposes.

\textbf{Team Size.} To proxy for the cost associated with worker departures, I use the number of team members overseen by a manager. The main analysis uses team size in levels. I also test robustness to using the logarithm of team size, as well as a binary indicator for managing a small team (bottom quartile of the distribution, defined as three or fewer members).

\textbf{Talent Visibility.}  I proxy for talent visibility using the firm’s internal classification of functional areas into high- and low-visibility categories. High-visibility functions include engineering, quality management, and project management, where individual performance (e.g., through patents or milestone completions) tends to be externally observable. In contrast, low-visibility functions—such as HR, finance, and logistics—offer fewer externally verifiable signals of individual contributions.

Appendix Table \ref{tab:incentives_summ} reports summary statistics for the three baseline measures used in the main specifications. The table shows substantial heterogeneity in hoarding incentives across managers. On average, managers receive roughly one-third of their compensation through performance-related pay, with a quarter receiving 50\% or more. The average team size is seven, and about 25\% of managers oversee teams of three or fewer. Just under half of managers work in low-visibility functions, highlighting that many operate in contexts where individual talent is less externally observable.

As a complement to the main results in Table \ref{fig:hoarding_predictions}, which rely on my baseline measures of hoarding incentives, Appendix Table \ref{tab:incentives_reg} presents regression estimates using alternative constructions of the incentive variables. Across all four columns, the estimated relationships are directionally consistent with the baseline results and quantitatively comparable when scaled for comparability, as I discuss below.

Columns 1 and 2 of Appendix Table \ref{tab:incentives_reg} present alternative measures of financial incentives. Column 1 of Appendix Table \ref{tab:incentives_reg} uses a binary indicator for whether a manager is classified by the firm as having a high share of  compensation based on team performance. Managers with high performance-related pay are 9.4 percentage points more likely to be hoarding-prone, consistent with the direction and magnitude of the baseline result in Table \ref{fig:hoarding_predictions}, where the predicted difference between the 90th and 10th percentile of the continuous measure is 13.0 percentage points. Column 2 of Appendix Table \ref{tab:incentives_reg}  uses the logarithm of performance-related pay and finds that a 1\% increase in pay is associated with a 0.041 percentage point increase in hoarding propensity, yielding a predicted difference between the 90th and 10th percentile of 8.3 percentage points.

Columns 3 and 4 of Appendix Table \ref{tab:incentives_reg}  focus on alternative proxies for the cost of worker departure. Column 3 of Appendix Table \ref{tab:incentives_reg}  uses a binary indicator for leading a small team (three or fewer members) and finds an 11.4 percentage point increase in hoarding propensity, similar to the predicted difference between the 90th and 10th percentile of 13.1 percentage points for the continuous team size measure in Table \ref{fig:hoarding_predictions}. Column 4 of Appendix Table \ref{tab:incentives_reg} uses the log of team size and finds that a 1\% increase in team size reduces the probability of hoarding-prone behavior by 0.10 percentage points, with a predicted  difference between the 90th and 10th percentile of 16.6 percentage points, again reflecting a similarly sized effect (note that the sign difference in the reported estimate reflects stronger incentives to hoard in smaller teams).

Overall, these results confirm that the main findings are robust to alternative specifications of the hoarding incentive measures. Both the signs and magnitudes are consistent with the baseline results presented in Table \ref{fig:hoarding_predictions}, supporting the conclusion that managerial incentives are systematically associated with observed hoarding behavior.

	\begin{table}[!h]
		\caption{Summary Statistics for Baseline Measures of Talent Hoarding Incentives}
		\bigskip
		\begin{center}
			\scalebox{1.0}{{
\def\sym#1{\ifmmode^{#1}\else\(^{#1}\)\fi}
\begin{tabular}{l*{1}{cccccc}}
\hline\hline
                                        &      Mean &    Std.dev     & p10&  p25  & p75 &p90 \\
\hline
Performance-related share of compensation             &  0.36&        0.24&   0.12    & 0.18&        0.49  &   0.81   \\
Number of workers in team  &        6.71&        3.98&  2.35 &      3.71&        8.83 & 12.18\\

Binary indicator for low visibility       &        0.45&        0.50& 0.00 &       0.00&        1.00 & 1.00 \\
\hline
Observations    &\multicolumn{6}{c}{6,xxx}        \\
\hline\hline
\multicolumn{7}{p{0.9\linewidth}}{\footnotesize \textit{Notes}: This table presents manager-level summary statistics for the three baseline measures of talent hoarding incentives: (1) the share of  total compensation that is performance-related, (2) the number of team members overseen by the manager, and (3) a binary indicator for low talent visibility, defined as working in a functional area with limited visibility of individual achievements.Talent visibility is classified according to the firm's internal categorization of functional areas. High-visibility functions include engineering, quality management, and project management, while low-visibility functions include areas such as human resources, finance, and logistics. See Appendix Section \ref{sec:hoarding_incentives_details} for details on the construction of these incentive measures. }\\
\end{tabular}
}

}
		\end{center}
	 \label{tab:incentives_summ}
	\end{table}

	\begin{table}
		\caption{Robustness to Alternative Incentive Measures}
		\bigskip
		\begin{center}
			\scalebox{1.0}{{
\def\sym#1{\ifmmode^{#1}\else\(^{#1}\)\fi}
\begin{tabular}{l*{2}{cc}}
	  \hline  \hline 
\underline{Outcome variable} & \multicolumn{4}{c}{Top tercile hoarding measure} \\ \cline{2-5}

                                       &\multicolumn{1}{c}{(1)}           &\multicolumn{1}{c}{(2)}     &\multicolumn{1}{c}{(3)}     &\multicolumn{1}{c}{(4)}              \\
\hline
 \\[0.1mm] 
1\{High financial incentive\}     &  0.0940   &    --& -- & -- \\
                            &    (0.014)         &    --& -- & -- \\      
                            \\[0.1mm]       

Log(financial incentive amount)  & --   &    0.0410 & -- & --\\
                            & --   &  (0.008)    & -- & --\\
                                       \\[0.1mm]  
                                       
1\{Small team\}     & -- & -- &    0.1144   &    --\\
  & -- & -- &     (0.014)     &  --   \\       
\\[0.1mm]       

Log(team size)      & -- & -- &  --   &    -0.1006\\
  & -- & -- &  --   &   (0.011)    \\       
\\[0.1mm]       
\hline
Outcome Mean            & 0.3000              &    0.3000         & 0.3000              &    0.3000                \\
Predicted difference p90-p10  & 0.0940& 0.0832 & 0.1144 & -0.1656 \\
Observations        &       6,xxx    &    6,xxx      &       6,xxx    &    6,xxx            \\

\hline \hline					\multicolumn{5}{p{0.8\linewidth}}{\footnotesize \textit{Notes}:This table examines the robustness of the main findings on the relationship between managerial hoarding incentives and hoarding behavior by using alternative measures of hoarding incentives.  Column 1 uses a binary indicator based on the firm’s internal classification of whether a manager receives a high share of compensation tied to team performance. Column 2 uses the logarithm of the total amount of performance-based pay to capture the absolute magnitude of financial incentives.
Column 3 uses a binary indicator for managers leading a small team, defined as three or fewer direct reports (bottom quartile of the team size distribution). Column 4 uses the logarithm of the number of direct reports. Each coefficient is estimated separately based on a linear model for managers’ hoarding propensity using an OLS regression of the indicator that managers are hoarding-prone, defined by their mean deviation between predicted and actual potential ratings being in the top tercile of the manager distribution, on one incentive measure at a time. The predicted difference  p90-p10 refers to the predicted difference in the outcome between the 90th and 10th percentiles of the continuous regressors (between values 1 and 0 for binary indicators), holding all other covariates at their mean values. See Table \ref{fig:hoarding_predictions} for my main results using the baseline incentive measures. See Appendix Section \ref{sec:hoarding_incentives_details} for definitions of the incentive measures and Appendix Table \ref{tab:incentives_summ} for their distribution in the sample. Controls: Female, age, German citizenship, firm tenure, educational qualification. Robust standard errors in parentheses.}
 \end{tabular}}

}
		\end{center}
	 \label{tab:incentives_reg}
	\end{table}

\clearpage
\newpage
    \subsubsection{Measures of Hoarding Actions Based on Potential Ratings}  
\label{sec:appendix_hoardingactions_cutoff}

In the main analysis in Section \ref{sec:predictions_managers}, I define a manager as hoarding-prone if the deviation between their actual and predicted potential ratings places them in the top tercile of the manager-level distribution (see Figure \ref{fig:talenthoarding_distribution}). This classification identifies managers who systematically underrate their workers in public potential ratings.

To assess the robustness of this outcome definition, Appendix Table \ref{tab:hoarding_predictions_outcome} presents results using three alternative specifications of the hoarding outcome. First, Columns 1 to 3 use a more selective threshold, classifying a manager as hoarding-prone if their deviation falls in the top quartile of the distribution. Columns 4 to 6 use a broader definition, focusing on whether the deviation exceeds the median. Finally, Columns 7 to 9 avoid discretization altogether by using the continuous deviation between actual and predicted potential ratings.

Across the different specifications, results remain consistent in sign, significance, and economic magnitude with those in the baseline model (Table \ref{fig:hoarding_predictions}). For example, in the main specification, a 1 percentage point increase in the performance-related share of compensation increases the probability of being classified as hoarding-prone by 0.19 percentage points (Column 1 of Table \ref{fig:hoarding_predictions}). In Appendix Table \ref{tab:hoarding_predictions_outcome}, this effect is similar across alternative outcome definitions: 0.16 percentage points using the top quartile threshold (Column 1), 0.24 percentage points using the above-median threshold (Column 4), and an increase of 0.091 units in the continuous hoarding measure (Column 7). Given that the standard deviation of the continuous hoarding measure across managers is 0.212 (see Appendix Table \ref{table:measure_summ}), the estimated effect of 0.091 units corresponds to approximately 0.43 standard deviations, which is roughly comparable to the magnitude implied by the binary classification.\footnote{For a rough comparison, the distance between the median of the distribution of the continuous hoarding measure (–0.006) and the top-tercile cutoff (0.1036) is approximately 0.11 units, or 0.52 standard deviations.}

Comparable patterns emerge for the other two incentive proxies. Larger team size is consistently associated with lower hoarding behavior.  For instance, in my main specification, a one-person increase in team size is associated with a 1.33 percentage point decrease in the probability of being hoarding-prone (Column 2 of Table \ref{fig:hoarding_predictions}), which is similar to the 1.42 percentage point decrease using the top quartile threshold (Column 2 of  Appendix Table \ref{tab:hoarding_predictions_outcome}). The magnitudes of the coefficients are smaller when using the above-median or continuous outcome, but the sign and significance persist (Columns 5 and 8 of  Appendix Table \ref{tab:hoarding_predictions_outcome}) Likewise, managers in low-visibility functional areas consistently exhibit higher hoarding behavior, with coefficients between 0.0143 and 0.0420 (Columns 3, 6, and 9 of Appendix Table \ref{tab:hoarding_predictions_outcome}), similar in direction and magnitude to the baseline estimate of 0.0397 (Column 3 of Table \ref{fig:hoarding_predictions}).

Taken together, these results confirm that the main findings are not sensitive to the specific cutoff or outcome construction. The sign, statistical significance, and the economic magnitude of the estimated relationships are broadly aligned across alternative definitions of the hoarding outcome. As an additional robustness check, Section \ref{sec:appendix_manager_training} presents results using an alternative measure of hoarding behavior based on managers’ suppression of access to high-visibility training opportunities.

\begin{landscape}
	\begin{table}
		\caption{Robustness Checks Using Alternative Hoarding Outcomes Based on Potential Ratings}
		\bigskip
		\begin{center}
			\scalebox{1.0}{{
\def\sym#1{\ifmmode^{#1}\else\(^{#1}\)\fi}
\begin{tabular}{lccccccccccc}
	  \hline  \hline

\underline{Outcome variable}:  &\multicolumn{3}{c}{Top quartile}  &  &\multicolumn{3}{c}{Above median}  &  &\multicolumn{3}{c}{Continuous} \\ \cline{2-4} \cline{6-8} \cline{10-12}
                                       &\multicolumn{1}{c}{(1)}           &\multicolumn{1}{c}{(2)}         &\multicolumn{1}{c}{(3)}  & &\multicolumn{1}{c}{(4)}  &\multicolumn{1}{c}{(5)}  &\multicolumn{1}{c}{(6)}  & &\multicolumn{1}{c}{(7)}  &\multicolumn{1}{c}{(8)}  &\multicolumn{1}{c}{(9)}        \\
\hline
 \\[0.1mm] 
Financial incentive (share)      &  0.1578     &  -- &  -- &&0.2424  &  -- &  -- &&  0.0912   &  -- &  -- \\
                            &  (0.031)        & -- &   --  & & (0.034) && -- & -- & (0.015)   &  -- &  --  \\       
                            \\[0.1mm]      
Team size      &  --   &  -0.0142 &  --&& -- & -0.0075 & -- & & -- & -0.0027 & --\\
                            &  --       &  (0.001)    &   -- & & --&  (0.002)   & --& & --& (0.001)  & --  \\       
                            \\[0.1mm]                               
Talent visibility     &  --    &  --&  0.0420 && -- & -- &  0.0341 & & --& -- & 0.0143 \\
                            & --         &   -- &  (0.012) & & -- & -- & (0.014)  & & -- & -- & (0.006) \\       
                           
\hline
Outcome Mean            &   0.2501                &  0.2501                    &     0.2501  &    &   0.4999                &  0.4999                      &  0.4999   &     &     0.0102           &   0.0102                  &      0.0102         \\
Predicted difference p90-p10  & 0.1085 & -0.1395 &0.0429 &&  0.1667 & -0.0736 &0.0341 &&  0.0627 & -0.0266 &0.0143 \\
Observations        &       6,xxx    &    6,xxx        &   6,xxx  &  &       6,xxx    &    6,xxx        &   6,xxx  &  &       6,xxx    &    6,xxx        &   6,xxx        \\

\hline \hline					\multicolumn{12}{p{0.9\linewidth}}{\footnotesize \textit{Notes}: This table examines the robustness of the main findings on the relationship between managerial hoarding incentives and hoarding behavior by using alternative specifications of the outcome variable capturing managerial hoarding behavior.
 Columns 1 to 3 use a binary indicator for whether the deviation between a manager’s actual and predicted potential ratings falls in the top quartile of the distribution, rather than the top tercile used in the baseline analysis (Table \ref{fig:hoarding_predictions}). Columns 4 to 6 use a binary indicator for whether the deviation is above the median. Columns 7 to 9 use the continuous deviation between actual and predicted potential ratings as the outcome variable. Each coefficient is estimated separately based on a linear regression of hoarding propensity on one incentive measure at a time. The incentive variables include the share of performance-based compensation (Columns 1, 4, and 7), team size (Columns 2, 5, and 8), and an indicator for low talent visibility (Columns 3, 6, and 9). The predicted difference  p90-p10 refers to the predicted difference in the outcome between the 90th and 10th percentiles of the continuous regressors (between values 1 and 0 for binary indicators), holding all other covariates at their mean values. See Table \ref{fig:hoarding_predictions} for baseline results using the top-tercile indicator as the outcome. See Appendix Section \ref{sec:hoarding_incentives_details} for definitions of the incentive measures and Appendix Table \ref{tab:incentives_summ} for their distribution in the sample. Controls: Female, age, German citizenship, firm tenure, educational qualification. Robust standard errors in parentheses.}
 \end{tabular}}

}
		\end{center}
	 \label{tab:hoarding_predictions_outcome}
	\end{table}
\end{landscape}
 
\clearpage
\newpage
    \subsubsection{Measures of Hoarding Actions Based on Training Assignments} 
    \label{sec:appendix_manager_training}

This section parallels the robustness analyses in the preceding appendix sections but uses an alternative measure of managerial hoarding actions: whether managers restrict access to high-visibility training programs. I first describe how this hoarding measure is constructed and then assess robustness to alternative definitions of both outcome and regressors.

\textbf{Construction of Hoarding Measure.}
In each quarter, roughly 2\% of workers in my sample are assigned by their manager to a high-visibility training. These trainings are prestigious in-person sessions organized by a centralized and well-recognized unit within the firm. They enhance a worker's visibility beyond the team: participant lists are widely circulated internally, and workers often list these trainings on their CVs. Survey responses suggest that access to these programs is commonly used as a mechanism through which managers can suppress worker visibility. For example, one worker describes ``[Supervisors] keep employees in their positions by preventing further development, rejecting training courses, and increasing workloads to prevent capacity for new tasks and development measures'' (Appendix Table \ref{table:survey_talenthoarding}).  

To capture this dimension of talent hoarding, I follow the approach in Section \ref{sec:th_measure} and estimate an OLS regression that predicts training assignment based on performance ratings and observable worker characteristics (Equation \ref{eqn:potential}), replacing the potential rating as the dependent variable. For each manager, I then compute the average deviation between actual and predicted training assignments across all workers and quarters. This approach  yields a manager-level hoarding measure. The distribution of this score is approximately centered around zero, with a median of $0.0062$ and a standard deviation of $0.052$. I classify a manager as hoarding-prone if this average deviation lies in the top tercile of the distribution (mean deviation above $0.021$), which is designed to reflect systematic under-assignment of workers to visibility-enhancing trainings. 

\textbf{Robustness to Alternative Incentive Measures.}
Appendix Table \ref{tab:incentives_training_reg} presents regression results using the training-based hoarding outcome while varying the construction of the incentive measures. 
Columns 1 and 2 of Appendix Table \ref{tab:incentives_training_reg}  present alternative measures of financial incentives. Column 1 of Appendix Table \ref{tab:incentives_training_reg}  uses a binary indicator for whether a manager is classified by the firm as having a high share of compensation based on team performance. Managers with high performance-related pay are 15.8 percentage points more likely to be hoarding-prone, consistent with the direction and magnitude of the baseline result in Appendix Table \ref{tab:hoarding_predictions_training}, where the predicted difference between the 90th and 10th percentile of the continuous performance-related pay measure is 13.8 percentage points. Column 2 of Appendix Table \ref{tab:incentives_training_reg}  uses the logarithm of performance-related pay and finds that a 1\% increase in pay is associated with a 0.040 percentage point increase in hoarding propensity, yielding a predicted difference between the 90th and 10th percentile of 8.2 percentage points.

Columns 3 and 4 of Appendix Table \ref{tab:incentives_training_reg} focus on alternative proxies for the cost of worker departure. Column 3 of Appendix Table \ref{tab:incentives_training_reg} uses a binary indicator for managing a small team (defined as three or fewer direct reports) and finds a 7.9 percentage point increase in hoarding propensity, somewhat smaller than, but directionally aligned with, the predicted difference between the 90th and 10th percentile of 14.0 percentage points for the baseline specification using the continuous measure of team size in Appendix Table \ref{tab:hoarding_predictions_training}. Column 4 of Appendix Table \ref{tab:incentives_training_reg} uses the logarithm of team size and finds that a 1\% increase in team size reduces the probability of hoarding-prone behavior by 0.098 percentage points, with a predicted difference between the 90th and 10th percentile of 16.1 percentage points, similar to the magnitude in the baseline specification (i.e., 14.0 percentage points). Note that the sign difference in the reported estimate reflects stronger incentives to hoard in smaller teams.

\textbf{Robustness to Alternative Outcome Definitions.}
Appendix Table \ref{tab:hoarding_predictions_training_rob} presents regressions using three alternative definitions of the training-based hoarding outcome. Columns 1 to 3 use a more selective threshold (top quartile), Columns 4 to 6 use a broader threshold (above median), and Columns 7 to 9 use the continuous deviation as the outcome.

Across alternative outcome specifications, the results remain consistent in direction, statistical significance, and economic magnitude relative to the baseline estimates in Appendix Table \ref{tab:hoarding_predictions_training}. For example, in the main model, a 1 percentage point increase in the share of performance-related compensation is associated with a 0.20 percentage point increase in the likelihood of being hoarding-prone (Column 1 of Appendix Table \ref{tab:hoarding_predictions_training}). In Appendix Table \ref{tab:hoarding_predictions_training_rob}, the corresponding effect is 0.22 percentage points when using a top quartile threshold (Column 1). The effects using the other two outcomes are smaller, but consistent in direction. Based on the above-median definition of the outcome (Column 4), I find a 0.05 percentage points effect. Using the continuous hoarding measure, I find a 0.0086 unit increase (Column 7). Given that the standard deviation of the continuous score is 0.0521, this estimate corresponds to approximately 0.17 standard deviations.\footnote{The distance between the median of the continuous hoarding measure based on training assignments (0.0062) and the top-tercile threshold (0.021) is 0.0148 units or 0.28 standard deviations, suggesting that the effect is roughly comparable to the magnitude implied by the binary classification.}

Results using the other two incentive proxies (i.e., team size and talent visibility) also yield estimates that are consistent with the baseline analysis. Larger team size is negatively associated with hoarding behavior in both cases: the predicted difference between the 90th and 10th percentile is –0.1334 when using the quartile cutoff (Column 2 of Appendix Table \ref{tab:hoarding_predictions_training_rob}) and -0.1267 when using the median (Column 5 of Appendix Table \ref{tab:hoarding_predictions_training_rob}), closely matching the baseline estimate of –0.1403 (Column 2 of Appendix Table \ref{tab:hoarding_predictions_training}). Similarly, managers in low-visibility functions show consistently higher hoarding behavior, with coefficients between 0.0261 and 0.0306 across binary specifications (Columns 3 and 6 of Appendix Table \ref{tab:hoarding_predictions_training_rob}).

Taken together, using a definition of talent hoarding based on training assignments rather than potential ratings yields similar results, corroborating the main analysis in Section \ref{sec:managers}.

	\begin{table}
		\caption{Robustness to Alternative Incentive Measures Using Training Assignment as Alternative Hoarding Outcome}
		\bigskip
		\begin{center}
			\scalebox{1.0}{{
\def\sym#1{\ifmmode^{#1}\else\(^{#1}\)\fi}
\begin{tabular}{l*{2}{cc}}
	  \hline  \hline 
\underline{Outcome variable} & \multicolumn{4}{c}{Top tercile training measure} \\ \cline{2-5}

                                       &\multicolumn{1}{c}{(1)}           &\multicolumn{1}{c}{(2)}     &\multicolumn{1}{c}{(3)}     &\multicolumn{1}{c}{(4)}              \\
\hline
 \\[0.1mm] 
1\{High financial incentive\}     &  0.1576    &    --& -- & -- \\
                            &    (0.014)         &    --& -- & -- \\      
                            \\[0.1mm]       

Log(financial incentive amount)  & --   &    0.0402& -- & --\\
                            & --   &  (0.008)    & -- & --\\
                                       \\[0.1mm]  
                                       
1\{Small team\}      & -- & -- &    0.792   &    --\\
  & -- & -- &     (0.014)     &  --   \\       
\\[0.1mm]       

Log(team size)      & -- & -- &  --   &    -0.0978\\
  & -- & -- &  --   &   (0.010)    \\       
\\[0.1mm]       
\hline
Outcome Mean            & 0.3001             &    0.3001         & 0.3001            &    0.3001               \\
Predicted difference p90-p10 & 0.1576& 0.0816 & 0.0792 & -0.1610 \\
Observations        &       6,xxx    &    6,xxx      &       6,xxx    &    6,xxx            \\

\hline \hline					\multicolumn{5}{p{0.8\linewidth}}{\footnotesize \textit{Notes}: This table examines the robustness of the main findings on the relationship between managerial hoarding incentives and hoarding behavior by using both alternative measures of hoarding incentives and an alternative outcome variable. Specifically, the outcome captures whether managers withhold access to high-visibility training opportunities from their workers, as described in Appendix Section \ref{sec:appendix_manager_training}. Column 1 uses a binary indicator based on the firm’s internal classification of whether a manager receives a high share of compensation tied to team performance. Column 2 uses the logarithm of the total amount of performance-based pay to capture the absolute magnitude of financial incentives. Column 3 includes a binary indicator for managers leading a small team, defined as three or fewer direct reports (bottom quartile of the team size distribution). Column 4 uses the logarithm of the number of direct reports. Each coefficient is estimated separately based on a linear model for managers’ hoarding propensity using an OLS regression of the alternative hoarding outcome on one incentive measure at a time. The predicted difference  p90-p10 refers to the predicted difference in the outcome between the 90th and 10th percentiles of the continuous regressors (between values 1 and 0 for binary indicators), holding all other covariates at their mean values. See Appendix Table \ref{tab:hoarding_predictions_training} for the main results using the baseline incentive measures and the same training-based hoarding outcome. See Appendix Section \ref{sec:hoarding_incentives_details} for definitions of the incentive measures and Appendix Table \ref{tab:incentives_summ} for their distribution in the sample. Controls: Female, age, German citizenship, firm tenure, educational qualification. Robust standard errors in parentheses.}
 \end{tabular}}

}
		\end{center}
	 \label{tab:incentives_training_reg}
	\end{table}

\begin{landscape}
	\begin{table}
		\caption{Robustness Checks Using Alternative Hoarding Outcomes Based on  Training Assignment}
		\bigskip
		\begin{center}
			\scalebox{1.0}{{
\def\sym#1{\ifmmode^{#1}\else\(^{#1}\)\fi}
\begin{tabular}{lccccccccccc}
	  \hline  \hline

\underline{Outcome variable}:  &\multicolumn{3}{c}{Top quartile}  &  &\multicolumn{3}{c}{Above median}  &  &\multicolumn{3}{c}{Continuous} \\ \cline{2-4} \cline{6-8} \cline{10-12}
                                       &\multicolumn{1}{c}{(1)}           &\multicolumn{1}{c}{(2)}         &\multicolumn{1}{c}{(3)}  & &\multicolumn{1}{c}{(4)}  &\multicolumn{1}{c}{(5)}  &\multicolumn{1}{c}{(6)}  & &\multicolumn{1}{c}{(7)}  &\multicolumn{1}{c}{(8)}  &\multicolumn{1}{c}{(9)}        \\
\hline
 \\[0.1mm] 
Financial incentive  (share)     &   0.2215   &  -- &  -- &&0.0471  &  -- &  -- &&   0.0086 &  -- &  -- \\
                            & (0.032)       & -- &   --  & & (0.035) & -- & -- & & (0.004)   &  -- &  -- \\       
                            \\[0.1mm]      
Team size      &  --   & -0.0136&  --&& -- &-0.0129& -- & & -- & -0.0004& --\\
                            &  --       &  (0.001)   &   -- & & --&  (0.002)   & --& & --& (0.000)  & --  \\       
                            \\[0.1mm]                               
Talent visibility     &  --    &  --&   0.0306 && -- & -- &     0.0261    & & --& -- &  0.0044 \\
                            & --         &   -- &  (0.012) & & -- & -- & (0.014)  & & -- & -- &  (0.001) \\       
                           
\hline
Outcome Mean            &   0.2501                &  0.2501                    &     0.2501  &    &   0.4999                &  0.4999                      &  0.4999   &     &       0.0062          &     0.0062                 &       0.0062       \\
Predicted difference p90-p10  & 0.1524& -0.1334  &0.0306 &&0.0324   &-0.1267  &0.0261&&0.0059   &-0.0039  & 0.0044 \\
Observations        &       6,xxx    &    6,xxx        &   6,xxx  &  &       6,xxx    &    6,xxx        &   6,xxx  &  &       6,xxx    &    6,xxx        &   6,xxx        \\
\hline \hline				\multicolumn{12}{p{0.9\linewidth}}{\footnotesize \textit{Notes}: This table examines the robustness of the main findings on the relationship between managerial hoarding incentives and hoarding behavior by using alternative specifications of the outcome variable, based on managers’ assignment of workers to high-visibility training opportunities. Columns 1 to 3 use a binary indicator for whether the deviation between a manager’s actual and predicted training assignments falls in the top quartile of the distribution, rather than the top tercile used in the baseline analysis (Appendix Table \ref{tab:hoarding_predictions_training}). Columns 4 to 6 use a binary indicator for whether the deviation is above the median. Columns 7 to 9 use the continuous deviation between actual and predicted training assignments as the outcome variable. Each coefficient is estimated separately based on a linear regression of hoarding propensity on one incentive measure at a time. The incentive variables include the share of performance-based compensation (Columns 1, 4, and 7), team size (Columns 2, 5, and 8), and an indicator for low talent visibility (Columns 3, 6, and 9). The predicted difference  p90-p10 refers to the predicted difference in the outcome between the 90th and 10th percentiles of the continuous regressors (between values 1 and 0 for binary indicators), holding all other covariates at their mean values. See Appendix Table \ref{tab:hoarding_predictions_training} for baseline results using the top-tercile indicator based on the training assignment outcome. See Appendix Section \ref{sec:appendix_manager_training} for more information on the construction of this hoarding measure. See Appendix Section \ref{sec:hoarding_incentives_details} for definitions of the incentive measures and Appendix Table \ref{tab:incentives_summ} for their distribution in the sample. Controls: Female, age, German citizenship, firm tenure, educational qualification. Robust standard errors in parentheses. }	
 \end{tabular}}

}
		\end{center}
\label{tab:hoarding_predictions_training_rob}
	\end{table}
\end{landscape}

\clearpage
 \newpage
    
    \subsubsection{Robustness Tests Based on the Survey Data}   
  \label{sec:robust_incentives_survey}  

 This section complements the preceding analysis based on administrative data by describing the construction and distribution of hoarding incentive measures in the survey data. It also discusses robustness to alternative specifications of these measures.

\textbf{Measures of Hoarding Incentives.} 
Due to data protection constraints, the survey could not directly elicit the share of performance-related compensation for each manager, as is possible in the administrative records. Instead, I rely on survey responses to Question Q.O, which captures each manager’s leadership type. The firm internally uses this classification to determine whether a manager receives a high or low share of compensation tied to team performance, making this measure conceptually comparable to the binary indicator in the administrative analysis (Column 1 of Appendix Table  \ref{tab:incentives_reg}). To construct a more granular continuous measure, I use the administrative data to calculate the average share and total amount of performance-related pay for each combination of leadership type and functional area. I then assign respondents a share of performance pay using their leadership type, which yields 22 unique values ranging from 0.21 to 0.63 (Appendix Table \ref{tab:incentives_summ_survey}).

Team size was reported in categorical buckets to preserve respondent anonymity. I convert these categories to continuous values by assigning midpoint values to each range, and use a value of 13 for the top-coded group (\textit{10 or more}), the average team size in that category from the administrative data. The third incentive measure, which is an indicator for working in a low-visibility functional area, is constructed analogously to the administrative analysis using the firm’s internal function classification. Appendix Table \ref{tab:incentives_summ_survey} reports summary statistics for these three incentive measures. I find substantial variation across managers, which is consistent with the patterns observed in the administrative data.

\textbf{Robustness Exercise.} Appendix Table \ref{table:survey_hoarding_predictions_robust} assesses the robustness of the survey-based findings by varying the definition of managerial hoarding incentives. The two outcome variables capture whether managers engage in hoarding behavior, based on survey questions described in Appendix Tables \ref{tab:QG} and \ref{tab:QD}. Columns 1 to 4 use an indicator for whether respondents report that managers feel the need to dissuade team members from pursuing external opportunities. Columns 5 to 8 use an indicator for whether the risk of losing talent discourages managers from investing in their workers' career development.

Columns 1, 2, 5, and 6 of Appendix Table \ref{table:survey_hoarding_predictions_robust} focus on financial incentives. Columns 1 and 5 use a binary indicator for whether a manager is classified by the firm as having a high share of compensation based on team performance. Managers with high performance-related pay are 8.4 percentage points and 8.5 percentage points more likely to report hoarding actions, broadly consistent with the direction and magnitude of the baseline result in Table \ref{table:survey_hoarding_predictions}, where the predicted difference between the 90th and 10th percentile of the continuous measure was 3.4 percentage points and 4.1 percentage points. Columns 2 and 6 of Appendix Table \ref{table:survey_hoarding_predictions_robust} use the logarithm of performance-related pay and finds that a 1\% increase in performance-related pay is associated with a 0.052 and 0.060 percentage point increase in reported hoarding behavior, yielding a predicted difference between the 90th and 10th percentile of 0.88 percentage points. Note that the magnitude of this effect is smaller likely due to the presence of measurement error in imputing the amount of financial incentive from the administrative data; nonetheless, despite the presence of measurement error, the effect is still positive and statistically significant.

Columns 3, 4, 7, and 8 of Appendix Table \ref{table:survey_hoarding_predictions_robust} focus on alternative proxies for the cost of worker departure. Columns 3 and 7 use a binary indicator for leading a small team (bottom quartile) and find an 8.7 percentage point and 7.7 percentage point increase in reported hoarding behavior, in line with the predicted difference between the 90th and 10th percentile of 4.6 percentage points and 7.5 percentage points for the continuous team size measure in Table \ref{table:survey_hoarding_predictions}. Columns 4 and 8 of Appendix Table \ref{table:survey_hoarding_predictions_robust} use the log of team size and find that a 1\% increase in team size reduces the probability of reported hoarding behavior by 0.033 percentage points and 0.041 percentage points, yielding a predicted difference between the 90th and 10th percentile of –7.2 percentage points and -9.1 percentage points. The sign of the coefficient in Columns 3 and 7 naturally flips, as smaller teams imply stronger incentives to hoard, whereas team size in levels or logs increases with weaker incentives.

Finally, effect sizes are remarkably stable across the two different survey-based measures of hoarding behavior. For example, comparing Column 3 and Column 7 of Appendix Table \ref{table:survey_hoarding_predictions_robust}, both specifications using the small-team indicator yield effects around 0.08. Likewise, the financial incentive variables show similar magnitudes across the two outcomes, reinforcing the robustness of the findings.

While the magnitudes of the estimated effects naturally differ depending on the specification of the incentive measures, I find that the sign and statistical significance of the results are highly robust across all models. The magnitudes are economically similar and consistently aligned with my interpretation that talent hoarding behaviors are more frequent among managers who face stronger incentives to retain talent.

	\begin{table}[!h]
		\caption{Summary Statistics for Survey-Based Measures of Talent Hoarding Incentives}
		\bigskip
		\begin{center}
			\scalebox{1.0}{{
\def\sym#1{\ifmmode^{#1}\else\(^{#1}\)\fi}
\begin{tabular}{l*{1}{cccccc}}
\hline\hline
                                              &      Mean &    Std.dev     & p10&  p25  & p75 &p90 \\
\hline
Performance-related share of compensation             &        0.25&        0.08& 0.22 &      0.22&        0.25 & 0.34\\
Number of workers in team    &        8.72&        4.44&1.50 &        4.50&       13.00& 13.00\\
Binary indicator for low visibility    &        0.37&        0.48& 0.00&         0.00&        1.00& 1.00\\
\hline
Observations    &\multicolumn{6}{c}{3,xxx}        \\
\hline\hline
\multicolumn{7}{p{0.9\linewidth}}{\footnotesize \textit{Notes}: This table reports manager-level summary statistics for the three baseline measures of hoarding incentives based on the survey data: share of managers' total compensation that is performance-related, the size of the team the manager is responsible for, and whether talent visibility is low in the manager's functional area. The share of performance-related compensation is constructed by assigning each respondent the average share for their leadership type and functional area, based on the firm's internal compensation records. Team size is measured as a continuous variable by assigning the midpoint of each categorical range provided in the survey’s four response buckets. Talent visibility is classified according to the firm's internal categorization of functional areas. High-visibility functions include engineering, quality management, and project management, while low-visibility functions include areas such as human resources, finance, and logistics. See Appendix Section  \ref{sec:robust_incentives_survey} for details on the construction of these incentive measures. }\\
\end{tabular}
}
}
		\end{center}
	 \label{tab:incentives_summ_survey}
	\end{table}

\begin{landscape}
    
	\begin{table}
		\caption{Robustness to Alternative Incentive Measures Using Survey Data}
		\bigskip
		\begin{center}
			\scalebox{0.9}{{
\def\sym#1{\ifmmode^{#1}\else\(^{#1}\)\fi}
\begin{tabular}{lccccccccc}
	  \hline  \hline

	 \underline{Outcome variable:}	   &\multicolumn{4}{c}{Dissuade workers from leaving} &   &\multicolumn{4}{c}{Risk of losing talent}   \\ \cline{2-5} \cline{7-10}
                                       &\multicolumn{1}{c}{(1)}           &\multicolumn{1}{c}{(2)}      &\multicolumn{1}{c}{(3)}    &\multicolumn{1}{c}{(4)} &        &\multicolumn{1}{c}{(5)}           &\multicolumn{1}{c}{(6)}      &\multicolumn{1}{c}{(7)}    &\multicolumn{1}{c}{(8)}     \\
\hline
 \\[0.1mm] 
1\{High financial incentive\}     &     0.0841    &   -- & -- & -- && 0.0851 & -- & -- & -- \\
                            &  (0.035)           & -- & -- & -- && (0.047)& -- & -- & --   \\       
                   [0.1mm]       
Log(financial incentive amount)     &    --    &   0.0517 & -- & -- && -- & 0.0597  & -- & -- \\
                            &  --         & (0.023)   & -- & -- && --- & (0.030) & -- & --   \\       
                   [0.1mm]       
1\{Small team\}     &   -- & -- &   0.0873     &  -- && -- & --- & 0.0769 & -- \\
                            &   -- & -- & (0.020)  & --        && -- & -- & (0.025) & --     \\       
                     [0.1mm]     
                     Log(team size)      &    -- & -- & --     &   -0.0333  && -- & -- & -- &-0.0419\\
                            &  -- & -- & -- &       (0.010)   && -- & -- & -- &(0.012)    \\       
                     [0.1mm]

\hline

Outcome Mean            &     0.7506           &       0.7506     &     0.7506           &   0.7506  && 0.4461    & 0.4461    & 0.4461    &    0.4461              \\
Predicted difference p90-p10  & 0.0841 &0.0088  &0.0873 &-0.0719  && 0.0851&0.0102 &0.0769 & -0.0905 \\
Observations        &       3,xxx    &    3,xxx    &       3,xxx    &    3,xxx   &&        3,xxx    &    3,xxx  &       3,xxx    &    3,xxx          \\

\hline \hline					\multicolumn{10}{p{0.8\linewidth}}{\footnotesize \textit{Notes}: This table examines the robustness of the survey-based findings on the relationship between managerial hoarding incentives and hoarding behavior by using alternative measures of hoarding incentives. The outcome variables are based on two distinct survey questions. Columns 1 to 4 use a binary indicator for whether respondents report that managers feel the need to dissuade workers from pursuing external opportunities due to immediate team needs. The survey question reads: \textit{How often might leaders find themselves in situations where they need to dissuade a team member
from exploring opportunities in another department due to immediate team needs or performance
goals?} See Appendix Table \ref{tab:QG} for the distribution of responses. Columns 5 to 8 use a binary indicator based on whether respondents report that the risk of losing talent discourages managers from investing in employee career development. The corresponding question reads: \textit{What are the reasons that may prevent leaders from investing time and effort towards their
employees’ career development?} See Appendix Table \ref{tab:QD} for response distributions. Each coefficient is estimated separately using an OLS regression of the respective outcome on a single incentive measure at a time. Columns 1 and 5 use a binary indicator for receiving a high share of compensation tied to team performance, based on the firm’s internal classification. Columns 2 and 6 use the log of total performance-based compensation. Columns 3 and 7 use a binary indicator for managing a small team (bottom quartile of the team size distribution), and Columns 4 and 8 use the log of team size.
The predicted difference  p90-p10 refers to the predicted difference in the outcome between the 90th and 10th percentiles of the continuous regressors (between values 1 and 0 for binary indicators), holding all other covariates at their mean values. See Table \ref{table:survey_hoarding_predictions} for the main results using the baseline incentive measures. See Appendix Section \ref{sec:robust_incentives_survey} for further detail on the construction of the incentive measures and Appendix Table \ref{tab:incentives_summ_survey} for their distribution in the survey sample. 
Controls: Female, age, German citizenship, firm tenure, educational qualification. Robust standard errors in parentheses.}
 \end{tabular}}

}
		\end{center}
		\label{table:survey_hoarding_predictions_robust}
	\end{table}
\end{landscape}

    \clearpage
 \newpage
    \subsubsection{Placebo Test for Survey-Based Analysis of Talent Hoarding}
    \label{sec:appendix_survey_placebo}

This appendix subsection assesses the robustness of the survey-based measure of talent hoarding used in the main analysis (Panel B of Table \ref{table:survey_hoarding_predictions}). This measure is based on whether managers indicate that they refrain from investing in employee development because of the ``risk of losing talent if employees become more attractive to others.'' This response captures the core logic of talent hoarding: managers may deliberately withhold developmental opportunities to avoid making valuable employees more mobile.

A potential concern is that the proxies I use for hoarding incentives---(1) a higher share of performance-related pay, (2) smaller team size, and (3) low talent visibility in the manager's functional area---could also correlate with broader organizational frictions unrelated to hoarding, such as resource constraints, informational gaps, or short-term performance pressures. To address this, I conduct a placebo test using the same survey question but focus on alternative response options. If my proxies simply capture general managerial challenges, we would expect them to correlate with a broad set of barriers to employee development. Appendix Table \ref{tab:QD} lists all available response options and their frequencies.

Appendix Table \ref{table:survey_placebo} presents the results. In each panel, Column 1 replicates the main result using ``risk of losing talent'' as the outcome. Columns 2 to 5 examine other responses: limited resources, lack of employee interest, lack of knowledge, and ``need to prioritize short-term targets over long-term development.'' Across all panels, I find a robust and specific association between the proxies for managers' hoarding incentives and the ``risk of losing talent'' response. In contrast, there is no consistent relationship between hoarding incentives and the other response options.

Panel A uses the share of performance-related pay as the incentive measure. Managers with higher performance-related compensation are more likely to report talent hoarding behavior. The coefficient for the short-termism response is positive but not statistically significant. This is consistent with the idea that a higher share of performance-related pay may be correlated with greater pressure to prioritize short-term outcomes.\footnote{\citet{garicano2016organizations} show that short-termism arises when firms are unable or unwilling to evaluate long-term contributions, causing managers to focus on immediate metrics (e.g., quarterly targets) at the expense of long-run value. This misalignment affects decisions such as underinvestment in R\&D and is conceptually distinct from talent hoarding. While short-termism results from performance evaluation systems that emphasize short-run outcomes, leading managers to deprioritize long-term value creation, talent hoarding stems from managers' incentive to retain high-performing employees within their own teams, even when those employees could create more value elsewhere in the firm.} For the remaining response options in Panel A, I find negative but statistically insignificant coefficients.

Panel B focuses on team size. Managers of smaller teams are more likely to report hoarding behavior. There is no evidence that team size predicts other development barriers other than a negative association with the ``lack of knowledge'' response. This may reflect the fact that larger teams expose managers to more varied development needs, increasing their awareness of training resources. However, neither of the other two incentive measures (Panels A and B) are correlated with lack of knowledge, suggesting that measures of talent hoarding do not systematically conflate talent hoarding and information constraints. 

Panel C examines talent visibility. Managers in low-visibility functional areas are more likely to report the risk of losing talent as a barrier to development. Interestingly, they are \textit{less} likely to cite other barriers, such as limited resources or lack of employee interest. These patterns suggest that managers in low-visibility areas are especially concerned about retention, relative to other broader organizational constraints. Moreover, these patterns further support the conclusion that my baseline measures of talent hoarding do systematically capture other organizational frictions.

It is also worth noting that short-termism, understood as a consequence of evaluation criteria set by upper management (\citealp{garicano2016organizations}), should not be expected to vary systematically with team size or visibility. These factors shape the incentives to hoard talent but do not determine the time horizon of managerial performance evaluation. Consistent with this, survey responses referencing short-term pressures often cite top-down demands, such as working in ``firefighting mode'' or facing ``quick-win requirements'' from senior leadership as underlying determinants. 

Overall, the placebo analysis supports the interpretation that my survey-based measure captures the relationship between hoarding incentives and hoarding behavior, rather than general organizational barriers.

	\begin{table}[!h]
	\caption{Heterogeneity in Reports of Organizational Barriers by Incentives}
	\bigskip
	\begin{center}
		\scalebox{0.9}{{
\def\sym#1{\ifmmode^{#1}\else\(^{#1}\)\fi}
\begin{tabular}{l*{5}{c}}
	  \hline  \hline 

 \multicolumn{6}{l}{\textbf{Panel A}: Financial incentive}  \\
	 
	   &\multicolumn{1}{c}{Risk of}           &\multicolumn{1}{c}{Prioritize short-}         &\multicolumn{1}{c}{Lack of}           &\multicolumn{1}{c}{Limited}         &\multicolumn{1}{c}{Lack of}    \\
	   	   &\multicolumn{1}{c}{losing talent}           &\multicolumn{1}{c}{term goals}         &\multicolumn{1}{c}{knowledge}           &\multicolumn{1}{c}{resources}         &\multicolumn{1}{c}{employee interest}    \\
	 \hline
                               &\multicolumn{1}{c}{(1)}           &\multicolumn{1}{c}{(2)}         &\multicolumn{1}{c}{(3)}           &\multicolumn{1}{c}{(4)}         &\multicolumn{1}{c}{(5)}    \\
\hline
 \\[0.1mm] 
Financial incentive (share)              &      0.3226 &      0.0912         &     -0.2507         &     -0.2040         &     -0.1701         \\
                    &     (0.156)         &     (0.147)         &     (0.151)         &     (0.147)         &     (0.147)         \\
\hline
Outcome Mean        &      0.4461         &      0.6625         &      0.5394         &      0.7476         &      0.4110         \\

Observations        &       3,xxx    &    3,xxx        &   3,xxx   &    3,xxx         &   3,xxx         \\
\hline
  \\[0.1mm]     
 \multicolumn{6}{l}{\textbf{Panel B}: Team size}  \\  
 &\multicolumn{1}{c}{Risk of}           &\multicolumn{1}{c}{Prioritize short-}         &\multicolumn{1}{c}{Lack of}           &\multicolumn{1}{c}{Limited}         &\multicolumn{1}{c}{Lack of}    \\
&\multicolumn{1}{c}{losing talent}           &\multicolumn{1}{c}{term goals}         &\multicolumn{1}{c}{knowledge}           &\multicolumn{1}{c}{resources}         &\multicolumn{1}{c}{employee interest}    \\
               &\multicolumn{1}{c}{(1)}           &\multicolumn{1}{c}{(2)}         &\multicolumn{1}{c}{(3)}           &\multicolumn{1}{c}{(4)}         &\multicolumn{1}{c}{(5)}    \\
\hline
 \\[0.1mm] 
Team size      &     -0.0065&     -0.0001         &     -0.0050  &      0.0010         &      0.0020         \\
                    &     (0.002)         &     (0.002)         &     (0.002)         &     (0.002)         &     (0.002)         \\
\hline
Outcome Mean        &      0.4461         &      0.6625         &      0.5394         &      0.7476         &      0.4110         \\

Observations        &       3,xxx      &    3,xxx         &   3,xxx         &    3,xxx         &   3,xxx     \\
\hline
\\[0.1mm]     
\multicolumn{6}{l}{\textbf{Panel C}: Talent visibility}  \\  
&\multicolumn{1}{c}{Risk of}           &\multicolumn{1}{c}{Prioritize short-}         &\multicolumn{1}{c}{Lack of}           &\multicolumn{1}{c}{Limited}         &\multicolumn{1}{c}{Lack of}    \\
&\multicolumn{1}{c}{losing talent}           &\multicolumn{1}{c}{term goals}         &\multicolumn{1}{c}{knowledge}           &\multicolumn{1}{c}{resources}         &\multicolumn{1}{c}{employee interest}    \\
&\multicolumn{1}{c}{(1)}           &\multicolumn{1}{c}{(2)}         &\multicolumn{1}{c}{(3)}           &\multicolumn{1}{c}{(4)}         &\multicolumn{1}{c}{(5)}    \\
\hline
\\[0.1mm] 
Talent visibility    &      0.1443 &      0.0753         &     -0.0098         &     -0.1053 &     -0.2462\\
                    &     (0.056)         &     (0.055)         &     (0.055)         &     (0.052)         &     (0.052)         \\
\hline
Outcome Mean        &      0.4461         &      0.6625         &      0.5394         &      0.7476         &      0.4110         \\

Observations        &       3,xxx      &    3,xxx         &   3,xxx         &    3,xxx         &   3,xxx     \\
 
\hline \hline					\multicolumn{6}{p{1.1\linewidth}}{\footnotesize \textit{Notes}:  This table probes robustness of my survey-based analysis of the impact of hoarding incentives on managers' reported talent hoarding by focusing on manager reports of alternative organizational barriers as alternative outcomes. While Panel B of Table \ref{table:survey_hoarding_predictions} shows that stronger hoarding incentives are related with a higher propensity to cite the risk of losing talent in response to the question ``What are the reasons that may prevent leaders from investing time and effort toward their employees’ career development?'', this table expands this analysis to include the other response options: the need to prioritize short-term targets (Column 2), managers' lack of knowledge about development opportunities (Column 3), limited resources (Column 4), and lack of employee interest (Column 5).  See Appendix Table \ref{tab:QD} for the raw response distribution.
Each coefficient is estimated separately based on a linear model for the outcome of interest using an OLS regression on the respective measure of hoarding incentives, which are the share of the performance-related portion of manager compensation (Panel A), team size (Panel B), and an indicator for being in an area with low talent visibility (Panel C).  See Appendix Section   \ref{sec:robust_incentives_survey}   for more information on how these incentive measures were constructed.  Controls: Female, age, German citizenship, firm tenure, educational qualification. Robust standard errors in parentheses.  N=3,xxx.}
 \end{tabular}}
}
	\end{center}
	\label{table:survey_placebo}
\end{table}

\clearpage
 \newpage
 \subsection{Robustness of Worker-Level Results} 
  \label{sec:appendix_robustness_rotation}
Section~\ref{sec:instrument} presents evidence that manager rotations increase workers’ likelihood of applying to internal job openings, consistent with a reduction in talent hoarding. This section probes the robustness of the worker-level findings. In particular, I assess the sensitivity of the results to (i) alternative control sets, (ii) alternative measures of hoarding incentives, and (iii) alternative definitions of worker outcomes. Finally, I explore potential alternative mechanisms that could explain the observed application responses.

     \subsubsection{Alternative Control Sets}  
   
    I first document that my worker-level results are not sensitive to a specific set of covariates. Column 1 of Appendix Table \ref{table:robustness_controls} shows the raw effect of manager rotations on applications and only controls for quarter fixed effects. Column 2 adds demographic controls (i.e., gender, age, German citizenship, educational qualifications, marital status, and family status). Column 3 controls for position characteristics instead of demographics (i.e., function, division, location). Column 4 uses my baseline set of controls, which include in addition to quarter fixed effects female, age, German citizenship, educational qualifications, marital status, family status, parental leave, firm tenure,  division, function, location, full-time, hours, and number of direct reports. Column 5 adds controls for worker evaluations (i.e., performance and potential ratings) and workers' past internal mobility. This exercise demonstrates that the estimated effect of manager rotations remains stable across specifications, regardless of the control set used. The fact that the effect of manager rotations on worker applications does not vary a lot based on the specific set of controls used provides further evidence in support of the assumption that manager rotations appear orthogonal to worker characteristics. 

    \subsubsection{Alternative Measures of Worker Quality} 
I assess whether the heterogeneity in rotation effects across worker subgroups is robust to alternative definitions of worker quality. The main analysis in Figure~	\ref{fig:mechanism_departurecost}	 uses a composite quality index based on predicted internal hiring probabilities. To test robustness, I replicate the analysis using two additional continuous measures: educational attainment (measured in years) and past performance ratings (measured in numbers ranging from 1 to 4). Each measure is standardized to have mean zero and standard deviation one, and I estimate separate regressions interacting the standardized variable with the manager rotation indicator.

\begin{figure}[!ht]
		\thisfloatpagestyle{plainlower}
		\caption{Robustness of Rotation Effects to Alternative Measures of Worker Quality}
		\begin{minipage}[b]{0.8\linewidth}
			\centering
			\includegraphics[scale=1.0]{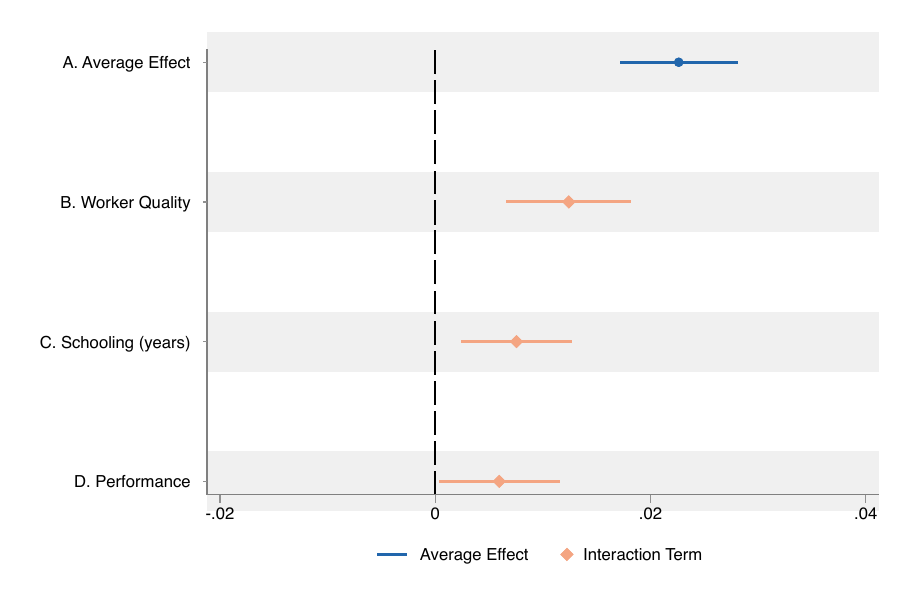}
		\end{minipage}	\\
		\label{fig:rob_workerquality}	
		\begin{minipage}[b]{1.0\linewidth}
			\footnotesize \textit{Notes}:    This figure examines whether the effect of manager rotations on worker applications varies consistently across alternative measures of worker quality. The outcome is an indicator for whether a worker applied to an internal job in a given quarter. Each coefficient stems from a separate regression based on Equation \ref{eqn:fs}.  Panel A shows the baseline average effect of a manager rotation (in blue). 
            The orange coefficients show the effect  of the rotation interacted with a standardized (mean-zero, SD=1) version of the respective proxy for hoarding incentives.   Panel B presents my baseline specification which uses a standardized composite index of worker quality (constructed as the predicted value from an OLS regression of internal hiring probabilities on worker characteristics). Panel C uses a standardized measure of years of schooling. Panel D uses a standardized measure of past performance ratings. Robust standard errors are clustered at the worker and rotation level. 95\%-level confidence intervals are shown. Controls include: gender, age, German citizenship, education, marital and family status, parental leave, firm tenure, division, function, location, full-time status, hours worked, number of direct reports, and quarter fixed effects. N = 3xx,xxx.
		\end{minipage}	
	\end{figure}

Appendix Figure~\ref{fig:rob_workerquality} presents the results. Panel A reports the baseline effect of manager rotations, which increase internal job applications by 2.3 percentage points. Panel B shows that a one-standard-deviation increase in the composite quality index is associated with an additional 1.2 percentage point increase in application likelihood ($p$ = 0.000). Panel C reports a similar result for educational attainment, with a 0.8 percentage point increase ($p$ = 0.004), and Panel D shows that past performance ratings predict an increase of 0.6 percentage points ($p$ = 0.039) following a rotation. Across all specifications, more productive workers experience significantly larger application responses to manager rotations, consistent with the prediction that they face stronger talent hoarding ex ante.

     \subsubsection{Alternative Specifications of Manager Hoarding Propensity Measure} 

     Next, I assess the robustness of the results to alternative ways of measuring managers’ talent hoarding propensity. The main analysis in Figure~\ref{fig:mechanism_departurecost} uses a continuous measure based on the average deviation between actual and predicted potential ratings. To test robustness, I discretize the distribution of this measure at various thresholds: the top tercile, the top quartile, and the top half,  which aligns with the definition of hoarding-prone managers used in Section~\ref{sec:managers}, facilitating comparability across empirical exercises. In each case, I estimate how the effect of manager rotations differs when the departing manager is classified as hoarding prone by including an interaction between the rotation indicator and an indicator for being above the respective threshold.

Across all specifications, the estimated application effects are larger when the departing manager is classified as hoarding prone, supporting the interpretation of this measure as capturing meaningful variation in talent hoarding behavior. While the confidence intervals in the discretized specifications are wider and include zero, reflecting the increased estimation noise, the magnitudes of the estimated effects are broadly comparable to those from the continuous specification. For example, the rotation effect increases by 1.3 percentage points for workers under managers in the top tercile ($p$ = 0.061) and 1.4 percentage points for workers under managers in the top quartile ($p$ = 0.056), versus 0.6 percentage points per standard deviation in the continuous specification ($p$= 0.033).

  \subsubsection{Alternative Measures of Higher-Level Positions}
While the main analysis defines higher-level positions as those associated with a pay increase, I obtain highly similar results when using a content-based definition of job hierarchy, based on the multidimensional ranking developed in \cite{brokenrung}. This alternative measure captures variation in formal responsibility by combining three features of a job: (i) the number of direct reports, (ii) the reporting distance to the CEO, and (iii) the level of managerial autonomy. Appendix Table~\ref{table:results_small} replicates the main analysis using this alternative hierarchy measure.

Across all columns, the magnitude and interpretation of the effects closely mirror the baseline effect. Column 1 of Appendix Table~\ref{table:results_small}  shows that manager rotations increase applications to higher-level positions by 1.13 percentage points when job level is defined by hierarchical rank, which is nearly identical to the baseline effect of 1.06 percentage points using the pay-based definition (Column 1 of Table~	\ref{table:results_all}, Panel B). Similarly, the effect of applying on being hired into a higher-level role is 46.8\% under the alternative hierarchy measure (Column 2 of Appendix Table~\ref{table:results_small} ), compared to 46.2\% in the baseline specification (Column 2 of Table~\ref{table:results_all}, Panel B). Finally, a sizable share of marginal applicants would go on to outperform their new peers: the effect on being hired and performing above the team average is 17.8\% under the alternative hierarchy definition (Column 3 of Appendix Table~\ref{table:results_small} ), versus 20.1\% in the baseline specification (Column 3 of Table~\ref{table:results_all}, Panel B). Taken together, these findings confirm that the core results are robust to using a more a hierarchy-based classification of job level.

	\begin{table}[!ht]
		\caption{Robustness in Application Effects}
		\begin{center}
			\scalebox{1.0}{{
\def\sym#1{\ifmmode^{#1}\else\(^{#1}\)\fi}
\begin{tabular}{l*{5}{c}}
\hline\hline
                 &\multicolumn{1}{c}{No} &   \multicolumn{1}{c}{Employee} &    \multicolumn{1}{c}{Position} & \multicolumn{1}{c}{Baseline}  &   \multicolumn{1}{c}{Baseline +} \\
                 
                 &\multicolumn{1}{c}{Controls} &   \multicolumn{1}{c}{Demographics} &   \multicolumn{1}{c}{Controls} &  \multicolumn{1}{c}{Controls}  &   \multicolumn{1}{c}{Rating + Mobility} \\
                        &\multicolumn{1}{c}{(1)}&\multicolumn{1}{c}{(2)}&\multicolumn{1}{c}{(3)}&\multicolumn{1}{c}{(4)}&\multicolumn{1}{c}{(5)}\\
\hline
Manager Rotation    &      0.0222&      0.0220&      0.0228&      0.0226&      0.0234\\
                    &     (0.003)         &     (0.003)         &     (0.003)         &     (0.003)         &     (0.003)         \\
\hline
Outcome Mean        &      0.0290         &      0.0290         &      0.0290         &      0.0290         &      0.0290         \\
Observations        &      3xx,xxx         &      3xx,xxx        &      3xx,xxx         &      3xx,xxx         &      3xx,xxx         \\

\hline\hline \multicolumn{6}{p{0.9\linewidth}}{\footnotesize \textit{Notes}: This table presents results from robustness exercises using different sets of controls to estimate the effect of manager rotations on worker applications. Column 1 presents the raw estimates, which only control for quarter fixed effects. Column 2 controls for quarter fixed effects and employee demographics, including gender, age, German citizenship, educational qualifications, marital status, and family status. Column 3 controls for quarter fixed effects and for employees' position characteristics (i.e., function, division, location). Column 4 contains my baseline set of controls: Female, age, German citizenship, educational qualifications, marital status, family status, parental leave, firm tenure, division, function, location, full-time, hours, number of direct reports, and quarter fixed effects.  Column 5 adds controls for worker evaluations (i.e., performance and potential ratings) and workers' past internal mobility. Robust standard errors are clustered at the worker and rotation level.}\\
\end{tabular}
}
}
		\end{center}
		\label{table:robustness_controls}
	\end{table}

	\FloatBarrier

	\FloatBarrier

    \begin{table}
		\caption{Robustness to Alternative Definition of Higher-Level Positions}
		\begin{center}
			\scalebox{1.0}{{
\def\sym#1{\ifmmode^{#1}\else\(^{#1}\)\fi}
\begin{tabular}{l*{3}{c}}
	  \hline  \hline 

	 & Applied & Hired & Hired + Perform  \\
  	  	  &\multicolumn{1}{c}{OLS}  	  	  &\multicolumn{1}{c}{IV} &  \multicolumn{1}{c}{IV} \\
                                       &\multicolumn{1}{c}{(1)}           &\multicolumn{1}{c}{(2)}         &\multicolumn{1}{c}{(3)}        \\
\hline
 \\[0.1mm] 
Manager Rotation      &     0.0113&        -             &       -              \\
&     (0.002)         &       -              &       -              \\
                            \\[0.1mm]       
                            
Applied      &           -          &      0.4675&      0.1782 \\
&          -           &     (0.079)         &     (0.060)         \\
                            \\[0.1mm]       
\hline
Outcome Mean       &      0.0118         &      0.0027         &      0.0013         \\
Observations        &       3xx,xxx          &  3xx,xxx        & 3xx,xxx        \\
\hline \hline					\multicolumn{4}{p{0.7\linewidth}}{\footnotesize \textit{Notes}:  This table reports the effects of manager rotations on worker career outcomes using an alternative classification of higher-level positions based on job hierarchy rankings. Positions are classified as higher-level if they have a hierarchy rank of 10 or above, following the hierarchy measure developed in \cite{brokenrung}, which incorporates number of direct reports, reporting distance to the CEO, and managerial autonomy. Each column presents results from a separate regression. Each coefficient is based on a separate regression. Column 1 reports the first-stage effect of manager rotation on applications  based on Equation \ref{eqn:fs}. Column 2  reports estimates from a two-stages least squares regression on landing a position that instruments for applying with manager rotation based on Equation \ref{eqn:iv} which represents the LATE. Column 3 estimates a similar two-stages least squares regression, but use an indicator for landing a position \textit{and} performing  better than the leave-out team average one year later as outcome variable. Robust standard errors are clustered at the worker and rotation level. Controls: Female, age, German citizenship, educational qualifications, marital status, family status, parental leave, firm tenure, division, function, location, full-time, hours, number of direct reports, and quarter fixed effects.}
 \end{tabular}}
}
		\end{center}
		\label{table:results_small}
	\end{table}

    \vspace{-0.3cm}
	 \clearpage
    \newpage
	\subsubsection{Alternative Mechanisms Underlying Rotation Design}
	\label{sec:confounder}
	\vspace{-0.2cm}
Having demonstrated that the observed application effects of manager rotations are robust across a wide range of empirical specifications and measurement approaches, I now turn to an important question: could these effects be driven by mechanisms other than a reduction in talent hoarding?
    
	The analysis in Section \ref{sec:mechanism} is consistent with talent hoarding being a key driver of worker responses to manager rotations; however, it remains possible that worker responses could also reflect factors unrelated to talent hoarding. For instance, managers and workers may jump ship because of bad news about the future outlook of the team, or may depart after the completion of a major milestone. Other alternative mechanisms such as worker loyalty, match effects, and role-model effects could also produce the observed worker responses. In this section, I conduct a series of additional robustness checks to assess the potential importance of such alternative mechanisms. 
	
	\vspace{2mm}
	\textit{Bad news about the future or unpleasant working conditions}.---
	I begin by testing whether a negative shock, such as bad news about the future outlook of the unit or unpleasant working conditions, could underlie the effect of manager rotations on worker applications. Under such a mechanism, one would expect to see effects on other team-level outcomes besides applications, such as team performance and morale. Appendix Figure \ref{fig:rotation_absent} estimates an event study around manager rotations and rejects economically significant changes in key outcomes, including team-level absenteeism and performance-related bonus pay, suggesting that rotations are not correlated with deteriorating conditions within the team. 
	
	Results from two placebo tests further contradict patterns that one would expect from  such shocks. First, under a correlated shock that makes managers and workers want to leave the team, one would expect increased worker applications even for managers' unsuccessful attempts to leave the team. However, Panel A of Appendix Figure \ref{fig:msw_robustness} finds no effect of managers' unsuccessful rotations on worker applications.  Second, because coworkers are more similar in their career opportunities to a worker, one would also expect the correlation between coworker rotations and worker applications in the presence of such a shock to be larger than that between manager rotations and worker applications. Panel B of Appendix Figure \ref{fig:msw_robustness} estimates the impacts of rotations of the most senior teammate and shows that these events have much smaller impacts on worker applications.
	
	In addition, if the documented increase in internal applications is driven by teammates jumping ship, one would expect to see at least some effect on external transitions. However, I do not find that manager rotations yield increases in exit from the firm (Panel B of Figure \ref{fig:msw_dynamics}). Moreover, if teammates want to jump ship because of the long-term future outlook of the team, one should expect the increase in applications to persist for several quarters, particularly given that in the internal labor market, it can take several quarters for suitable vacancies to become available. Instead, the observed application responses are transitory. It is also unclear why workers in smaller teams, those that are harder to replace, and those under hoarding-prone managers would be more susceptible to such a shock, as suggested by the heterogeneity results presented in Figure \ref{fig:mechanism_departurecost}.\footnote{Managers' hoarding propensities are quite stable over time and are unlikely to be driven by a correlated shock.} 
	
	\vspace{2mm}
	\textit{Completion of a milestone or project.}---If most teams were to work on fixed project schedules together, the completion of a milestone could be another key channel for correlated shocks. Several institutional features imply that such milestones are not likely to be relevant in this setting. First, manager rotations are not designed to coincide with increases in job vacancies or team dissolution that would indicate systematic rotation patterns due to common milestones in the firm. Second, the vast majority of positions at the firm are not project-based. Even in areas where project-based work is more common, projects are not normally staffed within a team, but often have employees from different teams in one project, meaning that it is not common that an entire team and their manager will finish a milestone at the same time. To empirically test for this channel, I exclude workers in jobs related to project management (where the bulk of project-based work occurs) from my event study analysis. This robustness test yields very similar application patterns (Panel B of Appendix Figure \ref{fig:rotation_project}), suggesting that my results are not likely to be driven by project cycles. These findings are also consistent with the absence of effects on external transitions. If milestone completion were to drive responses to rotations, one would expect some effect on external transitions.

\begin{figure}[p]
		\thisfloatpagestyle{plainlower}
		\caption{Effect of Manager Rotations by Workers' Exposure Length to Manager}
		\centering
		\includegraphics[scale=0.9]{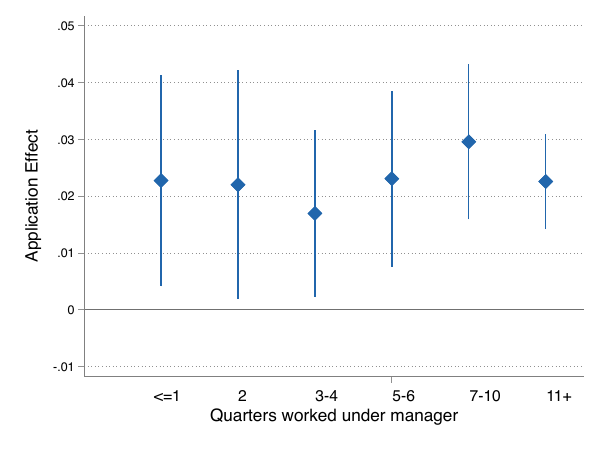}
		\label{fig:talenthoarding_attach}
		\begin{minipage}[b]{0.9\linewidth}
			\footnotesize \textit{Notes}: This figure assesses heterogeneity in the effect of manager rotation on applications by workers' length of exposure to the rotating manager. Each coefficient stems from a separate regression based on Equation \eqref{eqn:fs}.  Worker subgroups are defined by the number of quarters a worker has worked under the manager. Baseline application rates are 0.031 ($\leq$1), 0.029 ($2$), 0.032 (3-4), 0.033 (5-6), 0.030 (7-10), and 0.024 (11+). The total number of observations are 3x,xxx ($\leq$1), 3x,xxx (2), 5x,xxx (3-4), 3x,xxx (5-6), 5x,xxx (7-10) and 1xx,xxx (11+). The coefficients are not statistically distinguishable from each other (the lowest p-value from pairwise comparisons is 0.2180). Robust standard errors are clustered at the worker and rotation level and 95\%-level confidence intervals are displayed. Controls: Female, age, German citizenship, educational qualifications, marital status, family status, parental leave, firm tenure, division, function, location, full-time, hours, number of direct reports, and quarter fixed effects. 
		\end{minipage}	
	\end{figure}

	\vspace{2mm}
	\textit{Loyalty.}---In principle, manager rotations could induce workers to apply if they feel a strong sense of loyalty toward their manager. While loyalty undoubtedly represents an important part of interpersonal relations in the workplace, I do not find evidence that this mechanism drives the particular increase in applications around manager rotations. Loyalty is typically assumed to compound over time; however, Appendix Figure \ref{fig:talenthoarding_attach} finds no evidence that rotation effects vary by the length of time a worker was exposed to the rotating manager. Instead, rotations increase applications even for workers who have been exposed to their manager for one quarter or less. While one could have also expected that loyalty is particularly strong if the worker was personally hired by the manager, Appendix Table \ref{table:mechanism_internalexternal_reg} finds similar effects for workers who were hired by the rotating manager and those who were already in the team. Moreover, one would expect loyalty effects to also result in increases in external transitions. However, Panel B of Figure \ref{fig:msw_dynamics} shows that while internal and external transitions trend identically prior to the rotation, manager rotations do not yield an increase in worker's external transition out of the firm. 
	\vspace{2mm}
	
	\textit{Worker-Manager Specific Match Effects.}---Workers may have refrained from applying prior to a manager rotation because of match effects, which could occur because the manager provided favorable task assignments or offered better training opportunities. Several pieces of evidence suggest, however, that standard match effects cannot explain the increase in applications around a manager rotation. Similar to loyalty, match effects should increase worker applications for positions outside the team because the value of their default option falls when the manager leaves. However, I find no impact of manager rotations on external transitions. Under match effects, one would also expect rotations to be more impactful for workers who were hired by that manager, relative to workers who were already on the team when that manager arrived. This is because the manager would have had the ability to select new workers based on match quality. Instead, Appendix Table \ref{table:mechanism_internalexternal_reg} documents that manager rotations have very similar effects on both groups. While one would also expect that match effects compound over time, I find that application effects of manager rotations do not depend on workers' exposure time to their manager (Appendix Figure \ref{fig:talenthoarding_attach}). Finally, it is difficult to rationalize why match effects would give rise to the observed heterogeneity across different workers, for instance in terms of larger rotation effects for workers under hoarding-prone managers. Evidence discussed in Section \ref{sec:managers} suggests that match effects are likely smaller, not larger, under hoarding-prone managers, for example because they deter workers' participation in trainings. 
	\vspace{2mm}
	
	\textit{Salience and Role-Model Effects.}---Information-based mechanisms could also explain why rotations affect worker applications. For instance, managers who pursue rotations may generate information flows that make career planning more salient. However, since managers' unsuccessful rotation attempts do not increase applications even though workers likely find out about them, such a channel does not appear to operate through the salience of career planning. Successful manager rotations, on the other hand, could increase applications by providing information about the success likelihood of applications. If this were the case, one would expect that the increase in applications would persist, rather than quickly dissipate, given that few vacancies are available each quarter. Similarly, when I focus on manager rotations that are induced by the outgoing manager leaving the firm, which are less likely to increase information or salience about \textit{internal} career transitions, I find similar application patterns relative to my baseline results (Panel A of Appendix Figure \ref{fig:rotation_exit}).  
    
	Role-model effects are often found to be particularly impactful if role models are similar in attributes to affected individuals (\citealp{porter}, \citealp{meier}), but homophily does not appear to play an important role in my setting. First, motivated by the previous literature on homophily by gender, I show that the rotation of a manager with the same gender yields a similar and statistically indistinguishable application response relative to the rotation of a manager with the opposite gender (Panel A of Appendix Table \ref{table:incoming}).  Second, if observing others navigate their career is a key underlying factor of rotation effects, such a role-model effect should not be limited to managers. Since coworkers are more similar to workers than their managers with respect to their career opportunities, one would likely expect larger effects for observing coworkers rotate. However, Panel B  of Appendix Figure \ref{fig:msw_robustness} finds much smaller, rather than larger, effects of coworker rotations relative to manager rotations. 
    These findings indicate that any role-model effects that are at play likely operate through channels that differ from those documented in prior research.

  These additional analyses suggest that several plausible alternative explanations for the observed application effects around manager rotations are unlikely to be the dominant factors in my setting. Instead, the findings are in line with the view that reduced incentives to hoard talent are a primary driver of increased internal job applications.

	\FloatBarrier
	\begin{figure}[!ht]
		\caption{Placebo Tests for Effects of Manager Rotations}
		\centering
		\begin{minipage}[b]{0.8\linewidth}
			\centering
			\caption*{Panel A. Results by Outcome of Manager Application}
			\includegraphics[scale=0.7]{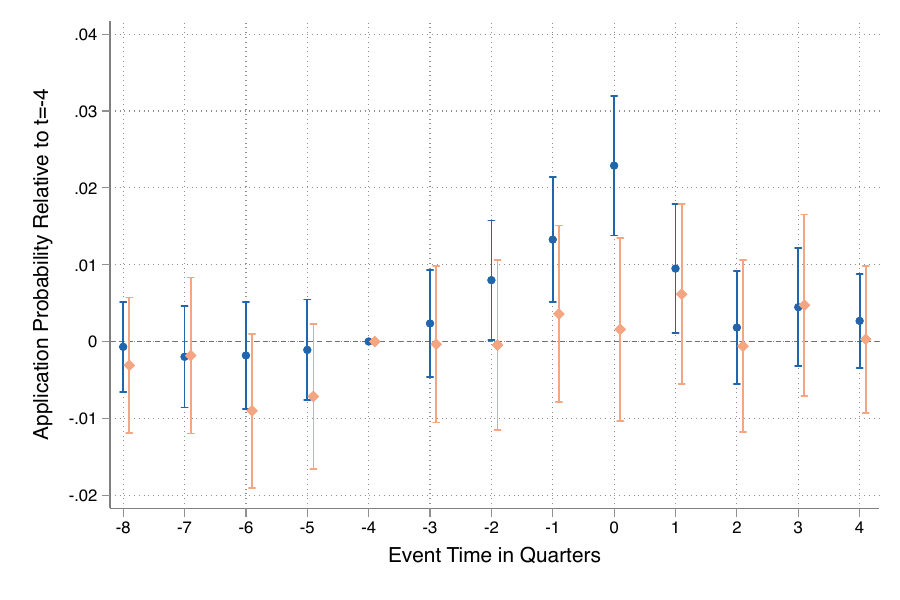} \\
			\vspace{-0.2cm}
			\includegraphics[scale=0.4]{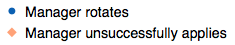}
		\end{minipage}	\\
		\vspace{2mm}
		\begin{minipage}[b]{0.8\linewidth}
			\centering
			\caption*{Panel B. Results by Type of Rotating Teammate}
			\includegraphics[scale=0.7]{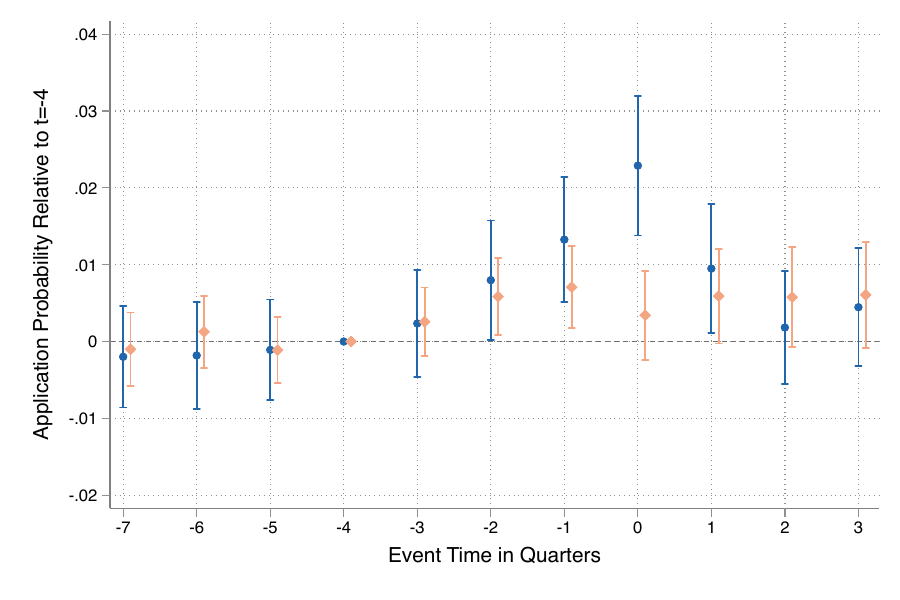} \\
			\vspace{-0.2cm}
			\includegraphics[scale=0.4]{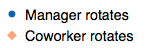}
		\end{minipage}	\\
		\vspace{2mm}
		\label{fig:msw_robustness}
		\begin{minipage}[b]{0.8\linewidth}
			\footnotesize \textit{Notes}: This figure presents placebo tests for manager rotations. Estimates stem from event study regressions, in which the outcome is an indicator that the worker applied to any internal position in a quarter and event time is defined relative to the occurrence of a rotation event. The specification includes worker and quarter fixed effects, but no other controls. I  bin event time dummy variables at $t=-8$ and $t=4$ and cluster standard errors at the worker and rotation level. 95\%-level confidence intervals are displayed. The mean application rate as of $t=-4$ is 0.027.  The sample includes those who have not experienced a manager rotation. Panel A compares a successful manager rotation (in blue) to a placebo event, in which a manager applied for an internal job rotation, but did not land the position and stayed in the team (in orange). Panel B compares a manager rotation  (in blue) to the rotation of the most senior coworker (in orange). 
		\end{minipage}	
	\end{figure}

     \begin{figure}[!ht]
		\thisfloatpagestyle{plainlower}
		\caption{Effects of Alternative Rotation Events}
		\vspace{2mm}
		\centering
		\begin{minipage}[b]{0.5\textwidth}
			\centering
			\caption*{Panel A.Manager Exits as Event}
			\includegraphics[scale=0.55]{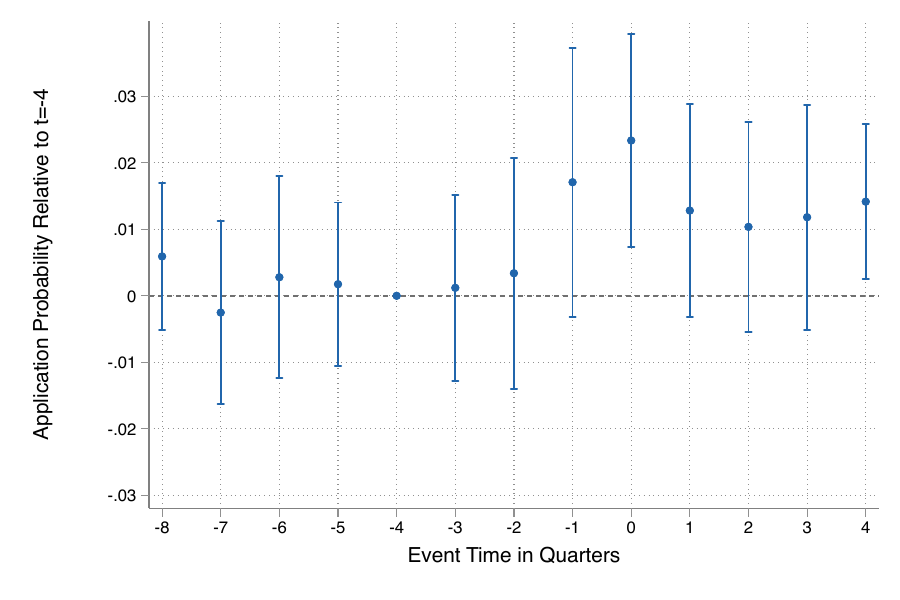}
		\end{minipage}	\\
		\begin{minipage}[b]{0.5\textwidth}
			\centering
			\caption*{Panel B. Sample Without Project Management}
	\includegraphics[scale=0.55]{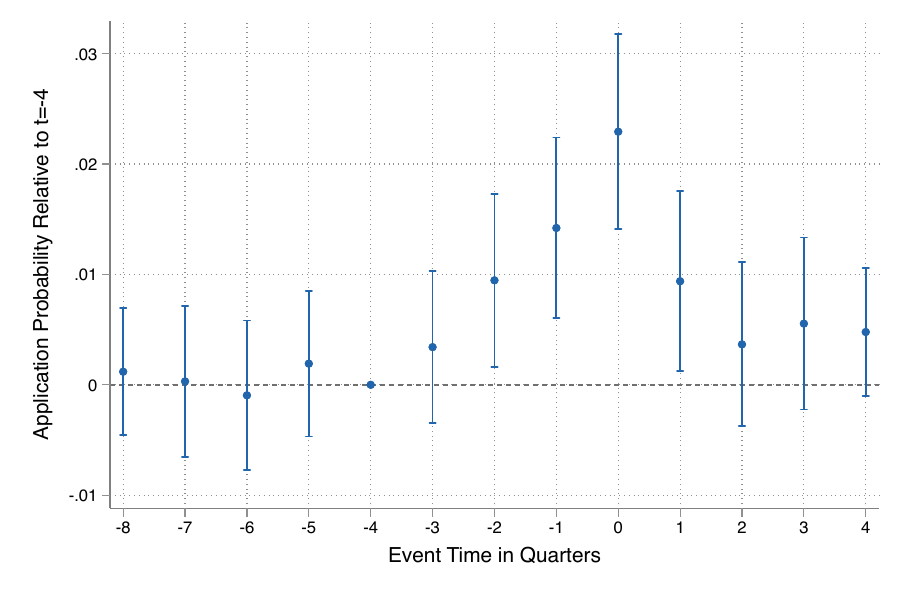} 
		\end{minipage}	
		\label{fig:rotation_exit} 	\label{fig:rotation_project}
		\begin{minipage}[b]{1.0\linewidth}
			\vspace{4mm}
			\footnotesize \textit{Notes}: This figure presents robustness tests for the mechanisms underlying manager rotations. Panel A focuses on manager rotations that were induced by the outgoing manager exiting the firm. Panel B focuses on manager rotations based on internal switches, but excludes workers from the analysis sample who work in project management positions and are most likely to work in projects with fixed milestones. Estimates stem from event study regressions, in which the outcome is an indicator that the worker applied in a quarter and event time is defined relative to the occurrence of a rotation event. The specification includes worker and quarter fixed effects, but no other controls. I  bin event time dummy variables at $t=-8$ and $t=4$ and cluster standard errors at the worker and rotation level. 95\%-level confidence intervals are displayed.  The mean application rate as of $t=-4$ is 0.027.  The sample includes those who have not experienced a manager rotation.
		\end{minipage}	
	\end{figure}

	\FloatBarrier
	 
\end{document}